\documentclass[english]{article}

\usepackage{amsmath}
\usepackage{amsthm}
\usepackage{amsfonts}

\usepackage{geometry}
\usepackage[english]{babel}
\usepackage{array}
\usepackage{float}
\usepackage{booktabs}
\usepackage{multirow}
\usepackage{graphicx}
\usepackage{rotfloat}
\usepackage{setspace}
\usepackage{hhline}
\usepackage{subfiles}
\usepackage{comment}

\usepackage[authoryear]{natbib}

\usepackage[bookmarks=true,bookmarksnumbered=true,bookmarksopen=false,
 breaklinks=false,pdfborder={0 0 0},pdfborderstyle={},backref=false,colorlinks=false]
 {hyperref}
\hypersetup{pdftitle={Regression Discontinuity Aggregation, with an Application to the Union Effects on Inequality},
 pdfauthor={K. Borusyak, M. Kolerman-Shemer},
 hidelinks}

\theoremstyle{plain}
\newtheorem{prop}{\protect\propositionname}
\theoremstyle{plain}
\newtheorem{assumption}{\protect\assumptionname}

\providecommand{\tabularnewline}{\\}
\providecommand{\assumptionname}{Assumption}
\providecommand{\propositionname}{Proposition}

\begin{document}
\title{Regression Discontinuity Aggregation,\\
with an Application to the Union Effects on Inequality}
\author{Kirill Borusyak\\
UC Berkeley\and Matan Kolerman-Shemer\\
The Hebrew University of Jerusalem\thanks{Borusyak: \protect\href{mailto:k.borusyak@berkeley.edu}{k.borusyak@berkeley.edu};
Kolerman-Shemer: \protect\href{mailto:matan.kolerman@mail.huji.ac.il}{matan.kolerman@mail.huji.ac.il}.
We thank Shmuel San for helpful suggestions and Jacob Lefler for excellent
research assistance.}}
\date{December 2024}
\maketitle
\begin{abstract}
We extend the regression discontinuity (RD) design to settings where
each unit's treatment status is an average or aggregate across multiple
discontinuity events. Such situations arise in many studies where
the outcome is measured at a higher level of spatial or temporal aggregation
(e.g., by state with district-level discontinuities) or when spillovers
from discontinuity events are of interest. We propose two novel estimation
procedures — one at the level at which the outcome is measured and
the other in the sample of discontinuities — and show that both identify
a local average causal effect under continuity assumptions similar
to those of standard RD designs. We apply these ideas to study the
effect of unionization on inequality in the United States. Using credible
variation from close unionization elections at the establishment level,
we show that a higher rate of newly unionized workers in a state-by-industry
cell reduces wage inequality within the cell.
\end{abstract}
\noindent
\global\long\def\expec#1{\mathbb{E}\left[#1\right]}%
\global\long\def\var#1{\mathrm{Var}\left[#1\right]}%
\global\long\def\cov#1{\mathrm{Cov}\left[#1\right]}%
\global\long\def\diag{\operatorname{diag}}%
\global\long\def\plim{\operatorname*{plim}}%
\global\long\def\E{\mathbb{E}}%
\global\long\def\Var{\mathrm{Var}}%
\global\long\def\one#1{\textbf{1}\left[#1\right]}%
\global\long\def\i{\textbf{i}}%
\newpage{}

\section{Introduction}

Regression discontinuity (RD) designs are a popular tool for causal
inference. In canonical sharp RD designs, the treatment is determined
by whether a “running variable” exceeds a fixed cutoff. For instance,
one may study whether unionization affects establishment outcomes,
leveraging the cutoff of 50\% votes in a union election \citep{dinardo2004economic,frandsen_2021}.
A local causal parameter is typically estimated via local linear regression
under the assumption that potential outcomes are continuous around
the cutoff. For standard RD methods to apply, each unit has to be
exposed to only one discontinuity.

In this paper, we extend the RD design to settings in which a unit
is exposed to multiple discontinuity events because the outcome is
defined at a higher level of aggregation (e.g., in space or time)
than the elections.\footnote{We use the term “elections” broadly to refer to events that
generate RD variation, as the majority of RD papers we cite are indeed
about elections of some sort.} The endogenous explanatory variable (“treatment”) in such settings
is typically a simple or weighted average or sum of the RD shocks
across multiple elections. We propose two new estimators in these
“RD aggregation” (RDA) settings. The “upper-level” estimator
is based on an instrumental variable (IV) regression at the level
at which the outcomes are measured, with a weighted average or sum
of the RD shocks from narrow elections as the instrument, while controlling
for similar weighted sums of the controls used in local linear estimation
(e.g., of the running variables). The “lower-level,” or “stacking,”
estimator is a fuzzy RD estimator in the sample of narrow elections,
with the unusual feature that the same values of the outcome and endogenous
variable are repeated for the elections affecting the same unit. We
show that, under assumptions similar to those of standard RD, both
estimators identify the same local average treatment effect. Moreover,
Monte Carlo simulations confirm that these estimators inherit attractive
bias-reduction properties of conventional RD methods. The setting
we consider nests many empirical literatures, and our ideas yield
practical insights for them. We then apply the proposed methods to
estimate the effects of unionization on earnings inequality at the
level of state-by-industry cells, with typical cells exposed to multiple
establishment-level unionization events.

The primary use cases for the RDA approach are empirical settings
in which RD events are aggregated across space, time, or in other
ways to construct a treatment variable at the level at which the outcome
is measured. Such settings are very common. First, a large group of
papers considers legislatures that consist of politicians elected
in single-seat constituencies. The researcher is interested in the
impact of the seat share of politicians who possess a specific characteristic:
e.g., gender\citep{clots2011women,clots2012female,bhalotra2014health,ben-porath2023},
religion \citep{bhalotra2014religion,bhalotra2021religion}, or affiliation
with a specific party \citep{folke2014shades,nellis2016parties,nellis2018secular,ben-porath2023}.
For instance, a state-level outcome is regressed on the overall seat
share of women; the seat share of women who won close races is used
as an IV. Second, in many cases a unit can be exposed to multiple
RD events over time (\citet{cellini2010value,dell2018nation,biasi2024works}).
In the RDA approach, the researcher can specify the treatment as the
total fraction of the period during which the unit is treated. Other
studies aggregate discontinuities across investors in a region, across
university courses taken by a student, across politicians that a firm
is linked to, across destinations that a city is connected to by a
direct flight, and across unionization elections to which we return
below.\footnote{Specifically, \citet{ater2021real} construct a regional housing supply
shock by aggregating across real estate investors. They leverage a
discontinuity in the capital gains tax law that pushes some investors
to sell their property. \citet{tan2023consequences} considers the
effects of the number of high letter grades of a university student,
as aggregating across individual courses. \citet{akey2015valuing}
studies firms that donate to multiple politicians and measures the
effects of a firm's political connectedness, as aggregating election
outcomes of those politicians. \citet{campante2018long} aggregate
discontinuities in flight availability between city pairs with respect
to distance to construct a shock to a city's overall connectivity.
\citet{Kolerman2023} and the application in this paper aggregate
establishment-level unionization data from close elections to estimate
the effects of unionization on political outcomes and inequality,
respectively.}

Our results also extend naturally to estimating spillover effects
from discontinuity events, such that each unit is exposed to multiple
elections of their neighbors (in a spatial or network sense). \citet{isen2014local},
for instance, leverages close referenda on tax changes in Ohio counties
to study the effects on neighboring counties. Similarly, \citet{dechezlepretre2023tax}
study the effects of tax breaks, which have discontinuous eligibility
rules, on technologically similar firms.

While the two estimators we propose are natural extensions of conventional
RD to RDA settings and they possess attractive statistical properties,
they differ from most \emph{ad hoc} estimation approaches used in
the existing empirical literature. Specifically, papers estimating
upper-level specifications do not include all local linear controls
aggregated in the way suggested by our upper-level solution. As a
result, they generally forfeit the reduced bias advantages of local
linear estimation that made it standard in conventional RD designs.
Conversely, studies using lower-level specifications include local
linear controls but, in many cases, use reduced-form estimation which
does not define a coherent causal model for the upper-level outcome.
Some of these papers also impose restrictions limiting the sample
that is not necessary with the RDA approach. Finally, some papers
resort to difference-in-differences and other non-RD methods in settings
where RD would be a standard tool for credible identification, if
not for the aggregated outcomes. The RDA approach makes RD identification
feasible in those settings.

We apply the RDA approach to study the union effects on earnings inequality.
This question is one of the oldest in labor economics, dating back
to at least \citet{lewis1963unionism}. It is also of central policy
importance in modern days: Barack Obama, for instance, noted that
\emph{“as union membership has fallen, inequality has risen”}
\citep{Obama2016}— a correlation observed in the data by \citet{card2001effect}.
Yet, establishing a causal link between unionization and inequality
has proved difficult. Recently, \citet{fortin2022right} point out
that \textit{“empirical evidence on the distributional impact of
{[}union{]} threat effects is limited by the challenge of finding
exogenous sources of variation in the rate of unionization.”} Similarly,
\citet{farber2021unions} for the most part use a selection-on-observables
designs. We make progress by leveraging close union elections in an
RDA design.

We find that a higher rate of unionization leads to a decline in various
measures of inequality at the level of state-by-industry cells: the
Gini index, the 90/10 ratio, the share of labor income by top 10\%
earners, and the variance in log wages. These impacts result primarily
from the decline of wages at the top of the wage distribution rather
than an increase at the bottom. While this result calls for further
investigation, we find evidence for one mechanism that can help explain
it: pension coverage increases in cells with higher unionization rates.
In a back-of-the-envelope calculation, we find that around 35\% of
the growth in the key within-cell inequality measures between 1970
and 2010 can be explained by the declining rates of unionization relative
to the 1960s. Our estimate of the union contribution is moderately
larger than the most comparable estimates from \citet{farber2021unions}.

We believe our empirical strategy for estimating the impacts of unionization
on aggregated outcomes can be helpful beyond the effects on inequality.
Some papers evaluated such effects by comparing neighboring counties
across the border of right-to-work laws, which were mostly passed
in the 1940s and 1950s and weakened unions: \citet{holmes1998effect},
\citet{feigenbaum2019bargaining}, and \citet{makridis2019right}
studied the impacts on productivity, manufacturing activity, and voting,
respectively. Other papers leveraged specific case studies that generate
exogenous variation in union strength when looking at outcomes such
as wages \citep{biasi2021labor} and the gender pay gap \citep{biasi2022flexible}.
In contrast, the RDA method allows researchers to leverage credible
RD variation from multiple decades and the whole common, which is
standard when studying the impacts on establishment-level outcomes.
Complementary work by \citet{Kolerman2023} has used the RDA method
to show that unionization makes local politics more Democratic in
the US.

Methodologically, our paper bridges the literatures on shift-share
instruments and RD designs. Our theoretical analysis rests on the
observation that both the treatment and the instrument in the RDA
setting are shift-share aggregates (i.e., weighted sums) of RD shocks.
We then leverage the results from \citet{borusyak2022quasi} to show
how upper-level IV regressions with the aggregated instrument and
controls are numerically equivalent to particular fuzzy RD regressions
at the lower level. This equivalence is useful in two ways: on one
hand, it brings upper-level regressions closer to standard fuzzy RD
and motivates our choice of upper-level controls as aggregations of
the standard controls from local linear estimation. On the other hand,
it motivates our lower-level estimator by showing that an appropriate
lower-level regression can identify a parameter at the upper level,
at which the causal effect is naturally defined.

The setting we study is distinct from “multi-score” RD settings
which also feature multiple running variables per unit (see \citet{cattaneo2022regression}
for a review). For instance, in spatial RD designs \citep[e.g.,][]{black1999better,dell2010persistent},
the latitude and longitude are the two running variables, and a geographic
boundary separates the treated and control areas. Another example
is when a student is treated if several tests scores (e.g., for math
and English) simultaneously exceed some cutoffs. Such designs feature
a single binary treatment (or instrument, in the fuzzy case), and
a simple transformation of the running variables reduces the problem
to a standard RD, in which the signed distance to the boundary is
the scalar running variable. In contrast, in our setting the instrument
aggregates RD shocks. While still a function of the underlying running
variables, it does not feature a single boundary, making existing
methods inapplicable.

A recent literature has examined certain RD settings with interference.
\citet{torrione2024regression} extend the multi-score approach, based
on the minimum Euclidian distance from the set of neighbors' running
variables to the boundary at which the spillover treatment, e.g. the
number or share of treated neighbors, switches between two prespecified
values. This allows them to identify more detailed spillover effects,
e.g. of having all treated neighbors vs. all untreated neighbors conditionally
on the unit itself being treated. We propose a very different estimation
approach which identifies a causal spillover parameter in a simpler
way, does not depend on the discreteness of the treatment, and applies
to a variety of settings beyond just spillovers. Other papers have
examined the case of interference occurring among units with similar
values of the running variable, as may be natural in spatial RD designs
(\citet{keele2018geographic}, \citet{auerbach2024regression}). In
our setting, even if the structure of spillovers is based on geography,
each unit has its own running variable, leading to different issues
and solutions.

While our RDA design considers treatments and instruments that are
shift-share aggregates of RD shocks, a recent literature has considered
more complex instruments constructed from RD variation, via the idea
of recentering. As \citet{borusyak2023nonrandom} pointed out, if
within some bandwidth around the cutoff the RD shocks are viewed as
completely random (as in the local randomization approach to RD; see
\citet{cattaneo2015randomization}), one can compute the expectation
of the instrument over the distribution of RD shocks and control for
it to isolate the exogenous variation. \citet{kott2022income} and
\citet{narita2021algorithm} compute the expected instrument by simulating
counterfactual realizations of RD shocks, while \citet{abdul2022breaking}
and \citet{angrist2024credible} compute it analytically. Our paper
focuses on the settings where the instrument is linear in the RD shocks
— which still covers a wide variety of empirical applications. Thanks
to this focus, we are able to leverage the local linear estimation
approach to RD and develop estimators with better statistical properties
than those based on local randomization.

The rest of this paper is organized as follows: in Section 2 we present
our theoretical framework and results in the context of cross-sectional
RD aggregation, as well as Monte Carlo simulations for the proposed
estimators. In Section 3, we discuss how our proposals differ from
the current empirical practice. In Section 4, we extend the analysis
to settings with spillovers and temporal aggregation. In Section 5,
we apply the RDA approach to estimate the impact of unions on US inequality.
Section 6 concludes.

\section{Theoretical Framework\protect\label{sec:Theoretical-Framework}}

In Section \ref{subsec:Setting}, we introduce the setting and compare
it with conventional RD designs. Sections \ref{subsec:Intuition-Upper-Level}
and \ref{subsec:Intuition-Lower-Level} propose two estimators that
we put forward and provide intuition why they may provide valid causal
estimates. Section \ref{subsec:Identification} formally shows how,
under conditions similar to those of standard fuzzy RD, these estimators
identify a local average causal effect, and Section \ref{subsec:Monte-Carlo}
establishes their attractive bias-reduction properties in a Monte
Carlo simulation.

\subsection{Setting\protect\label{subsec:Setting}}

We measure outcome $Y_{i}$ for a set of units $i=1,\dots,N$, which
we refer to as the “upper level.” Each unit consists of a set
$\mathcal{J}_{i}$ of “lower-level” subunits $j$ (with $\left|\mathcal{J}_{i}\right|=J_{i}>0$).
A subunit is characterized by a running variable $r_{j}$ and treatment
$z_{j}=\one{r_{j}\ge0}$ that happens whenever the running variable
exceeds the cutoff normalized to zero.\footnote{We use capital and small letters to denote unit- and subunit-level
variables, respectively.} We are interested in estimating the causal model
\begin{equation}
Y_{i}=\beta X_{i}+\varepsilon_{i},\label{eq:basic_causal}
\end{equation}
where $\varepsilon_{i}$ is the untreated potential outcome and $\beta$
is the causal effect (assumed homogeneous for now) of some upper-level
treatment $X_{i}$. Our benchmark case is when $X_{i}$ simply aggregates
$z_{j}$ across all relevant subunits,

\begin{equation}
X_{i}=\sum_{j\in\mathcal{J}_{i}}s_{j}z_{j},\label{eq:Xi_structure}
\end{equation}
but other treatments can be considered, too (see, e.g., Section \ref{subsec:Spillover-Settings}
for an example). Here $s_{j}>0$ are known importance weights which
may or may not be the same for all $j\in\mathcal{J}_{i}$ and may
or may not add up to one.\footnote{\label{fn:no-spillovers}We assume that the sets are non-overlapping
to simplify notation. The setting naturally extends to the case where
different units can be exposed to the same subunits, as when estimating
the spillovers of RD events; see Section \ref{subsec:Spillover-Settings}.} In the election example, $X_{i}$ may be the fraction of seats in
the state $i$ legislature represented by candidates from the focal
party. Here $z_{j}$ indicates the focal party's victory in district
$j$ and $s_{j}$ measures the seat share of the district $j$ in
the legislature. In the common case of one seat per district, $s_{j}$
is the inverse of the total number of seats.

The conventional sharp RD setting corresponds to the case where the
outcomes are measured at the same level as the running variables,
i.e. $i$ and $j$ are the same objects: $\mathcal{J}_{i}=\left\{ i\right\} $,
$s_{j}=1$, and $X_{i}=z_{i}$. In that scenario, the standard local
linear estimation approach restricts the sample to observations near
the cutoff ($\left|r_{j}\right|\le h$ for bandwidth $h$) and estimates
by ordinary least squares (OLS) the specification 
\begin{equation}
Y_{i}=\beta z_{j}+\gamma_{0}+\gamma_{1}r_{j}+\gamma_{2}r_{j}^{+}+\text{error}_{j},\qquad j\equiv i,\label{eq:sharpRD}
\end{equation}
where the controls $q_{j}=(1,r_{j},r_{j}^{+})$ include the intercept,
the running variable, and its interaction with being on the right
of the cutoff (i.e., $r_{j}^{+}=r_{j}\cdot\one{r_{j}\ge0}$). Sometimes
this regression is weighted by “kernel” weights that are larger
when $r_{j}$ is closer to the cutoff; importance weights are also
allowed.

But how can local variation in $r_{j}$ be leveraged to obtain a consistent
estimate of $\beta$ in (\ref{eq:basic_causal}) when each unit can
be exposed to multiple subunits, potentially with more than one near
the cutoff? We next propose two strategies, one at the upper level
and the other at the lower level, focusing on the intuitions before
proceeding to formal and simulation-based analysis of consistency
and bias.

\subsection{\protect\label{subsec:Intuition-Upper-Level}An Upper-Level Solution}

Our first proposal, at the level of upper-level of observations $i$,
involves instrumenting $X_{i}$ with its component arising from the
subunits whose running variables are near the cutoff:
\begin{equation}
Z_{i}=\sum_{j\in\mathcal{C}_{i}}s_{j}z_{j},\label{eq:Z}
\end{equation}
where $\mathcal{C}_{i}=\left\{ j\in\mathcal{J}_{i}\colon\ \one{\left|r_{j}\right|\le h}\right\} $
is the set of $i$'s subunits near the cutoff. In the election example,
$Z_{i}$ is the fraction of state legislature's seats represented
by a candidate from the focal party who won in a close election, out
of all seats.

We further propose to aggregate the vector of standard RD controls
$q_{j}$ to the $i$ level in the same way as the instrument is constructed,
\begin{equation}
Q_{i}=\sum_{j\in\mathcal{C}_{i}}s_{j}q_{j}=\bigl(\sum_{j\in\mathcal{C}_{i}}s_{j},\sum_{j\in\mathcal{C}_{i}}s_{j}r_{j},\sum_{j\in\mathcal{C}_{i}}s_{j}r_{j}^{+}\bigr)'.\label{eq:RDAcontrols}
\end{equation}
We label these three aggregated variables the \emph{RDA controls}.
The first of them is the total weight of subunits near the cutoff.
For instance, in the election example, it is the fraction of legislature
seats determined in a close election. The second is the average running
variable in narrow elections, rescaled by their total weight. And
the last one is the average running variable in narrow wins, rescaled
by the total weight of those narrow wins.

As a result, the upper-level RDA estimator corresponds to the instrumental
variable (IV) specification
\begin{equation}
Y_{i}=\beta X_{i}+\gamma_{0}\sum_{j\in\mathcal{C}_{i}}s_{j}+\gamma_{1}\sum_{j\in\mathcal{C}_{i}}s_{j}r_{j}+\gamma_{2}\sum_{j\in\mathcal{C}_{i}}s_{j}r_{j}^{+}+\tilde{\gamma}'\tilde{W}_{i}+\text{error}_{i},\label{eq:model-i}
\end{equation}
instrumented by $Z_{i}$, where $\tilde{W}_{i}$ includes the $i$-level
intercept and possibly additional variables that are continuous around
the cutoff. Its reduced-form parallels (\ref{eq:sharpRD}):
\begin{equation}
Y_{i}=\rho\sum_{j\in\mathcal{C}_{i}}s_{j}z_{j}+\gamma_{0}\sum_{j\in\mathcal{C}_{i}}s_{j}+\gamma_{1}\sum_{j\in\mathcal{C}_{i}}s_{j}r_{j}+\gamma_{2}\sum_{j\in\mathcal{C}_{i}}s_{j}r_{j}^{+}+\tilde{\gamma}'\tilde{W}_{i}+\text{error}_{i}.\label{eq:rf-i}
\end{equation}

While aggregating both the instrument $z_{j}$ and the covariates
$q_{j}$ in the same way is intuitive, why do we expect RDA to identify
a causal parameter? We provide intuition by showing that the IV estimator
from (\ref{eq:model-i}) is numerically equivalent to a standard fuzzy
RD estimator from a transformed regression, and can therefore be expected
to inherit some its properties.

Indeed, $Z_{i}$ can be viewed as a shift-share instrument constructed
from shifts $z_{j}$ and exposure shares $s_{j}$.\footnote{More precisely, the shares are $s_{j}$ for $j\in\mathcal{C}_{i}$
and zero for $j\not\in\mathcal{C}_{i}$. While shift-share instruments
usually involves shifts that are common to all observations, this
is not a requirement \citep[Appendix A.6]{borusyak2024practical}.} We can therefore leverage the numerical equivalence result of \citet[Prop. 1]{borusyak2022quasi}
to represent $\hat{\beta}$ as an IV estimator at the subunit level:
\begin{prop}
\label{prop:equivalence}For any unit-level variable $V_{i}$ let
$V_{i}^{\perp}$ denote the residual from a projection of $V_{i}$
on $W_{i}=\left(Q_{i},\tilde{W}_{i}\right)'$. Let $\i(j)$ denote
the unit to which subunit $j$ belongs. Then $\hat{\beta}$ from (\ref{eq:model-i})
equals the IV estimator of $\beta$ from a subunit-level specification
\begin{equation}
Y_{\i(j)}^{\perp}=\beta X_{\i(j)}^{\perp}+\lambda'q_{j}+\text{error}_{j},\label{eq:model-j-equiv}
\end{equation}
on the sample of $j\in\mathcal{C}\equiv\cup_{i}\mathcal{C}_{i}$,
where $X_{\i(j)}^{\perp}$ is instrumented by $z_{j}$ and the regression
is weighted by $s_{j}$.
\end{prop}
\begin{proof}
Follows from Proposition 1 in \citet{borusyak2022quasi} by observing
that the shock-level weight defined in that paper, $\sum_{i}\sum_{j\in\mathcal{C}}s_{j}\one{j\in\mathcal{C}_{i}}$,
simplifies to $s_{j}$, and that shock-level averages simplify to
$\bar{V}_{j}=\sum_{i}s_{j}\one{j\in\mathcal{C}_{i}}V_{i}/\sum_{i}s_{j}\one{j\in\mathcal{C}_{i}}=V_{\i(j)}$.
\end{proof}
The specification (\ref{eq:model-j-equiv}) is artificial but it is
useful because it maps directly to the standard fuzzy RD design: the
instrument is the subunit-level RD shock $z_{j}$ and the covariates
are $q_{j}=(1,r_{j},r_{j}^{+})$.\footnote{While in many fuzzy RD designs the treatment is binary, this is not
necessary \citep[e.g.,][Section VIII]{almond2010estimating}.} One can use the equivalent $j$-level representation of the upper-level
specification to choose bandwidth and to perform statistical inference
on the estimator, as in \citet{calonico2014robust}

\subsection{\protect\label{subsec:Intuition-Lower-Level}RD Stacking: A Lower-Level
Solution}

Our second proposal is to estimate a fuzzy RD specification on the
sample that stacks all subunits near the cutoff: 
\begin{equation}
Y_{\i(j)}=\beta X_{\i(j)}+\tilde{\gamma}'\tilde{W}_{\i(j)}+\lambda'q_{j}+\text{error}_{j},\qquad j\in\cup_{i}\mathcal{C}_{i}\label{eq:model-j-lowerlevel}
\end{equation}
with $X_{\i(j)}$ instrumented by the subunit-level treatment $z_{j}$
and using $s_{j}$ as importance weights. Kernel weights may be additionally
used, too.\footnote{Kernel weights should not be used when constructing $Z_{i}$ for the
upper-level specification. Correspondingly, they were not introduced
in Section \ref{subsec:Intuition-Upper-Level}.} Since (\ref{eq:model-j-lowerlevel}) is a conventional fuzzy RD specification,
it is clear it uses local variation around the cutoff.

The specification (\ref{eq:model-j-lowerlevel}) is peculiar: it provides
multiple expressions for $Y_{i}$ of the same unit. One may also find
it odd that the instrument $z_{j}$ varies among multiple observations
— subunits of the same unit — that mechanically have the same value
of the treatment $X_{\i(j)}$.

Despite this peculiarity, Proposition \ref{prop:equivalence} demonstrates
why specification (\ref{eq:model-j-lowerlevel}) may be helpful. It
only differs from specification (\ref{eq:model-j-lowerlevel}) by
having the outcome and treatment initially residualized on $W_{i}$.
These controls, including $Q_{i}$, are continuous around the cutoff
and therefore may be expected not to affect identification. Moreover,
since local linear controls $q_{j}$ enter both specifications, one
may hope that both approaches inherit the bias properties of standard
RD. Below we show that such hope is justified.

The stacking approach has an additional advantage over the upper-level
solution: since it is just a fuzzy RD specification with raw outcome
and treatment, it naturally lends itself to standard RD plots for
the reduced-form and first-stage of (\ref{eq:model-j-lowerlevel}).

\subsection{\protect\label{subsec:Identification}Identification of a Local Causal
Effect}

To describe the local parameter identified by RDA, we introduce the
following causal model with heterogeneous causal effects. For $\boldsymbol{z}_{i}=\left(z_{j}\right)_{j\in\mathcal{J}_{i}}$,
let $X_{i}(\boldsymbol{z}_{i})$ denote the potential treatment of
unit $i$ depending on the treatments of its subunits. When equation
(\ref{eq:Xi_structure}) holds, it defines $X_{i}(\boldsymbol{z}_{i})$,
but we allow for general $X_{i}$. We further make an exclusion restriction
that $\boldsymbol{z}_{i}$ only affects $Y_{i}$ via $X_{i}$, with
$Y_{i}(x)$ denoting potential outcomes.\footnote{Here we assume for clarity that both the first and second stages of
(\ref{eq:model-i}) are causal. Both assumptions can be relaxed as
in \citet{small2017instrumental} and Appendix A.1 of \citet{borusyak2022quasi}.} We further impose:
\begin{assumption}[Sampling of units]
The data are a random sample of $J_{i},\boldsymbol{r}_{i},\boldsymbol{z}_{i},\boldsymbol{s}_{i},\boldsymbol{q}_{i},X_{i},X_{i}(\cdot),\tilde{W}_{i},Y_{i},Y_{i}(\cdot)$
where bold symbols indicate $J_{i}$-dimensional vectors containing
similarly named subunit-level variables, and consistency conditions
hold: $z_{j}=\one{r_{j}\ge0}$, $X_{i}=X_{i}(\boldsymbol{z}_{i})$,
and $Y_{i}=Y_{i}(X_{i})$.
\end{assumption}
\begin{assumption}[Irrelevance of labels]
\label{assu:irrelevance-of-labels}For any $J$ in the support of
$J_{i}$, the distribution of $\left(\pi(\boldsymbol{r}_{i}),\pi(\boldsymbol{z}_{i}),\pi(\boldsymbol{s}_{i}),\right.$\\
$\pi(\boldsymbol{q}_{i}),X_{i},X_{i}(\pi(\cdot)),\tilde{W}_{i},Y_{i},Y_{i}(\cdot))\mid J_{i}=\bar{J}$
is the same for all permutations $\pi$ of $\bar{J}$ elements.
\end{assumption}
Under these assumptions, the distribution of unit-level data induces
a distribution of the relevant variables over subunits, such as the
joint distribution of $z_{j}$ and $ $$\boldsymbol{z}_{\i(j)-j}$
where $\left(z_{j},\boldsymbol{z}_{\i(j)-j}\right)\equiv\boldsymbol{z}_{\i(j)}$.
We then impose a generalization of the RD continuity assumption:
\begin{assumption}[Continuity]
(a) Expected reweighted potential treatment and outcome, $\expec{s_{j}X_{\i(j)}(z,\boldsymbol{z}_{-j})\mid r_{j}=r}$
and $\expec{s_{j}Y_{\i(j)}(X_{\i(j)}(z,\boldsymbol{z}_{\i(j)-j}))\mid r_{j}=r}$
are continuous in $r$ at $r=0$ for $z=0,1$.

(b) Expected importance weights $\expec{s_{j}\mid r_{j}=r}$ and the
cumulative distribution function of $\boldsymbol{z}_{\i(j)-j}\mid r_{j}=r$
are continuous in $r$ at $r=0$.
\end{assumption}
While this assumption is high-level, it will be satisfied when the
distributions of raw potential outcomes $Y_{\i(j)}(x)$, potential
treatments $X_{\i(j)}(z,\boldsymbol{z}_{\i(j)-j})$, importance weights,
and leave-out treatments $\boldsymbol{z}_{\i(j)-j}$ are continuous
around the cutoff for $r_{j}$. The last assumption is satisfied when
the vector of running variables $\boldsymbol{r}_{i}$ has joint density,
in particular ruling out the case where multiple running variables
are perfectly correlated.\footnote{However, when multiple running variables are perfectly correlated,
they can be combined into one without loss.}

\begin{assumption}[Density]
The density of $r_{j}$ is positive and continuous at 0.
\end{assumption}
\begin{assumption}[Monotonicity and first stage]
$X_{\i(j)}(1,\boldsymbol{z}_{\i(j),-j})\ge X_{\i(j)}\left(0,\boldsymbol{z}_{\i(j),-j}\right)$
a.s. and\\
$\expec{s_{j}\left(X_{\i(j)}(1,\boldsymbol{z}_{\i(j),-j})-X_{\i(j)}(0,\boldsymbol{z}_{\i(j),-j})\right)\mid r_{j}=0}>0$.
\end{assumption}
Assumption 5 is trivially satisfied when the treatment is defined
by equation (\ref{eq:Xi_structure}). For simplicity, we abstract
away from additional controls, setting $\tilde{W}_{i}=1$; \citet{calonico2019regression}
show that adding predetermined covariates in a linearly separable
way does not affect the RD estimand. The following proposition shows
that under these assumptions both of our proposals identify the same
local average treatment effect which puts more weight on units with
more numerous and more important subunits near the cutoff, as well
as subunits with a stronger first-stage impact on the treatment.
\begin{prop}
\label{prop:identification}Suppose Assumptions 1–5 hold. Then, as
$h\to0$, the estimand $\beta_{h}^{u}\equiv\cov{Y_{i}^{\perp},Z_{i}}/\cov{X_{i}^{\perp},Z_{i}}$
of (\ref{eq:model-i}) and the similarly defined estimand $\beta_{h}^{\ell}$
of (\ref{eq:model-j-lowerlevel}) converge to the same convexly-weighted
average of potential outcome slopes:
\begin{equation}
\beta_{0}\equiv\frac{\expec{s_{j}\cdot\left(Y_{\i(j)}\left(X_{\i(j)}(1,\boldsymbol{z}_{\i(j)-j})\right)-Y_{\i(j)}\left(X_{\i(j)}\left(0,\boldsymbol{z}_{\i(j)-j}\right)\right)\right)\mid r_{j}=0}}{\expec{s_{j}\cdot\left(X_{\i(j)}(1,\boldsymbol{z}_{\i(j)-j})-X_{\i(j)}\left(0,\boldsymbol{z}_{\i(j)-j}\right)\right)\mid r_{j}=0}}.\label{eq:estimand}
\end{equation}
\end{prop}
\begin{proof}
See Appendix \ref{sec:Appx-Proof}.
\end{proof}
We note that the proof makes it clear that, among the covariates,
only $\sum_{j\in\mathcal{C}_{i}}s_{j}$ in the upper-level specification
and the intercept in the lower-level specification are necessary for
identification. A parallel to the local randomization approach to
RD designs provides intuition for this conclusion: if the running
variables are viewed as drawn from a uniform distribution, controlling
for $\sum_{j\in\mathcal{C}_{i}}s_{j}$ is equivalent to controlling
for the expected value of $Z_{i}$, as in the \citet{borusyak2023nonrandom}
recentering procedure. This control isolates a unit's “luck”
in getting a higher-than-expected vs lower-than-expected value of
the instrument. While our identification result (or the local randomization
approach) do not provide a justification for including the aggregated
local linear controls, the next subsection shows that they are important
for reducing bias, as in standard RD designs based on the continuity
assumption.

We also note that the effective weights on causal effects in (\ref{eq:estimand})
diverge from the importance weights the regressions are estimated
with. Consider the case where $X_{i}$ is defined as in equation (\ref{eq:Xi_structure}),
such that the first-stage effect of $z_{j}$ equals $s_{j}$, and
the causal effects are heterogeneous across units but linear in $x$,
$Y_{i}(x)=Y_{i}(0)+\beta_{i}x$. Then the total weight placed on $\beta_{i}$
in (\ref{eq:estimand}) is $\sum_{j\in\mathcal{C}_{i}}s_{j}^{2}$.
When $s_{j}=1/J_{i}$, as when the treatment is the \emph{share }of
legislature seats won by certain candidates (e.g., female or Democratic),
the effective weight is smaller in larger legislatures, as the law
of large numbers eliminates much of the variation in the instrument.
If, however, $s_{j}=1$, as when the treatment is the \emph{number}
of such legislature seats, the effective weight is larger in larger
legislatures, as they will tend to have more individual seats with
narrow elections. Importance weights in both upper- and lower-level
regressions can be adjusted if other averages of causal effects are
of interest.

\subsection{\protect\label{subsec:Monte-Carlo}Bias and Efficiency: Monte Carlo
Simulations}

The key advantage of the two estimators we propose is that they inherit
the attractive bias and variance properties of standard RD estimators.
In particular, when the bandwidth shrinks to zero, their bias shrinks
at a quadratic rate or faster. We leave theoretical results establishing
this property to future drafts; for now, we illustrate the performance
of alternative estimators via a Monte Carlo simulation.

We report the performance of the upper-level IV estimator with specification
(\ref{eq:model-i}) and the lower-level IV estimator with specification
(\ref{eq:model-j-lowerlevel}).\footnote{We do not include any additional controls besides the intercept, setting
$\tilde{W}_{i}=1$.} We contrast them with a benchmark upper-level IV estimator which
includes the control necessary for identification, $\sum_{j\in\mathcal{C}_{i}}s_{j}$,
but not the other controls aimed at reducing bias. Section \ref{sec:Common-Practices}
we show that the latter estimator is commonly used in practice, and
it also aligns with the recentering approach based on the local randomization
view of RD.

In each simulation, we draw 1,000 units (cells), each with 5 subunits
(elections) with equal importance weights $s_{j}=1/5$. For each subunit,
we generate the running variable as $r_{j}=\rho\zeta_{\i(j)}+\sqrt{1-\rho^{2}\cdot}\xi_{j}$,
where $\zeta_{i}$ and $\xi_{j}$ are i.i.d. standard normal variables
and $\rho=0.5$ captures the correlation of running variables among
subunits of the same unit. The treatment variable is defined as the
share of treated subunits, $X_{i}=\sum_{j\in\mathcal{J}_{i}}s_{j}z_{j}$
for $z_{j}=\one{r_{j}>0}$. To focus on the bias and variance, we
assume the true causal effect is zero for all observations.

\begin{figure}
\caption{Monte Carlo Simulations\protect\label{fig:Monte-Carlo-Simulations}}
\medskip{}

\begin{centering}
\begin{tabular}{ccc}
{\small (a) Bias: Outcome \#1} & {\small (b) Bias: Outcome \#2} & {\small (c) Bias: Outcome \#3}\tabularnewline
{\small\includegraphics[width=0.3\textwidth]{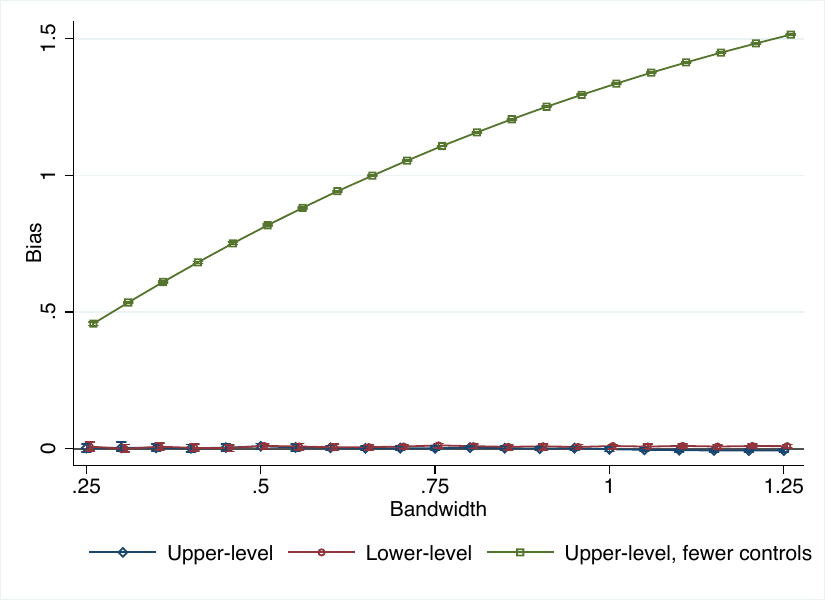}} & {\small\includegraphics[width=0.3\textwidth]{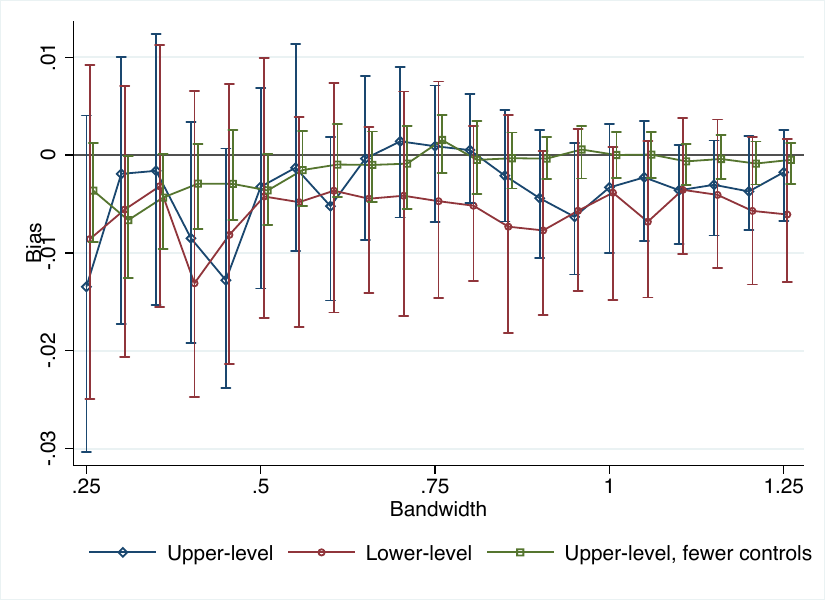}} & {\small\includegraphics[width=0.3\textwidth]{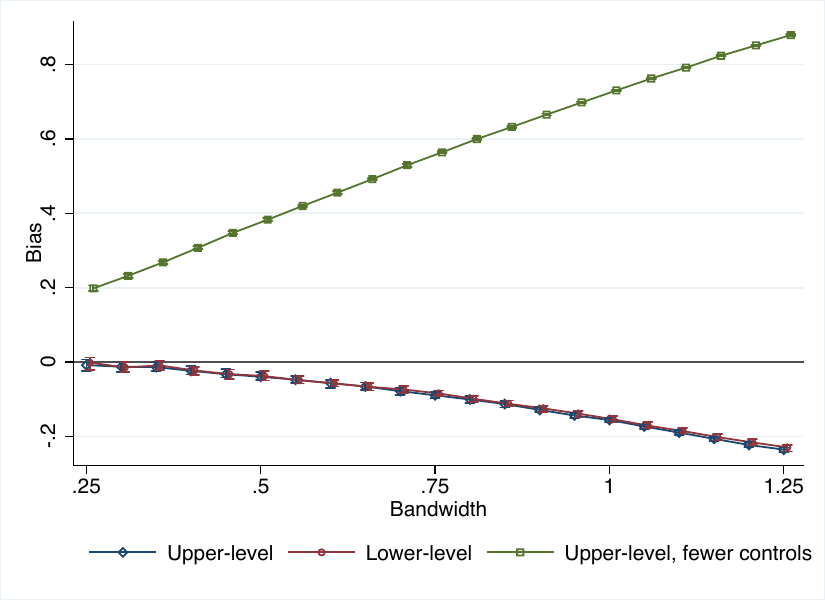}}\tabularnewline
 &  & \tabularnewline
{\small (d) Standard deviation: Outcome \#1} & {\small (e) Standard deviation: Outcome \#2} & {\small (f) Standard deviation: Outcome \#3}\tabularnewline
{\small\includegraphics[width=0.3\textwidth]{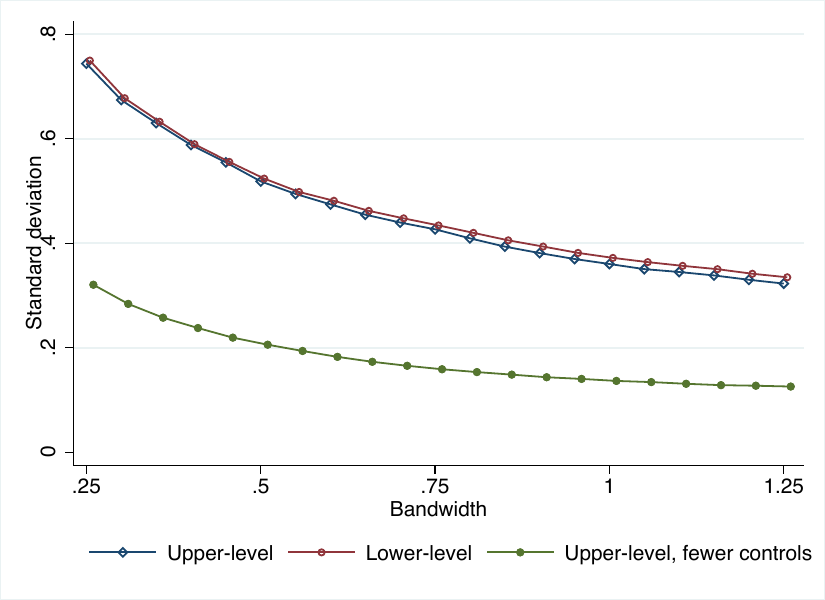}} & {\small\includegraphics[width=0.3\textwidth]{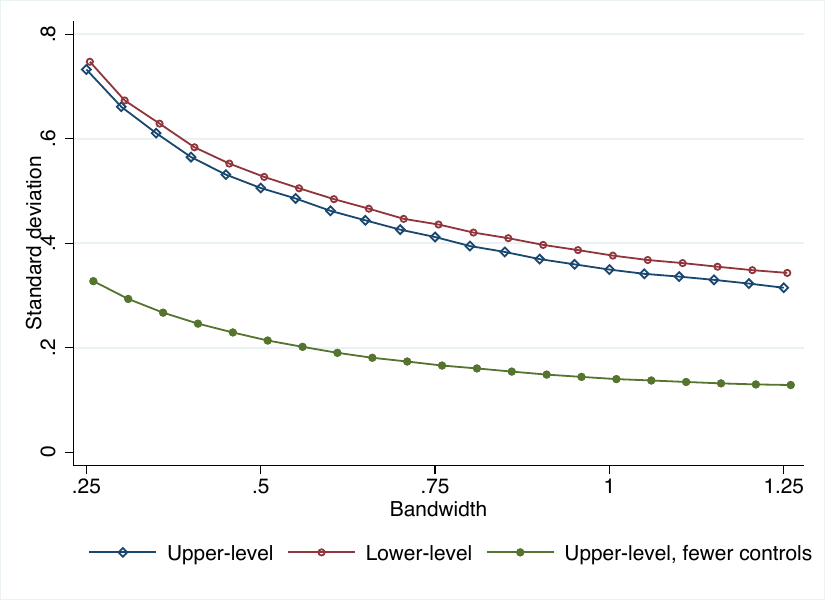}} & {\small\includegraphics[width=0.3\textwidth]{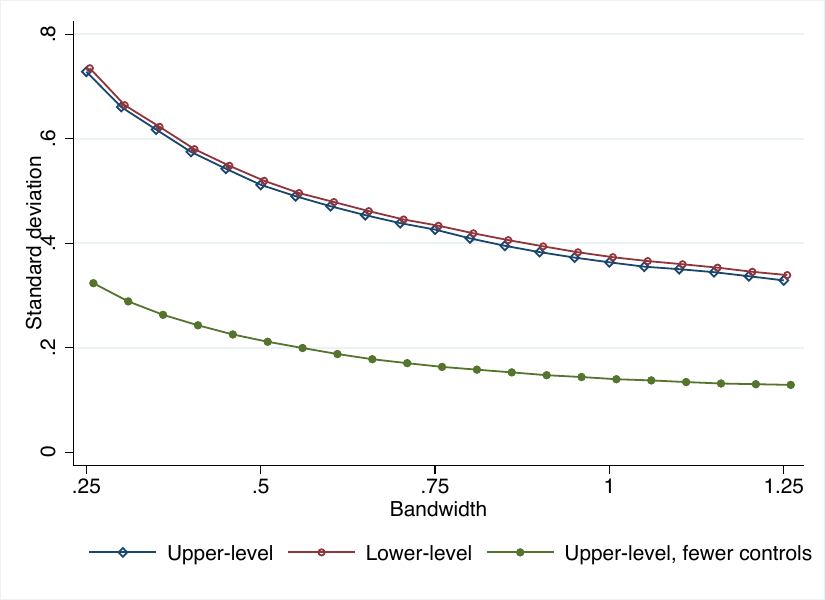}}\tabularnewline
\end{tabular}
\par\end{centering}
{\footnotesize\emph{Notes: }}{\footnotesize The top row reports, for
each bandwidth value, the median bias across 2,500 simulated datasets,
along with the 95\% bootstrap confidence interval for it based on
300 bootstrap draws. The bottom row reports standard deviations of
estimates across simulations. See main text for the data-generating
processes and definitions of the estimators.}{\footnotesize\par}
\end{figure}

The top row of Figure \ref{fig:Monte-Carlo-Simulations} reports the
median bias of the three estimators for a range of bandwidth values,
along with the bootstrap 95\% confidence interval for this median.\footnote{The range $h\in[0.25,1.25]$ corresponds to approximately $[20\%,80\%]$
of subunits included in the close-elections sample. The confidence
interval is hard to see on some panels because the number of simulations
is large enough for the median to be very precisely estimated.} The bias depends crucially on how potential outcomes vary with the
running variables of the subunits. We therefore consider several data-generating
processes for the outcome; focusing here on the median bias, we do
not include an independent error. Panel (a) considers a unit-level
outcome that is linear in the running variables of all subunits, $Y_{i}=\sum_{j\in\mathcal{J}_{i}}s_{j}r_{j}$.
In standard RD, that would eliminate bias completely, as long as the
local linear controls are included. In the RDA setup, this is not
the case, as the same unit can include both close and non-close elections.
Yet, the qualitative behavior is the same: the bias of our proposed
upper- and lower-level estimators is negligible compared to the benchmark
upper-level estimator with fewer controls (and the 95\% bootstrap
confidence interval includes zero for most bandwidths).\footnote{Another way to see this is that the reduced-form of our lower-level
IV regression is exactly unbiased, as $\expec{Y_{i}\mid r_{j}}$ is
linear in this simulation.} Panel (b) considers an outcome which depends on the running variables
nonlinearly but with the same curvature on either side of the cutoff,
$Y_{i}=\sum_{j\in\mathcal{J}_{i}}s_{j}r_{j}^{2}$. Here all three
estimators achieve negligible bias; note the scale of the vertical
axis as well as the 95\% confidence interval for the bias estimate
based on our 2,500 simulations. Finally, panel (c) considers the case
where curvature changes around the cutoff, $Y_{i}=\sum_{j\in\mathcal{J}_{i}}s_{j}r_{j}^{2}\one{r_{j}>0}$.
This is the case studied in the standard RD setup by, e.g., \citet{imbens2012optimal}
who show the importance of local linear controls in achieving quadratic
decay of the bias. Inheriting this property, the bias of our proposed
estimators is not only smaller than that for the benchmark one for
all bandwidths; its slope also approaches zero for smaller $h$, which
does not happen for the benchmark estimator.

In unreported simulations, we check robustness of these findings.
First, we generate heterogeneous importance weights across subunits
of the same unit (adding up to one). Second, while all outcomes in
our main simulation depend symmetrically on all running variables,
we also consider the outcome which equals the running variable of
one arbitrarily picked subunit. Third, we increase the number of units
from 1,000 to 2,000 or the number of subunits per unit from 5 to 10.
The bias patterns are unchanged.

The results so far have shown that, at least for the data-generating
processes we considered, the bias of the upper- and lower-level regressions
is essentially identical. There are, however, some differences in
efficiency. To assess them, we repeat the simulation for the same
outcomes adding \emph{i.i.d.} standard normal noise to capture the
idiosyncratic component of the error. The second row of Figure \ref{fig:Monte-Carlo-Simulations}
reports standard deviations of the estimators across simulations.
It shows that the upper-level regression is more efficient than the
lower-level one, albeit the difference is small. The intuition is
that aggregated controls better capture the correlation between potential
outcomes and the running variables. This is not an artifact of our
outcome definitions; rather, it follows from the irrelevance of labels
(Assumption \ref{assu:irrelevance-of-labels}): to the extent the
running variable of one subunit predicts the unit outcome, the appropriate
average of them is at least as predictive.

To sum up, the Monte Carlo simulations confirm that including the
aggregated local linear controls or estimating the effects at the
lower-level in a standard fuzzy RD both succeed at substantially reducing
bias. The upper-level estimator is also slightly more efficient, as
it is better able to reduce residual variance.

\section{Common Practices\protect\label{sec:Common-Practices}}

Many published articles feature designs suitable for RDA. We identified
over 50 such papers listed in Appendix Table \ref{table:rda=000020papers}.
In this section, we describe the empirical strategies employed by
the authors of those articles. We detail the main differences between
their approaches and ours, highlighting the potential gains from using
the estimators from Section \ref{sec:Theoretical-Framework}. Section
\ref{subsec:literature-DiD} discusses studies that, in the absence
of the RDA methodology, did not use RD variation at all. We then turn
to papers that used some RD specifications at the aggregated or disaggregated
levels in Sections \ref{subsec:literature-upper} and \ref{subsec:literature-lower},
respectively. Throughout the discussion, we generically refer to the
subunits as elections.

\subsection{Difference-in-Differences Specifications in RDA Settings\protect\label{subsec:literature-DiD}}

The first group of studies, listed in Panel A of Appendix Table \ref{table:rda=000020papers},
uses difference-in-differences and similar designs in settings where
RDA estimation is feasible and could provide more credible identification.
While the parallel trends assumption underlying difference-in-differences
analyses is often hard to justify \emph{ex ante}, continuity assumptions
underlying conventional RD and inherited by RDA are usually more convincing.

\citet[Tables 8--10]{besley2003political} consider the effects of
legislature composition — one of the prime applications of RDA. To
estimate the effects on policy choices of the fractions of female
and Democratic legislators in U.S. state houses, where most seats
are elected in single-seat races, they employ the two-way fixed effects
OLS specification with a time-varying treatment:
\[
Y_{it}=\beta X_{it}+\tilde{\gamma}'\tilde{W}_{it}+FE_{i}+FE_{t}+\text{error}_{it}.
\]
A credible instrument for $X_{it}$ could be obtained by leveraging
the structure of the treatment as aggregating discontinuity events.

Similar examples are found in difference-in-differences studies of
the effects of political connections on industries and firms. \citet[Table 10]{akcigit2023connecting}
estimate how a larger fraction of politically connected firms in an
industry — defined as firms whose employee has been elected in a local
council — affects industry outcomes. They use OLS estimation, despite
using conventional RD for firm-level outcomes. \citet{cooper2010corporate}
similarly consider the association between the number of politicians
supported by a firm who get elected and firm performance.

Finally, \citet{saez2019payroll} estimate the wage effects of a payroll
tax cut that applied to workers below the age of 25. Their difference-in-differences
specification compares firms with different shares of young workers,
under a parallel trends assumption. The RDA approach would instead
leverage the age cutoff, recognizing that a firm's share of young
workers is an aggregate of worker-level discontinuities. The RDA strategy
would thus entail comparing firms with workers age 25 or just above
to firms with workers just below 25, providing credible RD variation.

\subsection{Upper-Level Outcome and Upper-Level Specification\protect\label{subsec:literature-upper}}

We now consider 19 papers from Panel B of Appendix Table \ref{table:rda=000020papers}
where RD variation is used in some way and the regression analysis
is conducted at the upper-level ($i$-level in the notation of Section
\ref{sec:Theoretical-Framework}). These specifications can generally
be represented as

\begin{equation}
Y_{i}=\beta X_{i}+\tilde{\psi}'\tilde{Q}_{i}+\tilde{\gamma}'\tilde{W}_{i}+\text{error},\label{eq:rda_i_common}
\end{equation}
where $X_{i}$ is constructed from all subunits and is instrumented
by $Z_{i}=\sum_{j\in\mathcal{C}_{i}}s_{j}z_{j}$ constructed from
the subunits near the cutoff only. Furthermore, $\tilde{Q}_{i}$ represents
controls constructed in some way from subunit-level variables and
$\tilde{W}_{i}$ are other controls and fixed effects.

The key difference between our proposal and what is found in the literature
is in the set of controls $\tilde{Q}_{i}$. We discuss the three RDA
controls $Q_{i}$ in equation (\ref{eq:model-i}) in turn. The majority
of papers include the first control: the total weight of narrow elections.\footnote{Instead of controlling for it, some papers (e.g., \citet{folke2014shades},
\citet{azoulay2019public}), in the language of \citet{borusyak2023nonrandom},
recenter the instrument $Z_{i}$ by subtracting 0.5 times the total
weight of narrow elections.} One exception is \citet{valentim2024does} who instead control for
the weight of narrow elections \emph{below} the cutoff; this produces
a biased coefficient.\footnote{To see this bias, notice that $\beta\sum_{j\in\mathcal{C}_{i}}s_{j}z_{j}+\gamma_{0}\sum_{j\in\mathcal{C}_{i}}s_{j}=\left(\beta+\gamma_{0}\right)\sum_{j\in\mathcal{C}_{i}}s_{j}z_{j}+\gamma_{0}\sum_{j\in\mathcal{C}_{i}}s_{j}(1-z_{j})$.
The left-hand side here corresponds to (\ref{eq:model-i}), while
the right-hand side to the specification of \citet{valentim2024does}.
The reduced-form coefficient therefore identifies $\beta+\gamma_{0}$
instead of $\gamma_{0}$. \citet{akey2015valuing} (Table 7, column
2) employs a similar specification, except he interprets the second
coefficient as a causal effect of losing.}

The second RDA control — the aggregated running variable in narrow
elections — is not found in any of the papers we have reviewed. While
some papers have no analog to it (e.g., \citet{campante2018long,nellis2018secular}),
several variations have been employed. \citet{azoulay2019public}
take an unweighted average of running variables near the cutoff, while
the instrument is a sum of RD shocks weighted by some dollar amounts.
The same is true for \citet{folke2014shades} and \citet{merilainen2022political}
whose average additionally includes non-close elections. \citet{clots2012female}
and \citet{bhalotra2014religion,bhalotra2021religion} use a different
strategy: they control for the running variables of each subunit (including
with non-narrow elections) separately. This requires ordering the
subunits in an arbitrary way (such that changing the order would change
the IV estimate) and filling missing values as zeros for units that
have fewer subunits than others. Neither of these strategies appears
to have a clear justification.

Finally, we are not aware of any study that included any version of
the running variable aggregated among narrow wins only. Absent the
second and third RDA variables, the estimators used in the literature
do not inherit the properties of standard RD and therefore may not
enjoy the low bias advantages of local linear estimation.

We conclude by noting that, while most of the papers estimate (\ref{eq:rda_i_common})
by IV, \citet{akey2015valuing} considers its reduced-form instead.
We see several disadvantages to this approach. It is a less coherent
specification than IV since the explanatory variable depends on the
bandwidth choice, and thus is not an economic variable with independent
meaning. Practically speaking, if the fraction of narrow victories
is correlated with the fraction of non-narrow victories (as may happen
when the bandwidth is not infinitesimal), the reduced-form specification
is biased. And even if the correlation is small or absent, the reduced-form
specification is likely less efficient, as the error term includes
the effects of non-narrow victories.

\subsection{Upper-Level Outcome and Lower-Level Specification\protect\label{subsec:literature-lower}}

The final group of 8 papers (Panel C of Appendix Table \ref{table:rda=000020papers})
considers outcomes measured at the more aggregated level than the
discontinuity events but estimates certain regressions at the lower
level. To structure the discussion, recall that the stacking approach
of Section \ref{subsec:Intuition-Lower-Level} involves the sample
of all narrow elections and a fuzzy RD specification with the reduced-form
\begin{equation}
Y_{\i(j)}=\rho z_{j}+\lambda'q_{j}+\text{error},\label{eq:lower-level-RF}
\end{equation}
with $X_{\i(j)}$ as the endogenous variable in the structural equation,
and $s_{j}$ as weights. The current practice deviates in terms of
the sample restrictions and the use of OLS vs IV.

First, while some papers (e.g., \citet{de2020politics,de2024partisanship})
keep the full stacked sample of narrow elections, others restrict
the sample in different ways. \citet{beach2017gridlock} only keep
one narrow election for each unit, which has the running variable
closest to the cutoff (although this does not affect too many units
in their setting). \citet[Section III]{dell2018nation} only keep
the narrow election that happened chronologically first. These approaches
limit the amount of identifying variation, which our paper shows not
to be necessary.

Second, while some papers (\citet{carozzi2022political} and \citet{dell2018nation})
clearly introduce the upper-level treatment and use the IV specification
(\ref{eq:model-j-lowerlevel}), the majority of papers estimate the
reduced-form (\ref{eq:lower-level-RF}) by OLS. This has both conceptual
and practical issues. Conceptually, (\ref{eq:lower-level-RF}) is
not a coherent causal model: it provides multiple, mutually inconsistent,
expressions for the outcome, and the Stable Unit Treatment Value Assumption
(SUTVA) is clearly violated. The causal effects of an extra election
victory are also expected to vary substantially with $s_{j}$, limiting
external validity of the OLS coefficient. For instance, \citet{de2020politics}
find the effect of an extra Democrat to be much larger in small legislatures.
This issue would be avoided by an IV specification with the share
of Democrats in the legislature as the treatment. Practically, in
finite samples where the bandwidth is non-negligible, $z_{j}$ may
be non-trivially correlated with the treatment statuses of other subunits
(both narrow and others). This potentially results in a form of asymptotic
bias that doesn't arise in IV analyses.

\section{Extensions: Spillovers and Temporal Aggregation}

So far, we have considered settings in which outcomes are measured
at a higher level of cross-sectional (e.g., spatial) aggregation than
the RD events. In this section, we discuss two related settings to
which our RDA ideas extend. Section \ref{subsec:Spillover-Settings}
considers estimation of spillovers in RD settings, while Section \ref{subsec:Dynamic-RD}
focuses on aggregating RD events over time. We describe how the two
estimators from Section \ref{sec:Theoretical-Framework} apply in
those settings and compare them with current practice.

\subsection{Spillovers from RD Events\protect\label{subsec:Spillover-Settings}}

A number of studies consider spillovers from discontinuity events.
Instead of units and subunits, we now have outcome units $i$ and
intervention units $j$.\footnote{This terminology is common when analyzing experiments on bipartite
networks (e.g., \citet{zigler2021bipartite}).} We first discuss how our framework and solutions extend to this setting,
and then compare it with the current empirical practice.

In spillover cases, the endogenous variable $X_{i}$ is often the
number or fraction of treated neighbors, corresponding to the exogenous
(“contextual”) peer effects model. Thus, it still follows equation
(\ref{eq:Xi_structure}), with $\mathcal{J}_{i}$, reinterpreted as
the set of $i$'s peers. Alternatively, the endogenous linear-in-means
peer effects model is sometimes used, in which $X_{i}$ is the average
outcome of the peers \citep{manski1993identification}. In either
case, the instrument is still defined by (\ref{eq:Z}), with $\mathcal{C}_{i}$
reinterpreted as the set of $i$'s peers with narrow elections.

Our upper-level solution applies with no change, while the lower-level
solution can now take two forms. The stacked equation is now defined
at the bilateral level: on a sample of \emph{pairs} $\left\{ (i,j)\colon j\in\mathcal{C}_{i}\right\} $,
estimate
\begin{equation}
Y_{i}=\beta X_{i}+\gamma'\tilde{W}_{i}+\lambda'q_{j}+\text{error}_{ij},\label{eq:lowlevel-spillovers-bilateral}
\end{equation}
with $z_{j}$ as the IV for $X_{i}$ and $s_{j}$ as weights. This
is still a standard fuzzy RD specification with a continuous treatment,
albeit with both $i$ and $j$ potentially observed multiple times.
Absent additional $i$-level controls, i.e. if $\tilde{W}_{i}=0$,
one can also obtain the same estimate $\hat{\beta}$ by collapsing
(\ref{eq:lowlevel-spillovers-bilateral}) to the level of intervention
units, i.e. to the sample of narrow elections $j$, by averaging across
outcome units that are $j$'s neighbors: i.e., one can estimate
\begin{equation}
\bar{Y_{j}}=\beta\bar{X}_{j}+\lambda'q_{j}+\text{error}_{j},\label{eq:lowlevel-spillovers-collapsed}
\end{equation}
where $\bar{Y}_{j}=\frac{1}{N_{j}}\sum_{i\in\mathcal{I}_{j}}Y_{i}$
and $\bar{X}_{j}=\frac{1}{N_{j}}\sum_{i\in\mathcal{I}_{j}}X_{i}$
for $\mathcal{I}_{j}=\left\{ i\colon j\in\mathcal{J}_{i}\right\} $
and $N_{j}=\left|\mathcal{I}_{j}\right|$, instrumenting $\bar{X}_{j}$
with $z_{j}$ and using $s_{j}N_{j}$ as weights.\footnote{Note that $\bar{X}_{j}$ is a convoluted treatment: it is the average
treatment $X_{i}$ of $j$'s neighbors where $X_{i}$ itself is an
average or sum of its neighbors' RD shocks. The first-stage captures
that, when $j$ is treated, the average treatment exposure $X_{i}$
of $j$'s neighbors mechanically goes up.} Versions of all three approaches have been used previously; yet,
like in Sections \ref{subsec:literature-upper} and \ref{subsec:literature-lower},
none of them exactly coincides with our proposal.

Surprisingly, we have found only one paper that performs estimation
at the $i$ level \citep{mora2023peer}. This stands in contrast not
only with the many papers using the upper-level specification in the
aggregation setting of Section \ref{subsec:literature-upper}, but
more importantly with how peer effects are usually estimated in economics,
via either exogenous or endogenous peer effects models \citep{manski1993identification}.
Ultimately, $i$-level specifications are, in our view, the most natural
and coherent way to study $i$-level outcomes. Similar to papers in
Section \ref{subsec:literature-upper}, \citet{mora2023peer} do not
include the third RDA control of equation (\ref{eq:rf-i}) and only
include the second RDA control as a robustness check.\footnote{In addition, they limit the sample of outcome units to those outside
the bandwidth and, in the main analysis, define the treatment by averaging
only across the peers within the bandwidth. Our theory would apply
even with the full sample and a more natural definition of the treatment
based on all peers.}

All other papers use specifications at the bilateral level or averaged
by intervention unit $j$, and they are subject to the issues raised
in Section \ref{subsec:literature-lower}. The first issue involves
unnecessarily restricting the sample. We find this in \citet{dahl2014peer}
who entirely drop all units with more than one neighbor near the cutoff.
The second issue is that most papers consider reduced-form OLS regressions
on $z_{j}$, rather than introducing the treatment at the level of
outcome units (or perhaps as in (\ref{eq:lowlevel-spillovers-collapsed})).
The only exception we are aware of is \citet{dube2019fairness} who
estimate a bilateral IV specification similar to what we propose,
with the average status of the peers as the endogenous variable. Other
papers — e.g., \citet{baskaran2018does} and \citet{dechezlepretre2023tax}
at the bilateral level and \citet{santoleri2022causal} and \citet{bhalotra2018pathbreakers}
at the $j$ level — are subject to the conceptual and practical concerns
of reduced-form specifications discussed in Section \ref{subsec:literature-lower}.\footnote{These concerns also apply to papers that use IV estimation with a
$j$-level treatment, rather than our proposed $i$-level treatment.
\citet{isen2014local}, for instance, estimates the effect of $Y_{j}$
on the average outcome of $j$'s neighbors. This is in contract to
a conventional endogenous peer effects model that would estimate the
effect of the average outcome of $i$'s neighbors on $Y_{i}$. See
\citet{clark2009performance} and \citet{dahl2014peer} for related
examples.}

\subsection{RD Aggregation over Time\protect\label{subsec:Dynamic-RD}}

Another setting in which RDA ideas can be useful is when multiple
RD events are aggregated over time. For instance, one may specify
a state-level treatment as the share of years in the last decade when
a Democratic governor is in power (in a cross-section of states or
a panel with a rolling window). Our upper-level solution would entail
instrumenting this treatment with the share of years in which the
current governor is a Democrat who won in a narrow election, and controlling
for the share of years determined by narrow elections, along with
the other RDA controls. In turn, the lower-level solution would involve
stacking all narrow elections of the previous decade, repeating the
same outcome and aggregated treatment, and including standard RD controls.
An unusual feature of the dynamic setting is that earlier elections
can influence the outcomes of later elections, for instance via incumbency
advantage \citep{lee2008randomized}. Our solutions allow for this,
under a standard exclusion restriction.

A more complex version of this setting is when elections are not regular
and the previous election's result can affect whether election is
held in a future year. This is the case in the literature that studies
the effects of public school funding on, e.g., house prices and educational
outcomes. This literature uses RD variation from local referenda for
school district bonds that authorize capital spending (see \citet{jackson2024impacts}
for a review). If the bond is approved in one year, it is less likely
that another referendum will be held the following year. In our lower-level
solution, there will be endogenous selection of the set of narrow
referenda included in the stacked sample. Still, nearly random outcomes
of those referenda should make the estimator valid under an appropriate
exclusion restriction (that precludes, for instance, a direct effect
of holding a referendum on the outcome through an increase in media
attention). Similarly, in our upper-level solution, RDA controls,
such as the share of years with narrow referenda, may now resemble
“bad controls,” as they can be affected by the results of earlier
referenda while also related to the error term. Yet, it is nearly
by chance whether a narrow election results in the bond passing or
not, conditional on prior referendum margins. Based on Proposition
\ref{prop:equivalence}, we cautiously conjecture that the upper-level
solutions also continues to be valid.

The current practice has followed a different approach, proposed by
\citet{cellini2010value} in the school bond context. They offer several
estimators; the “one-step” dynamic RD specification most frequently
adopted in later studies is as follows:
\begin{equation}
Y_{it}=\sum_{h\in\mathcal{H}}\left(\beta_{h}x_{i,t-h}+\phi_{h}M_{i,t-h}+P\left(r_{i,t-h};\theta_{h}\right)\right)+\tilde{\gamma}'\tilde{W}_{it}+\text{error}_{it}.\label{eq:onestep-drd}
\end{equation}
Here the set of $\beta_{h}$ coefficients for a set $\mathcal{H}$
of lags ($h\ge0$) and possibly leads ($h<0$) aims to capture the
dynamic effects of school spending $x_{i,t-h}$. The spending $h$
periods ago is instrumented with the indicator of the bond approved
in a referendum, $z_{i,t-h}$ (and reduced-form specifications are
also considered). The set of controls includes the indicators $M_{i,t-h}$
that there was a referendum in year $t-h$ and global polynomials
$P(r_{i,t-h};\theta_{h})$ in the vote margin $r_{i,t-h}$ with coefficients
$\theta_{h}$ (where the vote margin is normalized to zero in years
without an election), as well as additional controls $\tilde{W}_{it}$.
\citet{cellini2010value} point out that their specification does
not allow earlier elections to affect the occurrence of future elections
or the future vote margin: if that happened, $M_{i,t-h}$ and $P(r_{i,t-h};\theta_{h})$
would be bad controls biasing the coefficients of interest $\beta_{h}$.\footnote{They also propose a “recursive” estimator which allows for some
of these endogenous responses, although \citet{hsu2021dynamic} clarified
that this is only true with homogeneous effects. The recursive estimator
has not been popular in later work, perhaps because it tends to be
noisier.}

We view our RDA solutions as complementary to the dynamic RD specification.
The key advantage of the \citet{cellini2010value} specification is
that it allows to trace the dynamics of the effects (and also do placebo
testing by looking at the lead coefficients). We note, however, that
the effect of cumulative spending is also of interest: for instance,
\citet[Table 4]{biasi2024works}, and \citet[Tables 4--5]{baron2022school}
similarly report the average effect of the bond approval across different
horizons. The key disadvantage of specification (\ref{eq:onestep-drd})
is the potential bias from bad controls, which may not arise with
RDA solutions, especially our lower-level specification. Additionally,
(\ref{eq:onestep-drd}) is based on global polynomial estimation,
which is generally avoided in more recent RD analyses for being noisy
and sensitive to the order of the polynomial \citep{gelman2019high},
while the RDA solutions follow the prevalent local linear estimation
approach.\footnote{A local polynomial version of the IV specification (\ref{eq:onestep-drd})
would replace $M_{i,t-h}$ with a dummy of holding a close referendum
at $t-h$, replace $P\left(r_{i,t-h};\theta_{h}\right)$ with a local
polynomial (filled in with a zero if there was no referendum at all
or the referendum was not close), and replace $z_{i,t-h}$ with dummies
of the bond approved in a close referendum as the set of instruments.
We are not aware of any paper that used this estimation procedure
or studied its properties formally.}

\section{Application: The Impact of Unions on US Inequality}

We now apply the RDA identification strategy to estimate the effect
of unions on wage inequality within state-by-industry cells in the
US. Section \ref{subsec:Application-Setting} introduces the empirical
setting and adapts the RDA design to this application. Section \ref{subsec:Application-Data}
describes the data construction and presents summary statistics. Section
\ref{subsec:Application-Balance} validates our research design in
a series of balance tests. Section \ref{subsec:Application-Results}
presents the empirical results, and Section \ref{subsec:Application-Magnitudes}
discusses what these results mean for the contribution of declining
unions to growing inequality in the US.

\subsection{\protect\label{subsec:Application-Setting}Empirical Setting and
Specifications}

Trade unions are significant labor market institutions in all Western
countries, including the US. Over the last 100 years, union density
in the US has followed an inverse U-shaped pattern: there was a sharp
increase in unionization during the Roosevelt era (1930–45), a relative
steady state of high unionization rates from 1945 to 1960, and a continuous
decline since the 1960s \citep[Figure 1]{farber2021unions}. The strong
negative correlation between this trend and the U-shaped pattern of
U.S. income inequality \citep[e.g.,][]{piketty2018distributional}
contributes to the widespread perception of a relationship between
the two phenomena.

Recent studies have explored the relationship between unionization
and inequality \citep{callaway2018unions,collins2019unions,fortin2021labor,farber2021unions}.
Although these studies report a significant negative association between
unions and inequality in a cross-section or a panel, a causal interpretation
of those results requires strong assumptions, as illustrated by the
quote in the Introduction. Notably, these studies have not adopted
the RD approach that is common in the analyses of the union effects
on establishment-level outcomes \citep{dinardo2004economic,sojourner2015impacts,frandsen_2021}.

Unionization in the US occurs at the level of a bargaining unit, defined
as “a group of two or more employees who share community of interest
and may reasonably be grouped together for purposes of collective
bargaining” \citep[p.12]{nlrb1997basic}. In most cases, a bargaining
unit corresponds to a single establishment, representing on average
close to half of the establishment's workforce \citep{frandsen_2021}.\footnote{To be precise, \citet[Table 1]{frandsen_2021} reports an average
of 93 votes and 254 total employees in his sample covering most unionization
elections between 1980–2009. Using data from all unionization elections
in this period, we found an average turnout rate of 85\% (as a fraction
of eligible voters), suggesting a typical bargaining unit size of
approximately 109 employees (93/0.85). This indicates that bargaining
units represented about 43\% (109/254) of the total workforce in establishments
holding elections.} The election typically offers two choices: for or against union formation.
To form a union, a majority of 50\% plus one out of the cast votes
is necessary, lending itself to an RD design.

We exploit the discontinuities from narrow union elections to estimate
union effects at the more aggregate level of state-industry cells.
Specifically, we define the outcome $\Delta Y_{sit}$ as the change
in an economic outcome (e.g., a measure of wage inequality) in state
$s$ and industry $i$ during decade $t$,\footnote{We use differenced outcome variables both to increase estimation power
and as a way to reduce potential bias from manipulated unionization
election results, as suggested by \citet{frandsen2017party,frandsen_2021}.
Additionally, using differenced outcome variables aligns with our
treatment—the rate of new unionization—representing most of the \emph{change}
in the share of unionized workers.} and the endogenous treatment $\text{NewUnions}_{sit}$ as the share
of newly unionized workers. The latter is measured as the ratio of
the number of workers in the region-industry cell who join unions
through unionization elections (both narrow and not) during the decade
and the workforce in the cell at the beginning of the decade.

Looking at outcomes more aggregated than establishments has two advantages.
First, it allows us to study the effects on outcomes, such as inequality,
that are typically defined at more aggregated levels. Second, even
for outcomes—such us average wages—that can be measured by establishment,
our estimates capture the effects due to the spillovers of unionization
on other establishments within the same cell. Establishments in the
same state and industry often compete for the same workers and therefore
may adjust wages or even enter or exit the market in response to another
establishment's unionization. Defining cells as a combination of state
and industry follows the approach of \citet{fortin2021labor,fortin2022right}.
Aggregating the analysis to purely geographic units — either states
\citep{farber2021unions} or economic areas similar to commuting zones
(such as Census State Economic Areas in \citet{collins2019unions})
— could help capture cross-industry spillovers, but such aggregation
is not feasible for our study. State-level aggregation would substantially
reduce statistical power, as typical cells would contain many close
elections, and the law of large numbers would eliminate most identifying
variation in our RDA approaches. Meanwhile, data limitations do not
allow us to achieve spatial resolution smaller than states.\footnote{For metropolitan areas, Census public files report geographic units
less aggregated than states. However, both the number of these units
and their boundaries change each decade, hindering the construction
of panel data and the use of differenced outcomes.}

Applying our upper-level IV solution from Section \ref{subsec:Intuition-Upper-Level}
to this setting, we estimate the following structural equation and
first-stage:

\begin{align}
\Delta Y_{sit} & =\beta\text{NewUnions}_{sit}+\lambda'Q{}_{sit}+FE_{st}+FE_{it}+\text{error}_{sit},\label{eq:ss_unions_upper}\\
\text{\text{NewUnions}}_{sit} & =\pi Z_{sit}+\rho'Q{}_{sit}+FE_{st}+FE_{it}+\text{error}_{sit}.\label{eq:fs_unions_upper}
\end{align}
Here the instrument $Z_{sit}$ is the employment share of the bargaining
units unionized in a narrow election in the state-industry workforce.
Formally, $Z_{sit}=\sum_{j\in\mathcal{C}_{sit}}s_{j}z_{j}$ where
$\mathcal{C}_{sit}$ is the set of close elections in the $si$ cell
during decade $t$, $z_{j}$ is the dummy that the vote share for
the union strictly exceeds 50\% in workplace $j$, and $s_{j}$ is
the number of eligible voters in election $j$ as a fraction of the
cell workforce at the beginning of the decade.\footnote{We distinguish between winning and obtaining a majority of the votes,
as in a very small fraction of cases the final result was revised
due to an appeal process. Our setting is “fuzzy” in this sense,
and we handle it similarly to the standard fuzzy RD design: our instrument
is defined based on dummies for crossing the 50\% cutoff, while the
treatment is based on dummies for union victory.} The covariates $Q_{sit}$ consist of the three RDA controls defined
by (\ref{eq:RDAcontrols}); in this context they represent the share
of workforce involved in narrow elections ($\sum_{j\in\mathcal{C}_{sit}}s_{j}$),
a rescaled average of the running variable $r_{k}$, defined as the
union vote share minus 0.5, in closed elections ($\sum_{j\in\mathcal{C}_{sit}}s_{j}r_{j}$),
and a rescaled average of the running variable in close wins ($\sum_{j\in\mathcal{C}_{sit}}s_{j}r_{j}^{+}$).
Finally, $FE_{st}$ and $FE_{it}$ are state-by-decade and industry-by-decade
fixed effects, respectively, included solely to increase estimation
power, as the RDA design alone should be sufficient to obtain causal
estimates.\footnote{We do not include state-by-industry fixed effects both because the
outcome is already measured in differences and because including such
fixed effects would result in a substantial reduction of the sample:
our industry definitions change after 2000, leading to only one post-2000
state-by-industry observation (see Section \ref{subsec:Application-Data}).} We also explore specifications with different sets of fixed effects.
To obtain more efficient and economically meaningful estimates, we
weight each observation by the cell employment at the beginning of
the decade.

We also apply the stacking approach of Section \ref{subsec:Intuition-Lower-Level},
which involves estimating the following structural equation and first-stage
in the sample of narrow unionization elections $j$:

\begin{align}
\Delta Y_{s(j)i(j)t(j)} & =\beta\text{NewUnions}_{s(j)i(j)t(j)}+\lambda'q_{j}+FE_{s(j)t(j)}+FE_{i(j)t(j)}+\text{error}_{j},\label{eq:ss_unions_lower}\\
\text{\text{NewUnions}}_{s(j)i(j)t(j)} & =\pi z_{j}+\rho'q_{j}+FE_{s(j)t(j)}+FE_{i(j)t(j)}+\text{error}_{j}.\label{eq:fs_unions_lower}
\end{align}
where $s(j),i(j),t(j)$ respectively indicate the state, industry,
and decade in which election $j$ took place and $q_{j}$ consist
of the standard RD controls: a constant, the vote margin, and that
margin interacted with the union getting a majority. The structural
equation is identical to the upper level specification, except for
the controls and the error term. Per Proposition \ref{prop:equivalence},
we weight observations by the number of eligible workers in each workplace
$j$.

\subsection{\protect\label{subsec:Application-Data}Data}

We collect data on unionization and labor market outcomes by state,
industry, and decade over five decades, from 1960s to 2000s.\footnote{This period was chosen based on the availability of unionization data,
which is accessible from 1961 onwards and contains industry variables
until the beginning of 2009.} The data on establishment-level union elections is from the National
Labor Review Board (NLRB), maintained by Henry Farber and John-Paul
Ferguson. For each election, we observe the state, industry, and year,
as well as the total number of eligible workers and the votes cast
for and against. We construct the treatment variable $\text{\text{NewUnions}}_{sit}$
using the results of all elections. The instrument, in turn, aggregates
only the elections that satisfy three criteria. First, it only includes
close elections, defined by the vote share for the union of $50\pm10$
percent (the smallest bandwidth employed in recently published papers
that employ the RD strategy in the context of US unions; \citet{knepper2020fringe,frandsen_2021}).\footnote{The 10p.p. bandwidth also closely aligns with the optimal bandwidth
for local linear estimation. Applying the \citet{calonico2014robust}
procedure to our lower-level specification across our main five inequality
outcomes, we find optimal bandwidths ranging from 9.5\% to 13.7\%,
with an average of 10.9\%. We also check robustness to using a 15\%
bandwidth which is also used by \citet{knepper2020fringe} and \citet{frandsen_2021}.} Second, we exclude tied elections and elections with a 1-vote margin,
as there is evidence of manipulation in such elections \citep{frandsen_2021}.
Finally, we drop elections with less than 20 votes, as is customary
in the unionization literature \citep{dinardo2004economic,lee2012long,frandsen_2021}.\footnote{During the period of our analysis, there are two types of unionization
elections: certification elections, where workers attempt to form
a new union, and decertification elections, where workers or employers
try to dissolve an existing union. The rules for both types of elections
are identical, as are the consequences: if the union receives a strict
majority of votes, it will be certified or re-certified; if not, it
will be decertified. The vast majority of elections are certification
elections, accounting for 88.2\% of elections and covering 90.2\%
of the workers. We do not distinguish between these two types of elections
in our analysis; e.g., $NewUnions_{sit}$ counts workers in both newly
certified and re-certified unions. Our results are robust to restricting
the set of elections used for the treatment and instrument construction
to certification elections only.}

Our wage inequality measures and other wage-related outcomes by state,
industry, and year are based on the decennial population census samples
from 1960 to 2000, supplemented by the American Community Survey for
2009–2013 that represents 2010. With some variation over the years,
these samples represent roughly 5\% of the US population. We include
observations from all 50 states, as well as from the District of Columbia,
defined as an additional state. We only include workers with a strong
attachment to the labor market, defined as working all 52 weeks during
the year and at least 20 hours per week on average.\footnote{We exclude from the sample workers who worked fewer than 52 weeks,
as they have a higher likelihood of having switched industries during
the past year, in which case we could attribute their annual income
to the wrong industry.} Two of our main inequality outcomes are computed using total labor
income: the Gini index and the top-10\% income share. For the other
measures, we calculate the average annual hourly pre-tax wage of each
worker by dividing the total annual wage income by the product of
the number of weeks employed and the average hours worked per week
last year. We then compute the college wage premium, the log 90-10
wage ratio, and the variance of log wages, again by state-industry
cell. The construction of the inequality measures is detailed in the
Data Appendix, following \citet{autor2008trends} and \citet{farber2021unions}
with minor modifications. For the inequality measures to be meaningful,
we restrict the sample to include only cells with more than 1,000
workers after applying census projection weights.\footnote{This entails dropping 24.9\% of the cells, but only 0.8\% of total
employment and 1.3\% of close unionization elections. We verify there
are no remaining cells with fewer than 10 raw observations. For measures
based on smaller samples, such as for the average college wage, we
further exclude cells with fewer than 10 relevant observations at
the beginning or at the end of the decade.}

We additionally use the Current Population Survey (CPS) Monthly and
Annual Social and Economic Supplement samples. We measure union density
in each cell and year based on the self-reported union coverage question.
We also construct two measures of worker benefits: indicators of a
pension plan and employer-sponsored health insurance coverage. These
questions are available prior from 1977–2014; to increase sample size
we combine CPS data from several years to represent each decennial
census year, as detailed in the Data Appendix. The cleaning process
of the CPS data is otherwise the same as for the census data.

A challenge in the data construction process was to harmonize industry
definitions over a long period and between different data sources.
The NLRB data is based on the SIC and NAICS schemes, depending on
the year, while the population census, ACS, and CPS use a different
classification. We created two distinct coding schemes. Until 2000,
our industry coding is based on the initial two digits of the SIC87
codes; from 2000 onwards, it relies on the initial three digits of
the NAICS97 industry classification. In both cases, we merged a few
categories, resulting in 70 industries categories pre-2000 and 73
(different) categories post-2000 that cover all private sectors.\footnote{The NLRB data do not cover public sector unionization and, thus, those
industries are excluded from our analysis. Our classification is more
detailed than 10–11 industries used by \citet{fortin2021labor,fortin2022right}.} To calculate differenced outcomes for each decade, we estimate all
measures for the year 2000 based on both schemes.\footnote{Since the 2000 census industry question was based on the NAICS industry
scheme, we relied on the IPUMS crosswalk to match observations to
the SIC-based census industries. While this could introduce noise
in the 1990–2000 differences, we find it reassuring that employment
changes do not exhibit substantially higher volatility than in the
decade before or after.} In total, we were able to match 93\% of unionization elections to
our industry codes and 99\% of individual census observations.

\begin{table}
\begin{centering}
\caption{Summary Statistics\protect\label{table:summary}}
\medskip{}
\par\end{centering}
\begin{centering}
{\small{}%
\begin{tabular}{lcccc}
\toprule 
{\small Variable} & {\small Mean} & {\small Mean($\Delta$)} & {\small Data Source} & {\small Time Range}\tabularnewline
\midrule 
\multicolumn{5}{c}{{\small\textbf{Panel A: Outcomes}}{\small\medskip{}
}}\tabularnewline
{\small Log avg hourly wage} & {\small 2.825} & {\small 0.043} & {\small Census} & {\small 1960-2010}\tabularnewline
 & {\small (0.284)} & {\small (0.119)} &  & \tabularnewline
{\small Log college premium} & {\small 0.505} & {\small 0.049} & {\small Census} & {\small 1960-2010}\tabularnewline
 & {\small (0.164)} & {\small (0.106)} &  & \tabularnewline
{\small Log 90/10 ratio} & {\small 1.314} & {\small 0.070} & {\small Census} & {\small 1960-2010}\tabularnewline
 & {\small (0.216)} & {\small (0.109)} &  & \tabularnewline
{\small Gini index} & {\small 0.347} & {\small 0.023} & {\small Census} & {\small 1960-2010}\tabularnewline
 & {\small (0.070)} & {\small (0.032)} &  & \tabularnewline
{\small Top 10\% share} & {\small 0.282} & {\small 0.017} & {\small Census} & {\small 1960-2010}\tabularnewline
 & {\small (0.058)} & {\small (0.032)} &  & \tabularnewline
{\small Var(log hourly wage)} & {\small 0.310} & {\small 0.032} & {\small Census} & {\small 1960-2010}\tabularnewline
 & {\small (0.097)} & {\small (0.048)} &  & \tabularnewline
{\small Share of union members} & {\small 0.102} & {\small -0.040} & {\small CPS} & {\small 1980-2010}\tabularnewline
 & {\small (0.125)} & {\small (0.078)} &  & \tabularnewline
{\small Pension coverage share} & {\small 0.635} & {\small -0.009} & {\small CPS} & {\small 1980-2010}\tabularnewline
 & {\small (0.188)} & {\small (0.119)} &  & \tabularnewline
{\small Employer-sponsored insurance share} & {\small 0.678} & {\small -0.062} & {\small CPS} & {\small 1980-2010}\tabularnewline
 & {\small (0.164)} & {\small (0.096)} &  & \tabularnewline
 &  &  &  & \tabularnewline
\multicolumn{5}{c}{{\small\textbf{Panel B: Other Cell-Level Measures}}{\small\medskip{}
}}\tabularnewline
{\small Log employment} & {\small 10.865} & {\small 0.170} & {\small Census} & {\small 1960-2010}\tabularnewline
 & {\small (1.254)} & {\small (0.309)} &  & \tabularnewline
{\small Share with college degree} & {\small 0.241} & {\small 0.041} & {\small Census} & {\small 1960-2010}\tabularnewline
 & {\small (0.177)} & {\small (0.041)} &  & \tabularnewline
{\small Share white} & {\small 0.835} & {\small -0.025} & {\small Census} & {\small 1960-2010}\tabularnewline
 & {\small (0.119)} & {\small (0.042)} &  & \tabularnewline
{\small Share male} & {\small 0.586} & {\small -0.017} & {\small Census} & {\small 1960-2010}\tabularnewline
 & {\small (0.223)} & {\small (0.041)} &  & \tabularnewline
 &  &  &  & \tabularnewline
\multicolumn{5}{c}{{\small\textbf{Panel C: RDA Variables (bandwidth = 10\%)}}{\small\medskip{}
}}\tabularnewline
{\small$100\times\text{NewUnions}_{sit}$ (Share newly unionized)} & {\small 2.475} &  & {\small NLRB} & {\small 1960-2010}\tabularnewline
 & {\small (5.035)} &  &  & \tabularnewline
{\small$100\times Z_{sit}$ (Share newly unionized in close elections)} & {\small 0.634} &  & {\small NLRB} & {\small 1960-2010}\tabularnewline
 & {\small (1.591)} &  &  & \tabularnewline
{\small$100\times\sum_{j\in\mathcal{C}_{sit}}s_{j}$ (Share involved
in close elections)} & {\small 1.530} &  & {\small NLRB} & {\small 1960-2010}\tabularnewline
 & {\small (3.342)} &  &  & \tabularnewline
{\small$100\times\sum_{j\in\mathcal{C}_{sit}}s_{j}r_{j}$ (Agg. vote
share in close elections)} & {\small -0.018} &  & {\small NLRB} & {\small 1960-2010}\tabularnewline
 & {\small (0.118)} &  &  & \tabularnewline
{\small$100\times\sum_{j\in\mathcal{C}_{sit}}s_{j}r_{j}^{+}$ (Agg.
vote share in close wins)} & {\small 0.033} &  & {\small NLRB} & {\small 1960-2010}\tabularnewline
 & {\small (0.083)} &  &  & \tabularnewline
{\small\# of unionization elections} & {\small 55.2} &  & {\small NLRB} & {\small 1960-2010}\tabularnewline
 & {\small (83.7)} &  &  & \tabularnewline
{\small\# of close unionization elections} & {\small 8.2} &  & {\small NLRB} & {\small 1960-2010}\tabularnewline
 & {\small (14.7)} &  &  & \tabularnewline
\bottomrule
\end{tabular}}\medskip{}
\par\end{centering}
{\footnotesize\emph{Notes:}}{\footnotesize{} The first column displays
the weighted averages and (in parentheses) standard deviations of
the end-of-decade levels of the variables, while the second column
reports these statistics for first differences over the decade. The
statistics are weighted by employment at the beginning of the decade.
Observations are state-industry cells based on SIC industry codes
pre-2000 and NAICS codes thereafter. Panel A summarizes the outcome
variables used in the analysis. Panel B provides additional covariates.
Panel C presents the RDA variables (scaled by a factor of 100) along
with the average numbers of all and close unionization elections within
each cell.}{\footnotesize\par}

{\footnotesize\emph{Sources: }}{\footnotesize Decennial Census, CPS
Annual and Monthly Social and Economic Supplement, NLRB data, and
authors' calculations.}{\footnotesize\par}
\end{table}

Table \ref{table:summary} presents summary statistics for the key
variables at the state-industry-census year level, both in levels
and first-differences. Panel A includes outcomes, while Panel B details
other cell-level statistics. One can see, in particular, that all
within-cell inequality measures show an increasing trend, mirroring
national inequality trends. Panel C reports the endogenous variable,
the instrument, and the RDA controls described above, as well as the
average numbers of all and narrow unionization elections. In an average
decade, there are 55 union elections per cell resulting in 2.5\% of
the workforce joining the union. An average of 8 elections per cell,
involving 0.6\% of the workforce, results in a narrow win that enters
our instrument construction. Appendix Table \ref{table:summary_lower}
reports summary statistics at the election level, showing in particular
that the number of elections has seen a continuous decline since 1970s
(which matches the negative trend in union membership reported in
Table \ref{table:summary}, Panel A, with the CPS data since 1980s).

While we only look at inequality within state-industry cells, this
type of inequality is central to both the level and growth in overall
inequality. To show this, Appendix Table \ref{table:VarDecomp} decomposes
the overall variance of log wages into within- and between-cell components
in each census year, broadly following \citet{helpman2016trade}.
Throughout the period, the within component of the variance is the
dominant one, accounting for 73–80\% of total inequality. Moreover,
the within component experienced a continuous rise, contributing significantly
to the overall increase in inequality. In contrast, the between-group
component remained relatively stable, decreasing in some periods and
only experiencing a notable rise from 2000 to 2010.\footnote{A concern in this analysis is that some of the changes in the variance
components are due to the changes in the definitions of industries
in the year 2000 from SIC based to NAICS based. To address this, we
conducted an identical analysis on with annual CPS samples in Appendix
Figure \ref{fig:Theil-Decomposition}. It indicates no discernible
changes in the within and between components between 2002 and 2003,
the year of the CPS industry scheme change, providing strong evidence
that this transition does not drive the trends in the components.
This analysis also replicates the findings from Appendix Table \ref{table:VarDecomp}
using different data, confirming the key role of within-cell inequality.}

\subsection{\protect\label{subsec:Application-Balance}Balance}

We now perform balance tests to illustrate how the RDA instrument
and controls help isolate idiosyncratic variation in the treatment.
Column 1 of Table \ref{table:balance} regresses the share of all
newly unionized workers (i.e., our treatment of interest) on pre-determined
covariates and lagged inequality measures from Table \ref{table:summary}
(excluding those which are not available for all decades). The treatment
is correlated with many observables, as evidenced by the partial $R^{2}$
of 13.5\% and a decisively rejecting $F$-test, which suggests that
the treatment is also likely correlated with unobservables. In column
2 we estimate the same regression for our instrument: the share of
workers who join a union through close elections; the conclusion is
unchanged. In column 3, the inclusion of the first RDA control that
represents the share of workers involved in narrow union elections
vastly changes the statistics. The partial $R^{2}$ decreases massively
to 0.1\%, the \emph{F}-statistic of the covariates' joint significance
shrinks from 185.6 to 3.5, and most coefficients become orders of
magnitude smaller. Yet, covariates are still jointly significant (with
a p-value of 0.01\%). When we finally add the two other RDA controls
to the regression in column 4, the partial $R^{2}$ becomes essentially
zero and, while one out of ten covariates is individually significant,
jointly they are not (p-value of 12.3\%). The lack of correlation
between the residual variation in our instrument and the observables
speaks in favor of the validity of our IV strategy. In Appendix Table
\ref{table:balance15} we replicate this analysis with a 15\% bandwidth
defining close elections, and the same patterns emerge.

\begin{table}
\begin{centering}
\caption{Instrument Balance\protect\label{table:balance}}
\medskip{}
\par\end{centering}
\begin{centering}
{\small{}%
\begin{tabular}{lccccc}
\toprule 
 & {\small$\text{NewUnions}$} &  & \multicolumn{3}{c}{{\small$Z$}}\tabularnewline
\cmidrule(l){2-2}\cmidrule(l){4-6}
 & {\small (1)} &  & {\small (2)} & {\small (3)} & {\small (4)}\tabularnewline
\midrule 
{\small Log avg hourly wage$_{t-1}$} & {\small 1.056} &  & {\small 0.153} & {\small -0.100} & {\small 0.009}\tabularnewline
 & {\small (0.517)} &  & {\small (0.095)} & {\small (0.055)} & {\small (0.024)}\tabularnewline
{\small Log college premium$_{t-1}$} & {\small 4.452} &  & {\small 1.812} & {\small 0.083} & {\small -0.006}\tabularnewline
 & {\small (0.405)} &  & {\small (0.133)} & {\small (0.079)} & {\small (0.034)}\tabularnewline
{\small Log 90/10 ratio$_{t-1}$} & {\small 3.895} &  & {\small 0.889} & {\small -0.010} & {\small 0.044}\tabularnewline
 & {\small (0.630)} &  & {\small (0.194)} & {\small (0.108)} & {\small (0.049)}\tabularnewline
{\small Gini index$_{t-1}$} & {\small -58.022} &  & {\small -14.715} & {\small -0.876} & {\small 0.060}\tabularnewline
 & {\small (6.424)} &  & {\small (1.434)} & {\small (0.836)} & {\small (0.358)}\tabularnewline
{\small Top 10\% share$_{t-1}$} & {\small 40.646} &  & {\small 9.672} & {\small 0.641} & {\small -0.005}\tabularnewline
 & {\small (6.474)} &  & {\small (1.344)} & {\small (0.789)} & {\small (0.340)}\tabularnewline
{\small Var(log hourly wage)$_{t-1}$} & {\small -7.027} &  & {\small -1.805} & {\small 0.012} & {\small -0.181}\tabularnewline
 & {\small (1.486)} &  & {\small (0.287)} & {\small (0.170)} & {\small (0.077)}\tabularnewline
{\small Log employment$_{t-1}$} & {\small -0.158} &  & {\small -0.060} & {\small 0.003} & {\small -0.004}\tabularnewline
 & {\small (0.051)} &  & {\small (0.015)} & {\small (0.008)} & {\small (0.003)}\tabularnewline
{\small Share with college degree$_{t-1}$} & {\small -4.510} &  & {\small -1.035} & {\small 0.129} & {\small 0.067}\tabularnewline
 & {\small (0.609)} &  & {\small (0.126)} & {\small (0.076)} & {\small (0.040)}\tabularnewline
{\small Share white$_{t-1}$} & {\small -0.527} &  & {\small -0.096} & {\small -0.257} & {\small -0.024}\tabularnewline
 & {\small (0.561)} &  & {\small (0.174)} & {\small (0.094)} & {\small (0.040)}\tabularnewline
{\small Share male$_{t-1}$} & {\small -0.869} &  & {\small -0.377} & {\small -0.015} & {\small 0.027}\tabularnewline
 & {\small (0.516)} &  & {\small (0.126)} & {\small (0.063)} & {\small (0.025)}\tabularnewline
 &  &  &  &  & \tabularnewline
{\small\# significant coefficients} & {\small 9} &  & {\small 9} & {\small 2} & {\small 1}\tabularnewline
{\small Partial $R^{2}$} & {\small 0.1353} &  & {\small 0.0959} & {\small 0.0012} & {\small 0.0001}\tabularnewline
{\small Partial $F$-statistic} & {\small 292.7} &  & {\small 185.6} & {\small 3.5} & {\small 1.5}\tabularnewline
{\small Partial $F$-test, p-value} & {\small 0.0000} &  & {\small 0.0000} & {\small 0.0001} & {\small 0.1234}\tabularnewline
 &  &  &  &  & \tabularnewline
{\small RDA controls} & {\small None} &  & {\small None} & {\small$\sum_{j\in\mathcal{C}_{sit}}s_{j}$} & {\small All}\tabularnewline
{\small Observations} & {\small 12,799} &  & {\small 12,799} & {\small 12,799} & {\small 12,799}\tabularnewline
\bottomrule
\end{tabular}}\medskip{}
\par\end{centering}
{\footnotesize\emph{Notes:}}{\footnotesize{} Balance tests based on
OLS estimates weighted by employment at the beginning of the decade.
The outcome variable in the first column is the share of all newly
unionized workers, while the next three columns focus on shares joining
through close elections, with a bandwidth of 10\% applied to define
close elections. Column 3 additionally incorporates the first RDA
control, $\sum_{j\in\mathcal{C}_{sit}}s_{j}$, and column 4 adds the
remaining RDA variables, $\sum_{j\in\mathcal{C}_{sit}}s_{j}r_{j}$
and $\sum_{j\in\mathcal{C}_{sit}}s_{j}r_{j}^{+}$. Heteroskedasticity-robust
standard errors are shown in parentheses.}{\footnotesize\par}

{\footnotesize\emph{Sources: }}{\footnotesize Decennial Census, NLRB
data, and authors' calculations.}{\footnotesize\par}
\end{table}

Appendix Table \ref{table:balance_rhs} performs a related pre-trend
and placebo analysis, using predetermined covariates and lagged outcomes,
both in levels and differences, as dependent variables when estimating
the reduced-form of our upper-level IV specification:

\begin{align}
Y_{sit} & =\rho Z_{sit}+\lambda'Q{}_{sit}+FE_{st}+FE_{it}+\text{error}_{sit}.\label{eq:rf_unions}
\end{align}
None of the 15 coefficients are significant at the 5\% level using
either conventional standard errors or robust bias-corrected confidence
intervals of \citet{calonico2014robust}.

\subsection{\protect\label{subsec:Application-Results}Results and Mechanisms}

\begin{sidewaystable}
\begin{centering}
\caption{The Impact of New Unionization on 10-Year Changes in Inequality\protect\label{table:unions_inequilty}}
\medskip{}
\par\end{centering}
\begin{centering}
{\small{}%
\begin{tabular}{lccccc}
\toprule 
 & {\small$\Delta$Log college premium} & {\small$\Delta$Log 90/10} & {\small$\Delta$Gini coeff.} & {\small$\Delta$Top 10\% share} & {\small$\Delta$Var(log wage)}\tabularnewline
 & {\small (1)} & {\small (2)} & {\small (3)} & {\small (4)} & {\small (5)}\tabularnewline
\midrule 
\multicolumn{6}{c}{{\small\textbf{Panel A: Upper-Level Estimator}}}\tabularnewline
{\small Share newly unionized} & {\small -0.314} & {\small -0.459} & {\small -0.176} & {\small -0.143} & {\small -0.255}\tabularnewline
 & {\small (0.285)} & {\small (0.225)} & {\small (0.061)} & {\small (0.055)} & {\small (0.084)}\tabularnewline
 & {\small{[}-0.921,0.416{]}} & {\small{[}-0.933,0.122{]}} & {\small{[}-0.325,-0.035{]}} & {\small{[}-0.279,-0.020{]}} & {\small{[}-0.463,-0.054{]}}\tabularnewline
 &  &  &  &  & \tabularnewline
{\small Mean outcome} & {\small 0.515} & {\small 1.318} & {\small 0.348} & {\small 0.282} & {\small 0.311}\tabularnewline
{\small Mean treatment} & {\small 2.65\%} & {\small 2.76\%} & {\small 2.76\%} & {\small 2.76\%} & {\small 2.76\%}\tabularnewline
{\small RDA controls} & {\small$\checkmark$} & {\small$\checkmark$} & {\small$\checkmark$} & {\small$\checkmark$} & {\small$\checkmark$}\tabularnewline
{\small Industry-decade and state-decade FE} & {\small$\checkmark$} & {\small$\checkmark$} & {\small$\checkmark$} & {\small$\checkmark$} & {\small$\checkmark$}\tabularnewline
{\small Observations} & {\small 10,512} & {\small 12,910} & {\small 12,910} & {\small 12,910} & {\small 12,910}\tabularnewline
 &  &  &  &  & \tabularnewline
\multicolumn{6}{c}{{\small\textbf{Panel B: Lower-Level Estimator}}}\tabularnewline
{\small Share newly unionized} & {\small -0.258} & {\small -0.266} & {\small -0.133} & {\small -0.115} & {\small -0.206}\tabularnewline
 & {\small (0.284)} & {\small (0.185)} & {\small (0.052)} & {\small (0.048)} & {\small (0.073)}\tabularnewline
 & {\small{[}-0.896,0.443{]}} & {\small{[}-0.735,0.140{]}} & {\small{[}-0.284,-0.035{]}} & {\small{[}-0.256,-0.026{]}} & {\small{[}-0.427,-0.075{]}}\tabularnewline
 &  &  &  &  & \tabularnewline
{\small Mean outcome} & {\small 0.492} & {\small 1.189} & {\small 0.305} & {\small 0.251} & {\small 0.254}\tabularnewline
{\small Mean treatment} & {\small 9.00\%} & {\small 9.54\%} & {\small 9.54\%} & {\small 9.54\%} & {\small 9.54\%}\tabularnewline
{\small RD controls} & {\small$\checkmark$} & {\small$\checkmark$} & {\small$\checkmark$} & {\small$\checkmark$} & {\small$\checkmark$}\tabularnewline
{\small Industry-decade and state-decade FE} & {\small$\checkmark$} & {\small$\checkmark$} & {\small$\checkmark$} & {\small$\checkmark$} & {\small$\checkmark$}\tabularnewline
{\small Observations} & {\small 29,146} & {\small 31,256} & {\small 31,256} & {\small 31,256} & {\small 31,256}\tabularnewline
\bottomrule
\end{tabular}}\medskip{}
\par\end{centering}
{\footnotesize\emph{Notes: }}{\footnotesize Panel A reports IV estimates
of (\ref{eq:ss_unions_upper}). Observations are defined as state-industry
cells by decade (for 1960–2010). The endogenous variable is the share
of workforce unionized through unionization elections, relative to
the beginning-of-decade cell workforce. The instrument is the share
of workforce unionized through close elections, defined by the 10\%
bandwidth for the vote share around the cutoff of 50\%. RDA controls
are as in equation (\ref{eq:rf_unions}), and state-year and industry-year
FEs are included. See Data Appendix for the outcome definitions and
details of the sample construction. Observations are weighted by employment
at the beginning of the decade. Panel B shows the coefficients obtained
from the lower-level IV specification (\ref{eq:ss_unions_lower}),
in the sample of close elections. The specification includes standard
RD controls and is weighted by the number of eligible workers. In
both panels, heteroskedasticity-robust standard errors are in parentheses,
and robust bias-corrected confidence intervals of \citet{calonico2014robust}
are in brackets (in Panel A, constructed via the equivalent lower-level
specification as in Proposition \ref{prop:equivalence}).}{\footnotesize\par}

{\footnotesize\emph{Sources:}}{\footnotesize{} Decennial Census and
NLRB unionization data; authors' calculations.}{\footnotesize\par}
\end{sidewaystable}

Table \ref{table:unions_inequilty} examines the impact of unionization
on five key measures of income inequality: the college premium, the
90-10 wage ratio, the Gini coefficient, the top-10\% income share,
and the variance of log hourly wages. Panel A shows the estimates
using the upper-level approach, while Panel B shows the coefficients
obtained at the lower level. Along with each estimate, we report robust
bias-corrected confidence intervals following \citet{calonico2014robust},
obtained using the \texttt{RDrobust} package in \texttt{R}.\footnote{To obtain confidence intervals for our upper-level approach, we use
the numerically equivalent lower-level specification, per Proposition
\ref{prop:equivalence}. We do not use clustered inference (e.g.,
by state or by industry) for several reasons: theoretically, election
shocks should be mutually independent; empirically, we confirm that
clustered standard errors do not substantially exceed heteroskedasticity-robust
ones; and pragmatically, the implementation of clustering in the \texttt{RDrobust}
package does not allow elections on the two sides of the cutoff to
be part of the same cluster.}

\begin{figure}
\caption{Lower-Level Specifications: RD Plots\protect\label{fig:RD-plots}}

\begin{centering}
\medskip{}
\par\end{centering}
\begin{centering}
\includegraphics[width=0.85\columnwidth]{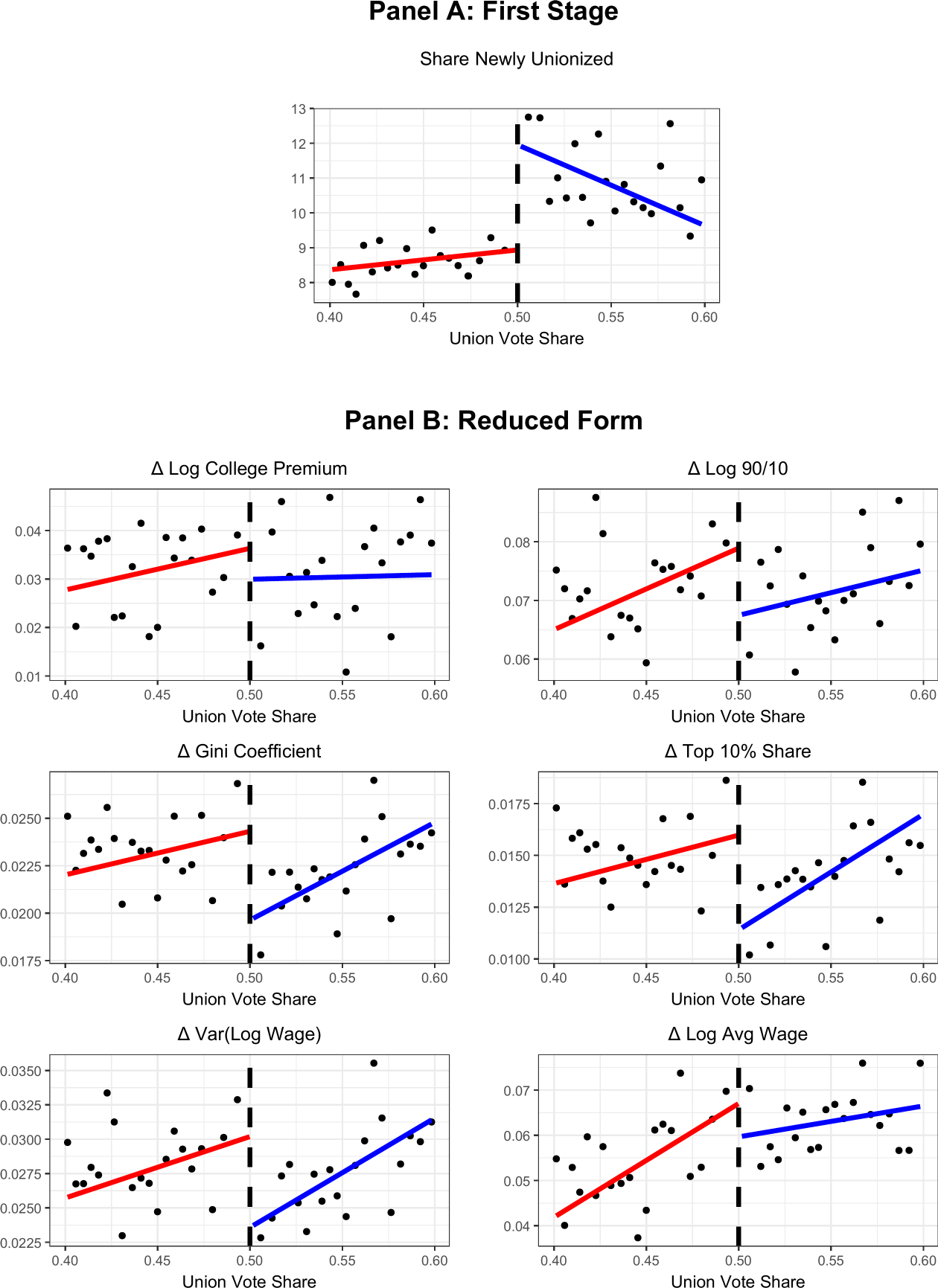}\medskip{}
\par\end{centering}
{\footnotesize\emph{Notes: }}{\footnotesize The first plot visualizes
the first-stage specification (\ref{eq:fs_unions_lower}) with no
fixed effects. The other plots visualize the reduced-form version
of (\ref{eq:ss_unions_lower}) with no fixed effects for the outcomes
included in Table \ref{table:unions_inequilty} and column 1 of Table
\ref{table:unions_inequilty=000020decomposition10}. The sample, weights,
and outcome definitions are the same as in Panel B of the relevant
tables. The x-axis represents the union vote share in a close union
election, while the y-axis shows the outcome in the corresponding
region-industry cell and year. The binned scatterplot shows 20 points
on each side of the cutoff, with each point representing the average
values for elections within specific ranges of the running variable.
These ranges were selected so that, on each side of the cutoff, each
point represents observations with the same sum of weights. The lines
correspond to the local linear estimates. “$\Delta$” stands
for the first difference over a decade.}{\footnotesize\par}

{\footnotesize\emph{Sources:}}{\footnotesize{} Decennial Census and
NLRB unionization data; authors' calculations.}{\footnotesize\par}
\end{figure}

The results reveal a negative effect of unionization on all measures
of inequality, with 95\% confidence intervals rejecting zero effects
for the last three inequality measures and with large but statistically
insignificant negative effects for the college premium and the 90-10
wage ratio. Specifically, upper-level estimates suggest that an increase
of 1 percentage point in new unionization reduces the college premium
by 0.31 log-points, the 90-10 ratio by 0.46 log-points, the Gini coefficient
by 0.018, the top-10 income share by 0.14 percentage points, and the
variance of log wages by 0.0025 (which is around 1\% of its mean).
Lower-level estimates are moderately smaller in magnitude but show
the same statistical significance. Figure \ref{fig:RD-plots} visualizes
these results with standard RD plots for the first-stage and reduced-form
lower-level specifications (excluding the fixed effects, to stay closer
to the raw data).

Appendix Table \ref{table:unions_inequilty-Robustness} reports several
robustness tests for these results, focusing on the upper-level analyses.
First, we replace our baseline bandwidth of 10\% with 15\% bandwidth,
a common \emph{ad hoc} choice in the literature (e.g., \citet{frandsen_2021,knepper2020fringe}).
Second, we exclude union decertification elections from the construction
of both the treatment and the instrument (along with the RDA controls).
Third, we replace state-year and industry-year fixed effects with
separate state, industry, and year and FEs, or with no FEs at all.
Finally, we perform estimation without initial employment weights.
All results are similar to those from Table \ref{table:unions_inequilty},
with coefficients from the unweighted estimation generally slightly
larger, possibly indicating a larger union effect on inequality in
smaller cells.

While we specified the treatment variable to be the rate of \emph{new}
unionization, the results would also be similar if we considered changes
in union density, i.e. in the overall share of unionized workers.
The latter measure would take into account differential employment
growth of unionized and non-unionized establishments, as well as establishment
entry and exit. Unfortunately, union density is only observed in the
CPS since 1977. Thus, in the spirit of two-sample IV \citep[ Ch.4.3]{angrist2009mostly},
we estimate the effect of new unionization rates (from the NLRB data)
on union density (from the CPS monthly samples) for the shorter period
of 1980–2010.\footnote{One may think of this analysis as a first-stage, except that this
is an IV specification leveraging close elections for the instrument
as before.} The coefficient is reported in the first column of Table \ref{table:benefits}.
This coefficient may be smaller than one due to various types of measurement
error: both on the right-hand side (discrepancies arising from out-of-state
workers, errors in coding the industry of unionization, or lower employment
growth in unionized establishments) and on the left-hand side (misclassification
of the union coverage variable in the CPS leading to a non-classical
measurement error).\footnote{Based on a 1977 validation survey, the estimate for the misclassification
rate in the union variable is 2.7\% \citep{card1996effect}.} Conversely, if unionized workplaces have higher employment growth,
the first-stage parameter might exceed one. In the data, we find estimates
ranging from 0.63 to 1.27, suggesting that our main IV estimates are
robust to using the change in the self-reported share of total unionized
workers as the endogenous variable.

\begin{table}
\begin{centering}
\caption{The Impact of New Unionization on 10-Year Changes in Union Density
and Fringe Benefits\protect\label{table:benefits}}
\medskip{}
\par\end{centering}
\begin{centering}
{\small{}%
\begin{tabular}{lccc}
\toprule 
 & {\small Union} & {\small Pension} & {\small Health insurance}\tabularnewline
 & {\small density} & {\small coverage rate} & {\small coverage rate}\tabularnewline
 & {\small (1)} & {\small (2)} & {\small (3)}\tabularnewline
\midrule 
\multicolumn{4}{c}{{\small\textbf{Panel A: Upper-Level Estimator}}}\tabularnewline
{\small Share newly unionized} & {\small 1.274} & 1.482 & {\small 0.063}\tabularnewline
 & {\small (0.551)} & {\small (0.780)} & {\small (0.466)}\tabularnewline
 & {\small{[}-0.043,2.490{]}} & {\small{[}-0.376,3.122{]}} & {\small{[}-1.010,0.986{]}}\tabularnewline
 &  &  & \tabularnewline
{\small Mean outcome} & {\small 0.105} & {\small 0.522} & {\small 0.680}\tabularnewline
{\small Mean treatment} & {\small 1.24\%} & {\small 1.49\%} & {\small 1.49\%}\tabularnewline
{\small RDA controls} & {\small$\checkmark$} & {\small$\checkmark$} & {\small$\checkmark$}\tabularnewline
{\small Industry-decade and state-decade FE} & {\small$\checkmark$} & {\small$\checkmark$} & {\small$\checkmark$}\tabularnewline
{\small Observations} & {\small 7,846} & {\small 6,630} & {\small 6,675}\tabularnewline
 &  &  & \tabularnewline
\multicolumn{4}{c}{{\small\textbf{Panel B: Lower-Level Estimator}}}\tabularnewline
{\small Share newly unionized} & {\small 0.634} & 0.858 & {\small 0.148}\tabularnewline
 & {\small (0.323)} & {\small (0.496)} & {\small (0.327)}\tabularnewline
 & {\small{[}-0.132,1.481{]}} & {\small{[}-0.285,2.090{]}} & {\small{[}-0.645,0.806{]}}\tabularnewline
 &  &  & \tabularnewline
{\small Mean outcome} & {\small 0.153} & {\small 0.569} & {\small 0.737}\tabularnewline
{\small Mean treatment} & {\small 3.90\%} & {\small 4.38\%} & {\small 4.43\%}\tabularnewline
{\small RDA controls} & {\small$\checkmark$} & {\small$\checkmark$} & {\small$\checkmark$}\tabularnewline
{\small Industry-decade and state-decade FE} & {\small$\checkmark$} & {\small$\checkmark$} & {\small$\checkmark$}\tabularnewline
{\small Observations} & {\small 14,119} & {\small 13,715} & {\small 13,743}\tabularnewline
\bottomrule
\end{tabular}}{\small\par}
\par\end{centering}
\medskip{}

{\footnotesize\emph{Notes:}}{\footnotesize{} This table reports IV estimates
from the same specifications as in Table \ref{table:unions_inequilty}
but for different outcomes: union density, pension coverage rate,
and employment-sponsored insurance coverage rate. The outcomes are
based on the CPS data and measured for the period from 1980–2010.}{\footnotesize\par}

{\footnotesize\emph{Sources:}}{\footnotesize{} Decennial Census, CPS
Annual and Monthly Social and Economic Supplement, NLRB unionization
data; authors' calculations.}{\footnotesize\par}
\end{table}

\begin{sidewaystable}
\begin{centering}
\caption{The Impact of Unions on 10-Year Wage Changes\protect\label{table:unions_inequilty=000020decomposition10}}
\medskip{}
\par\end{centering}
\begin{centering}
{\small{}%
\begin{tabular}{lcccccccc}
\toprule 
 & {\small All workers} & {\small College} & {\small High school} & {\small P90} & {\small P50} & {\small P10} & {\small Mgmt} & {\small Non-Mgmt}\tabularnewline
 & {\small (1)} & {\small (2)} & {\small (3)} & {\small (4)} & {\small (5)} & {\small (6)} & {\small (7)} & {\small (8)}\tabularnewline
\midrule 
\multicolumn{9}{c}{{\small\textbf{Panel A: Upper-Level Estimator}}}\tabularnewline
{\small Share newly unionized} & {\small -0.348} & {\small -0.449} & {\small -0.105} & {\small -0.282} & {\small -0.129} & {\small 0.176} & {\small -0.917} & {\small -0.181}\tabularnewline
 & {\small (0.150)} & {\small (0.349)} & {\small (0.142)} & {\small (0.180)} & {\small (0.145)} & {\small (0.166)} & {\small (0.399)} & {\small (0.134)}\tabularnewline
 & {\small{[}-0.762,-0.032{]}} & {\small{[}-1.221,0.468{]}} & {\small{[}-0.442,0.227{]}} & {\small{[}-0.735,0.138{]}} & {\small{[}-0.533,0.188{]}} & {\small{[}-0.283,0.497{]}} & {\small{[}-1.854,0.058{]}} & {\small{[}-0.561,0.091{]}}\tabularnewline
 &  &  &  &  &  &  &  & \tabularnewline
{\small Mean outcome} & {\small 2.832} & {\small 0.545} & {\small -0.019} & {\small 3.305} & {\small 2.637} & {\small 1.986} & {\small 0.507} & {\small 0.106}\tabularnewline
{\small Mean treatment} & {\small 2.76\%} & {\small 2.65\%} & {\small 2.76\%} & {\small 2.76\%} & {\small 2.76\%} & {\small 2.76\%} & {\small 2.71\%} & {\small 2.76\%}\tabularnewline
{\small RDA controls} & {\small$\checkmark$} & {\small$\checkmark$} & {\small$\checkmark$} & {\small$\checkmark$} & {\small$\checkmark$} & {\small$\checkmark$} & {\small$\checkmark$} & {\small$\checkmark$}\tabularnewline
{\small Industry-decade and state-decade FE} & {\small$\checkmark$} & {\small$\checkmark$} & {\small$\checkmark$} & {\small$\checkmark$} & {\small$\checkmark$} & {\small$\checkmark$} & {\small$\checkmark$} & {\small$\checkmark$}\tabularnewline
{\small Observations} & {\small 12,910} & {\small 10,541} & {\small 12,873} & {\small 12,910} & {\small 12,910} & {\small 12,910} & {\small 10,638} & {\small 12,910}\tabularnewline
 &  &  &  &  &  &  &  & \tabularnewline
\multicolumn{8}{c}{{\small\textbf{Panel B: Lower-Level Estimator}}} & \tabularnewline
{\small Share newly unionized} & {\small -0.303} & {\small -0.380} & {\small -0.119} & {\small -0.158} & {\small -0.173} & {\small 0.108} & {\small -0.924} & {\small -0.130}\tabularnewline
 & {\small (0.128)} & {\small (0.351)} & {\small (0.121)} & {\small (0.151)} & {\small (0.127)} & {\small (0.136)} & {\small (0.386)} & {\small (0.111)}\tabularnewline
 & {\small{[}-0.725,-0.100{]}} & {\small{[}-1.237,0.433{]}} & {\small{[}-0.44,0.134{]}} & {\small{[}-0.584,0.146{]}} & {\small{[}-0.547,0.061{]}} & {\small{[}-0.241,0.395{]}} & {\small{[}-1.946,-0.127{]}} & {\small{[}-0.492,0.048{]}}\tabularnewline
 &  &  &  &  &  &  &  & \tabularnewline
{\small Mean outcome} & {\small 2.809} & {\small 0.525} & {\small -0.010} & {\small 3.253} & {\small 2.658} & {\small 2.064} & {\small 0.468} & {\small 0.057}\tabularnewline
{\small Mean treatment} & {\small 9.54\%} & {\small 9.00\%} & {\small 9.54\%} & {\small 9.54\%} & {\small 9.54\%} & {\small 9.54\%} & {\small 9.20\%} & {\small 9.54\%}\tabularnewline
{\small RDA controls} & {\small$\checkmark$} & {\small$\checkmark$} & {\small$\checkmark$} & {\small$\checkmark$} & {\small$\checkmark$} & {\small$\checkmark$} & {\small$\checkmark$} & {\small$\checkmark$}\tabularnewline
{\small Industry-decade and state-decade FE} & {\small$\checkmark$} & {\small$\checkmark$} & {\small$\checkmark$} & {\small$\checkmark$} & {\small$\checkmark$} & {\small$\checkmark$} & {\small$\checkmark$} & {\small$\checkmark$}\tabularnewline
{\small Observations} & {\small 31,256} & {\small 29,148} & {\small 31,250} & {\small 31,256} & {\small 31,256} & {\small 31,256} & {\small 30,171} & {\small 31,256}\tabularnewline
\bottomrule
\end{tabular}}\medskip{}
\par\end{centering}
{\footnotesize\emph{Notes: }}{\footnotesize This table reports IV estimates
from the same specifications as in Table \ref{table:unions_inequilty}
but for different outcomes. The outcome in column 1 is the change
in log average wage in the sample of workers. In columns 2–3 and 7–8,
the outcome is the change in the log average wage among specific groups
of workers: those with exactly a college degree, exactly a high school
degree, managerial occupations, and non-managerial occupations, respectively.
State-industry cells with fewer than ten observations for the relevant
group at the beginning or end of the decade are excluded. In columns
4–6, the outcome is the change in the weighted 90th, 50th, and 10th
percentile of the distribution of log hourly wages within the cell.
Observations are weighted by total cell-level employment at the beginning
of the decade in all columns.}{\footnotesize\par}

{\footnotesize\emph{Sources:}}{\footnotesize{} Decennial Census and
NLRB unionization data; authors' calculations.}{\footnotesize\par}
\end{sidewaystable}

Table \ref{table:unions_inequilty=000020decomposition10} investigates
the sources of the inequality reduction due to unionization, finding
that it originates primarily from the decline in earnings among high-earners,
rather than an increase at the bottom of the earnings distribution.
Column 1 reports a large negative and marginally significant effect
on average wages.\footnote{This large negative effect is unusual relative to the other estimates
of a union wage premium at the worker or establishment level. \citet{card1996effect}
and \citet[Figure 5]{farber2021unions} find a positive effect using
a selection-on-observables design at the individual level. Staggered
difference-in-differences estimates from \citet{de2020two} also at
the individual level and RD estimates by \citet{dinardo2004economic}
and \citet{frandsen_2021} at the establishment level point out to
a null or slightly negative effect. Our estimate may differ for several
reasons. First, we incorporate spillover effects across establishments
in the same state and industry. Second, our estimates incorporate
spillovers on non-unionized employees, such as management (including
even managers located at a different establishment), while worker-level
estimates do not. Finally, like any RD-based estimate, we only use
local variation from close elections.}$^{,}$\footnote{A caveat to this finding is that it may partially result from worker
selection, as unionization has unequal effects on employment across
worker groups with different earnings potential, as studied in Appendix
Table \ref{table:emp}. There is no clear negative effect on overall
employment (Column 1), nor on the employment of high-school graduates
and non-college workers more broadly (Columns 2–3). In contrast, employment
of workers with exactly a college degree or those with college or
more education declines substantially in response to unionization
(Columns 4–5, marginally insignificant). This finding aligns with
\citet{frandsen_2021} who documents similar compositional effects
at the establishment level, where high-paid workers tend to leave
following unionization. Such composition effects can appear as a reduction
in average wages. One may expect the composition effects to be smaller
for columns 2–3 and 7–8 of Table \ref{table:unions_inequilty=000020decomposition10}
that investigate less heterogeneous worker populations.}The following columns of Table \ref{table:unions_inequilty=000020decomposition10}
measure the impacts of unionization on average wages of different
worker groups and parts of the wage distribution. Columns 2 and 4
indicate a negative (albeit not statistically significant) impact
of unions on wages among the higher-earning segments of the labor
force: college graduates and 90th percentile earners, respectively;
column 7 indicates a large and marginally significant negative effect
on managers' earnings. In contrast, columns 3, 5, and 8 show negative
but small and statistically insignificant wage impacts on the lower
segments: high school graduates, 50th percentile earners, and non-managers,
respectively. Column 6 finds a small and insignificant increase in
wages at the 10th percentile of the distribution. By construction,
the more negative effects at the top are consistent with our main
finding of the negative impact of unions on inequality.\footnote{Appendix Table \ref{table:robustness_decomposition} conducts several
robustness checks for these results. The overall conclusion of a negative
effect on average wages driven by the high-earners is preserved, although
the negative effect on the managerial wages tends to be weaker in
the robustness analyses.}

The null, or even negative, effects on median and average wages, is
puzzling and may raise questions about the motivations for unionization
and the resistance of employers to unionization. We shed light on
a possible explanation by considering the effects of unionization
on fringe benefits—health insurance and pensions coverage—in columns
2 and 3 of Table \ref{table:benefits}. Since both the outcomes and
the endogenous variable are measured in percentages, the coefficient
is interpreted as the number of covered workers induced by one new
union member. For health insurance, the effect is small and insignificant.
However, for pension coverage, we estimate that each new union member
corresponds to an increase of 1.48 pension holders (standard error
of 0.78) using the upper-level instrument and 0.86 (SE of 0.50) when
the lower level-instrument is employed. Although the coefficients
are noisy, the fact that they are around one or even larger suggests
a substantial spillover effect of unionization across establishments
within state-industry cells, given that some of the unionized establishments
had pension plans prior to unionization. This is consistent with the
recent findings by \citet{knepper2020fringe} that unionization of
establishments within firms has significant impacts on employer pension
contributions on other establishments of the same firm. Our analysis
may also suggest spillovers to other firms within the same state and
industry.

\subsection{\protect\label{subsec:Application-Magnitudes}How Much Did Declining
Unions Contribute to Growing Inequality?}

\begin{figure}
\caption{The Counterfactual with Unionization Rates Remaining at the 1960s
Level\protect\label{fig:counterfactual}}

\begin{centering}
\medskip{}
\par\end{centering}
\begin{centering}
\includegraphics[width=0.9\columnwidth]{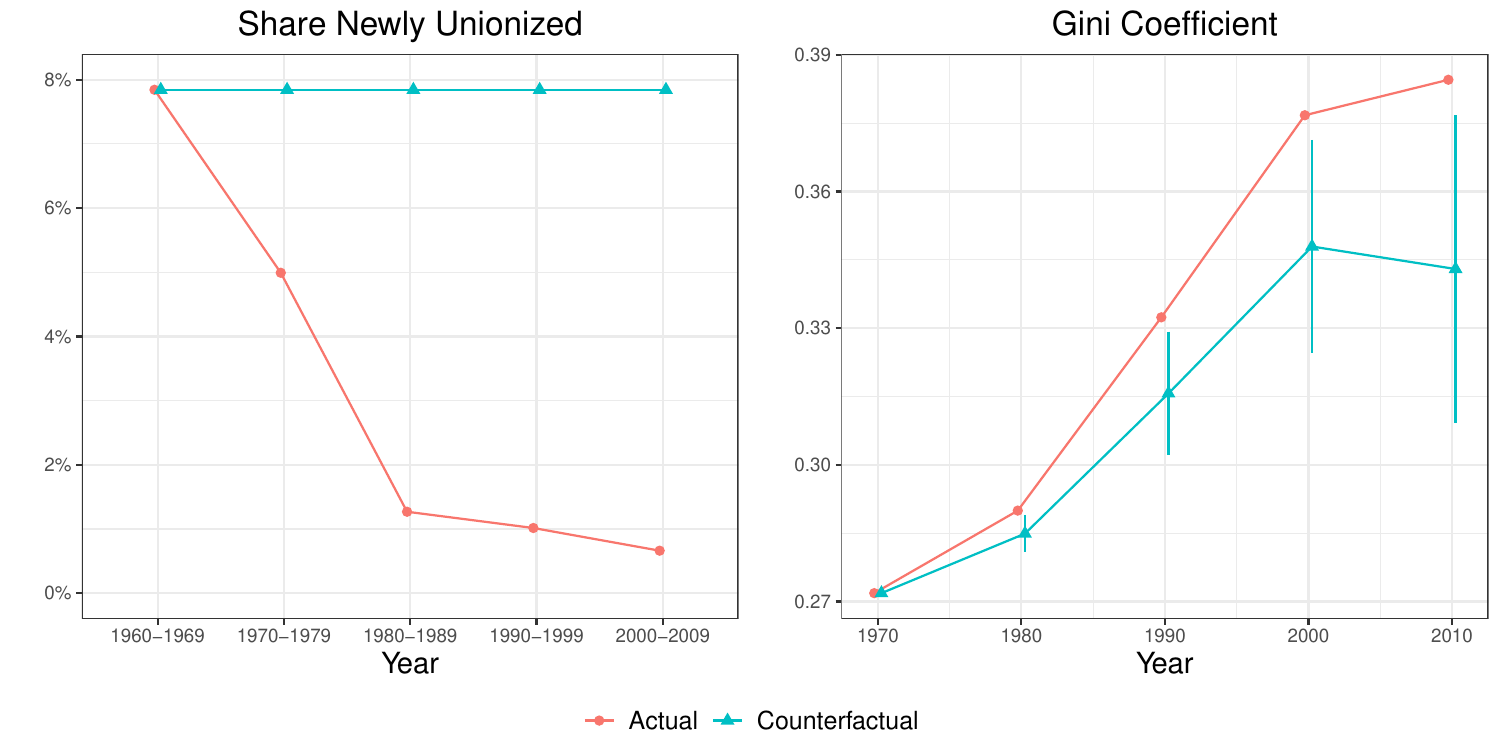}\medskip{}
\par\end{centering}
{\footnotesize\emph{Notes:}}{\footnotesize{} The first panel presents
the actual and counterfactual rates of newly unionized workers for
each decade between 1960 and 2010. The actual rate is calculated as
the share of workers joining unions through unionization elections
minus the share leaving due to union decertification elections. The
counterfactual rate assumes that the rate remained constant at its
1960s level. The second panel presents the actual and counterfactual
Gini coefficients of income inequality within state-industry cells,
averaged across cells (with beginning-of-decade employment weights).
The counterfactual is calculated by adding to the actual Gini the
product of the unionization effect estimated in Table \ref{table:unions_inequilty},
column 3, and the cumulative shortfall in the rate of new unionization
between the actual and counterfactual scenarios. Confidence intervals
for the counterfactual Gini are based on the 95\% robust bias-correct
confidence intervals for the treatment effect.}{\footnotesize\par}

{\footnotesize\emph{Sources:}}{\footnotesize{} Decennial Census, NLRB
unionization data; authors' calculations.}{\footnotesize\par}
\end{figure}

Rates of new unionization in the private sector in the US experienced
a massive and continuous decline since the 1970s, as can be seen in
the first panel of Figure \ref{fig:counterfactual}. In the 1960s,
the net flow of unionized workers through unionization elections was
7.8\%. This is the result of the share of workers joining unions through
unionization elections (8.3\%), minus some workers leaving unions
through union decertification elections. The net flow declined consistently,
reaching less than 1\% in the 2000s.

Using the results in Table \ref{table:unions_inequilty}, we conduct
a back-of-the-envelope calculation of how much of the growth in US
inequality during the period of our analysis can be attributed to
this sharp decline. To this end, we compute counterfactual inequality
measures that may have occurred if new unionization rates had remained
at their 1960s levels. Specifically, we add to each actual inequality
measure the product of the relevant treatment effect and the cumulative
shortfall in new unionization resulting from the sharp decline after
the 1960s.

Like any back-of-the-envelope calculation, ours requires some caveats.
The estimates from Table \ref{table:unions_inequilty} capture a short-term
effect and are identified using local variation from narrow elections.
Here, we extrapolate these effects to the long term, assuming that
the effect of new unionization during a decade on the level of inequality
by the end of the decade equal the effect on inequality at later dates,
too. Our analysis is also limited to assessing the effects of unions
on inequality within state-industry cells. However, as shown above,
within-cell inequality accounts for the majority of overall inequality
both in levels and in changes.\footnote{Another limitation is that we assess only the effect of new unionization,
not the full impact of changes in union density, which may also result
from the closure or growth of unionized establishments. However, this
concern is mitigated by our earlier finding that the effect of new
unionization on union density is approximately equal to 1.}

With these caveats, the second panel of Figure \ref{table:unions_inequilty}
presents the actual and counterfactual average within-cell Gini coefficients
by census year. The gap between the actual and counterfactual inequality
is small in 1980 and 1990 but becomes substantial by 2000 and 2010.
Appendix Figure \ref{fig:counterfactual_apdx} reports similar results
for the other inequality measures.

Table \ref{table:Counterfactuals} summarizes all results of this
back-of-the-envelope exercise showing a sizable contribution of declining
unions to the growing within-cell inequality between 1970 and 2010.
Unions account for 34-38\% of the changes in the Gini coefficient,
the 90/10 ratio, the top 10\% share, and the variance of log wages.
The contribution to the log college premium is even higher (at 61\%)
but imprecisely estimated. These estimates are moderately larger than
the \citet[Table 1]{farber2021unions} estimates of the contribution
of unions to within-state inequality in the US between 1968 and 2014.

\begin{table}
\begin{centering}
\caption{Counterfactual Inequality Measures\protect\label{table:Counterfactuals}}
\medskip{}
\par\end{centering}
\begin{centering}
\begin{tabular}{lcccccccc}
\toprule 
\multirow{2}{*}{Inequality measure} & \multicolumn{3}{c}{Actual} &  & \multicolumn{2}{c}{Counterfactual} &  & Union contribution\tabularnewline
\cmidrule{2-4}\cmidrule{6-7}
 & 1970 & 2010 & Change &  & 2010 & Change &  & as \% of change\tabularnewline
\midrule 
Log college premium & 0.430 & 0.579 & 0.149 &  & 0.489 & 0.059 &  & 60.5\%\tabularnewline
 &  &  &  &  & (0.062) &  &  & (41.8\%)\tabularnewline
Log 90/10 ratio & 1.113 & 1.428 & 0.314 &  & 1.322 & 0.208 &  & 33.7\%\tabularnewline
 &  &  &  &  & (0.058) &  &  & (18.3\%)\tabularnewline
Gini coefficient & 0.272 & 0.385 & 0.113 &  & 0.346 & 0.074 &  & 34.5\%\tabularnewline
 &  &  &  &  & (0.015) &  &  & (13.6\%)\tabularnewline
Top 10\% share & 0.226 & 0.310 & 0.084 &  & 0.278 & 0.052 &  & 38.3\%\tabularnewline
 &  &  &  &  & (0.014) &  &  & (17.3\%)\tabularnewline
Var(Log wage) & 0.220 & 0.362 & 0.143 &  & 0.310 & 0.091 &  & 36.4\%\tabularnewline
 &  &  &  &  & (0.021) &  &  & (15.0\%)\tabularnewline
\bottomrule
\end{tabular}
\par\end{centering}
\medskip{}

{\footnotesize\emph{Notes:}}{\footnotesize{} This table presents statistics
regarding the five measures of wage inequality within state-industry
cells used in Table \ref{table:unions_inequilty}. The 2010 counterfactual
measures are calculated by adding to each actual inequality measure
the product of the relevant treatment effect and the cumulative shortfall
in new unionization resulting from the sharp decline since the 1960s.
“Change” refers to the change from the actual measure in 1970
to the actual or counterfactual measure in 2010. The “Union Contribution”
is calculated as one minus the counterfactual change in each measure
divided by the actual change. The standard errors in parentheses are
based on the standard errors for the corresponding estimates in Table
\ref{table:unions_inequilty}.}{\footnotesize\par}

{\footnotesize\emph{Sources:}}{\footnotesize{} Decennial Census, NLRB
unionization data; authors' calculations.}{\footnotesize\par}
\end{table}

\section{Conclusion}

We introduced a novel extension to the regression discontinuity design,
termed Regression Discontinuity Aggregation (RDA), which aims at identifying
local average causal effects in contexts where each unit may be exposed
to multiple discontinuity events. The RDA approach is applicable when
the outcome is defined at a higher level of aggregation (e.g., in
space or time) than the discontinuity events or when spillovers of
discontinuity events are of interest. Our theoretical analysis rests
on the observation that an aggregation of RD events can be viewed
as a special case of a shift-share variable. This connection allowed
us to build two estimators for causal effects in RDA settings that
build on local linear estimation in conventional sharp and fuzzy RD
designs and inherit their bias-reduction properties.

To illustrate the usefulness of this approach, we applied the proposed
estimation techniques to the effects of unionization on earnings inequality
at the level of state-by-industry cells. We found that increased rates
of new unionization significantly reduce inequality, primarily by
lowering top-tier wages rather than raising lower-end wages. As a
result, the decline of unions since 1970s can explain a substantial
fraction of increased inequality. We additionally provide evidence
that unionization leads to increased pension coverage, potentially
in non-unionized establishments, too. Our findings provide causal
evidence supporting common perceptions about the effects of unionization
on inequality that were previously difficult to obtain as credibly.

\newpage
\bibliographystyle{plainnat}  
\bibliography{references}

\begin{thebibliography}{112}
\providecommand{\natexlab}[1]{#1}
\providecommand{\url}[1]{\texttt{#1}}
\expandafter\ifx\csname urlstyle\endcsname\relax
  \providecommand{\doi}[1]{doi: #1}\else
  \providecommand{\doi}{doi: \begingroup \urlstyle{rm}\Url}\fi

\bibitem[Abdulkadiroglu et~al.(2022)Abdulkadiroglu, Angrist, Narita, and
  Pathak]{abdul2022breaking}
Atila Abdulkadiroglu, Joshua~D Angrist, Yusuke Narita, and Parag Pathak.
\newblock Breaking ties: Regression discontinuity design meets market design.
\newblock \emph{Econometrica}, 90\penalty0 (1):\penalty0 117--151, 2022.

\bibitem[Akcigit et~al.(2023)Akcigit, Baslandze, and
  Lotti]{akcigit2023connecting}
Ufuk Akcigit, Salom{\'e} Baslandze, and Francesca Lotti.
\newblock Connecting to power: political connections, innovation, and firm
  dynamics.
\newblock \emph{Econometrica}, 91\penalty0 (2):\penalty0 529--564, 2023.

\bibitem[Akey(2015)]{akey2015valuing}
Pat Akey.
\newblock Valuing changes in political networks: Evidence from campaign
  contributions to close congressional elections.
\newblock \emph{The Review of Financial Studies}, 28\penalty0 (11):\penalty0
  3188--3223, 2015.

\bibitem[Almond et~al.(2010)Almond, Doyle~Jr, Kowalski, and
  Williams]{almond2010estimating}
Douglas Almond, Joseph~J Doyle~Jr, Amanda~E Kowalski, and Heidi Williams.
\newblock Estimating marginal returns to medical care: Evidence from at-risk
  newborns.
\newblock \emph{The Quarterly Journal of Economics}, 125\penalty0 (2):\penalty0
  591--634, 2010.

\bibitem[Angrist et~al.(2024)Angrist, Hull, Pathak, and
  Walters]{angrist2024credible}
Joshua Angrist, Peter Hull, Parag~A Pathak, and Christopher Walters.
\newblock Credible school value-added with undersubscribed school lotteries.
\newblock \emph{Review of Economics and Statistics}, 106\penalty0 (1):\penalty0
  1--19, 2024.

\bibitem[Angrist and Pischke(2009)]{angrist2009mostly}
Joshua~D Angrist and J{\"o}rn-Steffen Pischke.
\newblock \emph{Mostly harmless econometrics: An empiricist's companion}.
\newblock Princeton university press, 2009.

\bibitem[Ater et~al.(2021)Ater, Elster, and Hoffmann]{ater2021real}
Itai Ater, Yael Elster, and Eran~B Hoffmann.
\newblock Real-estate investors, house prices and rents: Evidence from
  capital-gains tax changes.
\newblock \emph{Maurice Falk Institute for Economic Research in Israel.
  Discussion paper series}, \penalty0 (1):\penalty0 1--33, 2021.

\bibitem[Auerbach et~al.(2024)Auerbach, Cai, and Rafi]{auerbach2024regression}
Eric Auerbach, Yong Cai, and Ahnaf Rafi.
\newblock Regression discontinuity design with spillovers.
\newblock \emph{arXiv preprint arXiv:2404.06471}, 2024.

\bibitem[Autor et~al.(2019)Autor, Dorn, and Hanson]{autor2019work}
David Autor, David Dorn, and Gordon Hanson.
\newblock When work disappears: Manufacturing decline and the falling marriage
  market value of young men.
\newblock \emph{American Economic Review: Insights}, 1\penalty0 (2):\penalty0
  161--178, 2019.

\bibitem[Autor et~al.(2008)Autor, Katz, and Kearney]{autor2008trends}
David~H Autor, Lawrence~F Katz, and Melissa~S Kearney.
\newblock Trends in us wage inequality: Revising the revisionists.
\newblock \emph{The Review of economics and statistics}, 90\penalty0
  (2):\penalty0 300--323, 2008.

\bibitem[Azoulay et~al.(2019)Azoulay, Graff~Zivin, Li, and
  Sampat]{azoulay2019public}
Pierre Azoulay, Joshua~S Graff~Zivin, Danielle Li, and Bhaven~N Sampat.
\newblock Public r\&d investments and private-sector patenting: evidence from
  nih funding rules.
\newblock \emph{The Review of economic studies}, 86\penalty0 (1):\penalty0
  117--152, 2019.

\bibitem[Bahar et~al.(2023)Bahar, Choudhury, Kim, and Koo]{bahar2023innovation}
Dany Bahar, Prithwiraj Choudhury, Do~Yoon Kim, and Wesley~W Koo.
\newblock Innovation on wings: Nonstop flights and firm innovation in the
  global context.
\newblock \emph{Management Science}, 69\penalty0 (10):\penalty0 6202--6223,
  2023.

\bibitem[Baron(2022)]{baron2022school}
E~Jason Baron.
\newblock School spending and student outcomes: Evidence from revenue limit
  elections in wisconsin.
\newblock \emph{American Economic Journal: Economic Policy}, 14\penalty0
  (1):\penalty0 1--39, 2022.

\bibitem[Baskaran and Hessami(2018)]{baskaran2018does}
Thushyanthan Baskaran and Zohal Hessami.
\newblock Does the election of a female leader clear the way for more women in
  politics?
\newblock \emph{American Economic Journal: Economic Policy}, 10\penalty0
  (3):\penalty0 95--121, 2018.

\bibitem[Baskaran and Hessami(2023)]{baskaran2023women}
Thushyanthan Baskaran and Zohal Hessami.
\newblock Women in political bodies as policymakers.
\newblock \emph{Review of Economics and Statistics}, pages 1--46, 2023.

\bibitem[Baskaran et~al.(2024{\natexlab{a}})Baskaran, Bhalotra, Min, and
  Uppal]{baskaran2024women}
Thushyanthan Baskaran, Sonia Bhalotra, Brian Min, and Yogesh Uppal.
\newblock Women legislators and economic performance.
\newblock \emph{Journal of Economic Growth}, 29\penalty0 (2):\penalty0
  151--214, 2024{\natexlab{a}}.

\bibitem[Baskaran et~al.(2024{\natexlab{b}})Baskaran, Hessami, and
  Schirner]{baskaran2024young}
Thushyanthan Baskaran, Zohal Hessami, and Sebastian Schirner.
\newblock Young versus old politicians and public spending priorities.
\newblock \emph{Journal of Economic Behavior \& Organization}, 225:\penalty0
  88--106, 2024{\natexlab{b}}.

\bibitem[Beach and Jones(2017)]{beach2017gridlock}
Brian Beach and Daniel~B Jones.
\newblock Gridlock: Ethnic diversity in government and the provision of public
  goods.
\newblock \emph{American Economic Journal: Economic Policy}, 9\penalty0
  (1):\penalty0 112--136, 2017.

\bibitem[Ben-Porath and Kolerman-Shemer(2024)]{ben-porath2023}
Netanel Ben-Porath and Matan Kolerman-Shemer.
\newblock Political priorities in a competitive landscape: Female leaders and
  the prevalence of abortions, 2024.

\bibitem[Besley and Case(2003)]{besley2003political}
Timothy Besley and Anne Case.
\newblock Political institutions and policy choices: evidence from the united
  states.
\newblock \emph{Journal of Economic Literature}, 41\penalty0 (1):\penalty0
  7--73, 2003.

\bibitem[Bhalotra and Clots-Figueras(2014)]{bhalotra2014health}
Sonia Bhalotra and Irma Clots-Figueras.
\newblock Health and the political agency of women.
\newblock \emph{American Economic Journal: Economic Policy}, 6\penalty0
  (2):\penalty0 164--197, 2014.

\bibitem[Bhalotra et~al.(2014)Bhalotra, Clots-Figueras, Cassan, and
  Iyer]{bhalotra2014religion}
Sonia Bhalotra, Irma Clots-Figueras, Guilhem Cassan, and Lakshmi Iyer.
\newblock Religion, politician identity and development outcomes: Evidence from
  india.
\newblock \emph{Journal of Economic Behavior \& Organization}, 104:\penalty0
  4--17, 2014.

\bibitem[Bhalotra et~al.(2018)Bhalotra, Clots-Figueras, and
  Iyer]{bhalotra2018pathbreakers}
Sonia Bhalotra, Irma Clots-Figueras, and Lakshmi Iyer.
\newblock Pathbreakers? women's electoral success and future political
  participation.
\newblock \emph{The Economic Journal}, 128\penalty0 (613):\penalty0 1844--1878,
  2018.

\bibitem[Bhalotra et~al.(2021)Bhalotra, Clots-Figueras, and
  Iyer]{bhalotra2021religion}
Sonia Bhalotra, Irma Clots-Figueras, and Lakshmi Iyer.
\newblock Religion and abortion: The role of politician identity.
\newblock \emph{Journal of Development Economics}, 153:\penalty0 102746, 2021.

\bibitem[Biasi(2021)]{biasi2021labor}
Barbara Biasi.
\newblock The labor market for teachers under different pay schemes.
\newblock \emph{American Economic Journal: Economic Policy}, 13\penalty0
  (3):\penalty0 63--102, 2021.

\bibitem[Biasi and Sarsons(2022)]{biasi2022flexible}
Barbara Biasi and Heather Sarsons.
\newblock Flexible wages, bargaining, and the gender gap.
\newblock \emph{The Quarterly Journal of Economics}, 137\penalty0 (1):\penalty0
  215--266, 2022.

\bibitem[Biasi et~al.(2024)Biasi, Lafortune, and
  Sch{\"o}nholzer]{biasi2024works}
Barbara Biasi, Julien~M Lafortune, and David Sch{\"o}nholzer.
\newblock What works and for whom? effectiveness and efficiency of school
  capital investments across the us.
\newblock Technical report, National Bureau of Economic Research, 2024.

\bibitem[Black(1999)]{black1999better}
Sandra~E Black.
\newblock Do better schools matter? parental valuation of elementary education.
\newblock \emph{The Quarterly Journal of Economics}, 114\penalty0 (2):\penalty0
  577--599, 1999.

\bibitem[Borusyak and Hull(2023)]{borusyak2023nonrandom}
Kirill Borusyak and Peter Hull.
\newblock Nonrandom exposure to exogenous shocks.
\newblock \emph{Econometrica}, 91\penalty0 (6):\penalty0 2155--2185, 2023.

\bibitem[Borusyak et~al.(2022)Borusyak, Hull, and Jaravel]{borusyak2022quasi}
Kirill Borusyak, Peter Hull, and Xavier Jaravel.
\newblock Quasi-experimental shift-share research designs.
\newblock \emph{The Review of Economic Studies}, 89\penalty0 (1):\penalty0
  181--213, 2022.

\bibitem[Borusyak et~al.(2024)Borusyak, Hull, and
  Jaravel]{borusyak2024practical}
Kirill Borusyak, Peter Hull, and Xavier Jaravel.
\newblock A practical guide to shift-share instruments.
\newblock Technical report, National Bureau of Economic Research, 2024.

\bibitem[Callaway and Collins(2018)]{callaway2018unions}
Brantly Callaway and William~J Collins.
\newblock Unions, workers, and wages at the peak of the american labor
  movement.
\newblock \emph{Explorations in Economic History}, 68:\penalty0 95--118, 2018.

\bibitem[Calonico et~al.(2014)Calonico, Cattaneo, and
  Titiunik]{calonico2014robust}
Sebastian Calonico, Matias~D Cattaneo, and Rocio Titiunik.
\newblock Robust nonparametric confidence intervals for
  regression-discontinuity designs.
\newblock \emph{Econometrica}, 82\penalty0 (6):\penalty0 2295--2326, 2014.

\bibitem[Calonico et~al.(2019)Calonico, Cattaneo, Farrell, and
  Titiunik]{calonico2019regression}
Sebastian Calonico, Matias~D Cattaneo, Max~H Farrell, and Rocio Titiunik.
\newblock Regression discontinuity designs using covariates.
\newblock \emph{Review of Economics and Statistics}, 101\penalty0 (3):\penalty0
  442--451, 2019.

\bibitem[Campante and Yanagizawa-Drott(2018)]{campante2018long}
Filipe Campante and David Yanagizawa-Drott.
\newblock Long-range growth: economic development in the global network of air
  links.
\newblock \emph{The Quarterly Journal of Economics}, 133\penalty0 (3):\penalty0
  1395--1458, 2018.

\bibitem[Card(1996)]{card1996effect}
David Card.
\newblock The effect of unions on the structure of wages: A longitudinal
  analysis.
\newblock \emph{Econometrica}, pages 957--979, 1996.

\bibitem[Card(2001)]{card2001effect}
David Card.
\newblock The effect of unions on wage inequality in the us labor market.
\newblock \emph{ILR Review}, 54\penalty0 (2):\penalty0 296--315, 2001.

\bibitem[Carozzi et~al.(2022)Carozzi, Cipullo, and
  Repetto]{carozzi2022political}
Felipe Carozzi, Davide Cipullo, and Luca Repetto.
\newblock Political fragmentation and government stability: Evidence from local
  governments in spain.
\newblock \emph{American Economic Journal: Applied Economics}, 14\penalty0
  (2):\penalty0 23--50, 2022.

\bibitem[Cattaneo and Titiunik(2022)]{cattaneo2022regression}
Matias~D Cattaneo and Rocio Titiunik.
\newblock Regression discontinuity designs.
\newblock \emph{Annual Review of Economics}, 14:\penalty0 821--851, 2022.

\bibitem[Cattaneo et~al.(2015)Cattaneo, Frandsen, and
  Titiunik]{cattaneo2015randomization}
Matias~D Cattaneo, Brigham~R Frandsen, and Rocio Titiunik.
\newblock Randomization inference in the regression discontinuity design: An
  application to party advantages in the us senate.
\newblock \emph{Journal of Causal Inference}, 3\penalty0 (1):\penalty0 1--24,
  2015.

\bibitem[Cellini et~al.(2010)Cellini, Ferreira, and
  Rothstein]{cellini2010value}
Stephanie~Riegg Cellini, Fernando Ferreira, and Jesse Rothstein.
\newblock The value of school facility investments: Evidence from a dynamic
  regression discontinuity design.
\newblock \emph{The Quarterly Journal of Economics}, 125\penalty0 (1):\penalty0
  215--261, 2010.

\bibitem[Census(1993)]{united19931987}
US~Census.
\newblock \emph{1987 Economic Censuses: Report series. Release 1E. Technical
  documentation. CD-ROM volume 1E}.
\newblock Number pt. 1 in 1987 Economic Censuses: Report Series. Release 1E.
  Technical Documentation. CD-ROM Volume 1E. U.S. Department of Commerce,
  Economics and Statistics Administration, Bureau of the Census, 1993.
\newblock URL \url{https://books.google.co.il/books?id=Co7iMN_5WNwC}.

\bibitem[Clark(2009)]{clark2009performance}
Damon Clark.
\newblock The performance and competitive effects of school autonomy.
\newblock \emph{Journal of political Economy}, 117\penalty0 (4):\penalty0
  745--783, 2009.

\bibitem[Clots-Figueras(2011)]{clots2011women}
Irma Clots-Figueras.
\newblock Women in politics: Evidence from the indian states.
\newblock \emph{Journal of public Economics}, 95\penalty0 (7-8):\penalty0
  664--690, 2011.

\bibitem[Clots-Figueras(2012)]{clots2012female}
Irma Clots-Figueras.
\newblock Are female leaders good for education? evidence from india.
\newblock \emph{American Economic Journal: Applied Economics}, 4\penalty0
  (1):\penalty0 212--44, 2012.

\bibitem[Collins and Niemesh(2019)]{collins2019unions}
William~J Collins and Gregory~T Niemesh.
\newblock Unions and the great compression of wage inequality in the us at
  mid-century: evidence from local labour markets.
\newblock \emph{The Economic History Review}, 72\penalty0 (2):\penalty0
  691--715, 2019.

\bibitem[Cooper et~al.(2010)Cooper, Gulen, and
  Ovtchinnikov]{cooper2010corporate}
Michael~J Cooper, Huseyin Gulen, and Alexei~V Ovtchinnikov.
\newblock Corporate political contributions and stock returns.
\newblock \emph{The Journal of Finance}, 65\penalty0 (2):\penalty0 687--724,
  2010.

\bibitem[Dahl et~al.(2014)Dahl, L{\o}ken, and Mogstad]{dahl2014peer}
Gordon~B Dahl, Katrine~V L{\o}ken, and Magne Mogstad.
\newblock Peer effects in program participation.
\newblock \emph{American Economic Review}, 104\penalty0 (7):\penalty0
  2049--2074, 2014.

\bibitem[Darolia(2013)]{darolia2013integrity}
Rajeev Darolia.
\newblock Integrity versus access? the effect of federal financial aid
  availability on postsecondary enrollment.
\newblock \emph{Journal of Public Economics}, 106:\penalty0 101--114, 2013.

\bibitem[de~Benedictis-Kessner and Warshaw(2020)]{de2020politics}
Justin de~Benedictis-Kessner and Christopher Warshaw.
\newblock Politics in forgotten governments: the partisan composition of county
  legislatures and county fiscal policies.
\newblock \emph{The Journal of Politics}, 82\penalty0 (2):\penalty0 460--475,
  2020.

\bibitem[de~Benedictis-Kessner et~al.(2024)de~Benedictis-Kessner, Jones, and
  Warshaw]{de2024partisanship}
Justin de~Benedictis-Kessner, Daniel Jones, and Christopher Warshaw.
\newblock How partisanship in cities influences housing policy.
\newblock \emph{American Journal of Political Science}, 2024.

\bibitem[De~Chaisemartin and d'Haultfoeuille(2020)]{de2020two}
Cl{\'e}ment De~Chaisemartin and Xavier d'Haultfoeuille.
\newblock Two-way fixed effects estimators with heterogeneous treatment
  effects.
\newblock \emph{American Economic Review}, 110\penalty0 (9):\penalty0
  2964--2996, 2020.

\bibitem[Dechezlepr{\^e}tre et~al.(2023)Dechezlepr{\^e}tre, Eini{\"o}, Martin,
  Nguyen, and Van~Reenen]{dechezlepretre2023tax}
Antoine Dechezlepr{\^e}tre, Elias Eini{\"o}, Ralf Martin, Kieu-Trang Nguyen,
  and John Van~Reenen.
\newblock Do tax incentives increase firm innovation? an rd design for r\&d,
  patents, and spillovers.
\newblock \emph{American Economic Journal: Economic Policy}, 15\penalty0
  (4):\penalty0 486--521, 2023.

\bibitem[Dell(2010)]{dell2010persistent}
Melissa Dell.
\newblock The persistent effects of peru's mining mita.
\newblock \emph{Econometrica}, 78\penalty0 (6):\penalty0 1863--1903, 2010.

\bibitem[Dell and Querubin(2018)]{dell2018nation}
Melissa Dell and Pablo Querubin.
\newblock Nation building through foreign intervention: Evidence from
  discontinuities in military strategies.
\newblock \emph{The Quarterly Journal of Economics}, 133\penalty0 (2):\penalty0
  701--764, 2018.

\bibitem[DiNardo and Lee(2004)]{dinardo2004economic}
John DiNardo and David~S Lee.
\newblock Economic impacts of new unionization on private sector employers:
  1984--2001.
\newblock \emph{The Quarterly Journal of Economics}, 119\penalty0 (4):\penalty0
  1383--1441, 2004.

\bibitem[Dube et~al.(2019)Dube, Giuliano, and Leonard]{dube2019fairness}
Arindrajit Dube, Laura Giuliano, and Jonathan Leonard.
\newblock Fairness and frictions: The impact of unequal raises on quit
  behavior.
\newblock \emph{American Economic Review}, 109\penalty0 (2):\penalty0 620--663,
  2019.

\bibitem[Eckert et~al.(2021)Eckert, Fort, Schott, and Yang]{eckert2021imputing}
Fabian Eckert, Teresa~C. Fort, Peter~K. Schott, and Natalie~J. Yang.
\newblock Imputing missing values in the us census bureau's county business
  patterns.
\newblock Technical report, National Bureau of Economic Research, 2021.

\bibitem[Farber et~al.(2021)Farber, Herbst, Kuziemko, and
  Naidu]{farber2021unions}
Henry~S Farber, Daniel Herbst, Ilyana Kuziemko, and Suresh Naidu.
\newblock Unions and inequality over the twentieth century: New evidence from
  survey data.
\newblock \emph{The Quarterly Journal of Economics}, 136\penalty0 (3):\penalty0
  1325--1385, 2021.

\bibitem[Feigenbaum et~al.(2019)Feigenbaum, Hertel-Fernandez, and
  Williamson]{feigenbaum2019bargaining}
James Feigenbaum, Alexander Hertel-Fernandez, and Vanessa Williamson.
\newblock From the bargaining table to the ballot box: Downstream effects of
  right-to-work laws.
\newblock Technical report, Working Paper, 2019.

\bibitem[Ferreira and Gyourko(2014)]{ferreira2014does}
Fernando Ferreira and Joseph Gyourko.
\newblock Does gender matter for political leadership? the case of us mayors.
\newblock \emph{Journal of Public Economics}, 112:\penalty0 24--39, 2014.

\bibitem[Folke(2014)]{folke2014shades}
Olle Folke.
\newblock Shades of brown and green: party effects in proportional election
  systems.
\newblock \emph{Journal of the European Economic Association}, 12\penalty0
  (5):\penalty0 1361--1395, 2014.

\bibitem[Fortin et~al.(2021)Fortin, Lemieux, and Lloyd]{fortin2021labor}
Nicole Fortin, Thomas Lemieux, and Neil Lloyd.
\newblock Labor market institutions and the distribution of wages: The role of
  spillover effects.
\newblock \emph{Journal of Labor Economics}, 39\penalty0 (S2):\penalty0
  S369--S412, 2021.

\bibitem[Fortin et~al.(2022)Fortin, Lemieux, and Lloyd]{fortin2022right}
Nicole Fortin, Thomas Lemieux, and Neil Lloyd.
\newblock Right-to-work laws, unionization, and wage setting.
\newblock Technical report, National Bureau of Economic Research, 2022.

\bibitem[Frandsen(2017)]{frandsen2017party}
Brigham~R. Frandsen.
\newblock Party bias in union representation elections: Testing for
  manipulation in the regression discontinuity design when the running variable
  is discrete.
\newblock In \emph{Regression discontinuity designs}. Emerald Publishing
  Limited, 2017.

\bibitem[Frandsen(2021)]{frandsen_2021}
Brigham~R. Frandsen.
\newblock The surprising impacts of unionization: Evidence from matched
  employer-employee data.
\newblock \emph{Journal of Labor Economics}, 39\penalty0 (4):\penalty0
  861--894, 2021.
\newblock \doi{10.1086/711852}.

\bibitem[Fu and Gregory(2019)]{fu2019estimation}
Chao Fu and Jesse Gregory.
\newblock Estimation of an equilibrium model with externalities: Post-disaster
  neighborhood rebuilding.
\newblock \emph{Econometrica}, 87\penalty0 (2):\penalty0 387--421, 2019.

\bibitem[Gelman and Imbens(2019)]{gelman2019high}
Andrew Gelman and Guido Imbens.
\newblock Why high-order polynomials should not be used in regression
  discontinuity designs.
\newblock \emph{Journal of Business \& Economic Statistics}, 37\penalty0
  (3):\penalty0 447--456, 2019.

\bibitem[Geys et~al.(2024)Geys, Murdoch, and S{\o}rensen]{geys2024public}
Benny Geys, Zuzana Murdoch, and Rune~J S{\o}rensen.
\newblock Public employees as elected politicians: Assessing direct and
  indirect substantive effects of passive representation.
\newblock \emph{The Journal of Politics}, 86\penalty0 (1):\penalty0 170--182,
  2024.

\bibitem[Helpman et~al.(2016)Helpman, Itskhoki, Muendler, and
  Redding]{helpman2016trade}
Elhanan Helpman, Oleg Itskhoki, Marc-Andreas Muendler, and Stephen~J Redding.
\newblock Trade and inequality: From theory to estimation.
\newblock \emph{The Review of Economic Studies}, 84\penalty0 (1):\penalty0
  357--405, 2016.

\bibitem[Holmes(1998)]{holmes1998effect}
Thomas~J Holmes.
\newblock The effect of state policies on the location of manufacturing:
  Evidence from state borders.
\newblock \emph{Journal of political Economy}, 106\penalty0 (4):\penalty0
  667--705, 1998.

\bibitem[Hsu and Shen(2021)]{hsu2021dynamic}
YC~Hsu and S~Shen.
\newblock Dynamic regression discontinuity under treatment effect
  heterogeneity.
\newblock \emph{Work. Pap., Univ. Calif., Davis}, 2021.

\bibitem[Hyytinen et~al.(2018)Hyytinen, Meril{\"a}inen, Saarimaa, Toivanen, and
  Tukiainen]{hyytinen2018public}
Ari Hyytinen, Jaakko Meril{\"a}inen, Tuukka Saarimaa, Otto Toivanen, and Janne
  Tukiainen.
\newblock Public employees as politicians: Evidence from close elections.
\newblock \emph{American Political Science Review}, 112\penalty0 (1):\penalty0
  68--81, 2018.

\bibitem[Imbens and Kalyanaraman(2012)]{imbens2012optimal}
Guido Imbens and Karthik Kalyanaraman.
\newblock Optimal bandwidth choice for the regression discontinuity estimator.
\newblock \emph{The Review of economic studies}, 79\penalty0 (3):\penalty0
  933--959, 2012.

\bibitem[{IPUMS USA}(2024)]{ipums_1970_countygroups}
{IPUMS USA}.
\newblock 1970 county group definitions: Composition file, 2024.
\newblock URL \url{https://usa.ipums.org/usa/volii/t1970maps.shtml}.
\newblock Geographic Tools Documentation.

\bibitem[Isen(2014)]{isen2014local}
Adam Isen.
\newblock Do local government fiscal spillovers exist? evidence from counties,
  municipalities, and school districts.
\newblock \emph{Journal of Public Economics}, 110:\penalty0 57--73, 2014.

\bibitem[Jackson and Mackevicius(2024)]{jackson2024impacts}
C~Kirabo Jackson and Claire~L Mackevicius.
\newblock What impacts can we expect from school spending policy? evidence from
  evaluations in the united states.
\newblock \emph{American Economic Journal: Applied Economics}, 16\penalty0
  (1):\penalty0 412--446, 2024.

\bibitem[Johnson(2020)]{johnson2020regulation}
Matthew~S Johnson.
\newblock Regulation by shaming: Deterrence effects of publicizing violations
  of workplace safety and health laws.
\newblock \emph{American economic review}, 110\penalty0 (6):\penalty0
  1866--1904, 2020.

\bibitem[Jones et~al.(2022)Jones, Kondylis, Loeser, and
  Magruder]{jones2022factor}
Maria Jones, Florence Kondylis, John Loeser, and Jeremy Magruder.
\newblock Factor market failures and the adoption of irrigation in rwanda.
\newblock \emph{American Economic Review}, 112\penalty0 (7):\penalty0
  2316--2352, 2022.

\bibitem[Keele and Titiunik(2018)]{keele2018geographic}
Luke Keele and Roc{\'\i}o Titiunik.
\newblock Geographic natural experiments with interference: The effect of
  all-mail voting on turnout in colorado.
\newblock \emph{CESifo Economic Studies}, 64\penalty0 (2):\penalty0 127--149,
  2018.

\bibitem[Knepper(2020)]{knepper2020fringe}
Matthew Knepper.
\newblock From the fringe to the fore: Labor unions and employee compensation.
\newblock \emph{Review of Economics and Statistics}, 102\penalty0 (1):\penalty0
  98--112, 2020.

\bibitem[Kolerman-Shemer(2023)]{Kolerman2023}
Matan Kolerman-Shemer.
\newblock Political impacts of trade unions, 2023.

\bibitem[Kott(2022)]{kott2022income}
Assaf Kott.
\newblock Income shocks, school choice, and long-term outcomes: Lessons from
  child allowances in israel.
\newblock 2022.

\bibitem[Kuliomina(2021)]{kuliomina2021personal}
Jekaterina Kuliomina.
\newblock Do personal characteristics of councilors affect municipal budget
  allocation?
\newblock \emph{European Journal of Political Economy}, 70:\penalty0 102034,
  2021.

\bibitem[Lang(2018)]{lang2018assessing}
Corey Lang.
\newblock Assessing the efficiency of local open space provision.
\newblock \emph{Journal of Public Economics}, 158:\penalty0 12--24, 2018.

\bibitem[Lee(2008)]{lee2008randomized}
David~S Lee.
\newblock Randomized experiments from non-random selection in us house
  elections.
\newblock \emph{Journal of Econometrics}, 142\penalty0 (2):\penalty0 675--697,
  2008.

\bibitem[Lee and Mas(2012)]{lee2012long}
David~S Lee and Alexandre Mas.
\newblock Long-run impacts of unions on firms: New evidence from financial
  markets, 1961--1999.
\newblock \emph{The Quarterly Journal of Economics}, 127\penalty0 (1):\penalty0
  333--378, 2012.

\bibitem[Lewis(1963)]{lewis1963unionism}
H.~Gregg Lewis.
\newblock \emph{Unionism and Relative Wages in the United States}.
\newblock University of Chicago Press, Chicago, 1963.

\bibitem[Makridis(2019)]{makridis2019right}
Christos~Andreas Makridis.
\newblock Do right-to-work laws work? evidence on individuals' well-being and
  economic sentiment.
\newblock \emph{The Journal of Law and Economics}, 62\penalty0 (4):\penalty0
  713--745, 2019.

\bibitem[Manski(1993)]{manski1993identification}
Charles~F Manski.
\newblock Identification of endogenous social effects: The reflection problem.
\newblock \emph{The review of economic studies}, 60\penalty0 (3):\penalty0
  531--542, 1993.

\bibitem[Martorell et~al.(2016)Martorell, Stange, and
  McFarlin~Jr]{martorell2016investing}
Paco Martorell, Kevin Stange, and Isaac McFarlin~Jr.
\newblock Investing in schools: capital spending, facility conditions, and
  student achievement.
\newblock \emph{Journal of Public Economics}, 140:\penalty0 13--29, 2016.

\bibitem[Meril{\"a}inen(2022)]{merilainen2022political}
Jaakko Meril{\"a}inen.
\newblock Political selection and economic policy.
\newblock \emph{The Economic Journal}, 132\penalty0 (648):\penalty0 3020--3046,
  2022.

\bibitem[Mora-Garc{\'\i}a and Rau(2023)]{mora2023peer}
Claudio~A Mora-Garc{\'\i}a and Tom{\'a}s Rau.
\newblock Peer effects in the adoption of a youth employment subsidy.
\newblock \emph{Review of Economics and Statistics}, 105\penalty0 (3):\penalty0
  614--625, 2023.

\bibitem[Narita and Yata(2023)]{narita2021algorithm}
Yusuke Narita and Kohei Yata.
\newblock Algorithm is experiment: Machine learning, market design, and policy
  eligibility rules.
\newblock \emph{arXiv preprint arXiv:2104.12909}, 2023.

\bibitem[{National Labor Relations Board}(1997)]{nlrb1997basic}
{National Labor Relations Board}.
\newblock Basic guide to the national labor relations act: General principles
  of law under the statute and procedures of the national labor relations
  board.
\newblock Government document, National Labor Relations Board, Washington, DC,
  1997.
\newblock Originally issued in 1962, revised edition.

\bibitem[Nellis and Siddiqui(2018)]{nellis2018secular}
Gareth Nellis and Niloufer Siddiqui.
\newblock Secular party rule and religious violence in pakistan.
\newblock \emph{American Political Science Review}, 112\penalty0 (1):\penalty0
  49--67, 2018.

\bibitem[Nellis et~al.(2016)Nellis, Weaver, Rosenzweig,
  et~al.]{nellis2016parties}
Gareth Nellis, Michael Weaver, Steven~C Rosenzweig, et~al.
\newblock Do parties matter for ethnic violence? evidence from india.
\newblock \emph{Quarterly Journal of Political Science}, 11\penalty0
  (3):\penalty0 249--277, 2016.

\bibitem[Obama(2016)]{Obama2016}
Barack Obama.
\newblock President obama's open letter to america's hardworking men and women.
\newblock
  \url{https://obamawhitehouse.archives.gov/blog/2016/09/04/president-obama-letter-americas-hardworking-men-and-women},
  September 2016.
\newblock At 8:00 PM ET.

\bibitem[Piketty et~al.(2018)Piketty, Saez, and
  Zucman]{piketty2018distributional}
Thomas Piketty, Emmanuel Saez, and Gabriel Zucman.
\newblock Distributional national accounts: methods and estimates for the
  united states.
\newblock \emph{The Quarterly Journal of Economics}, 133\penalty0 (2):\penalty0
  553--609, 2018.

\bibitem[Priyanka(2020)]{priyanka2020female}
Sadia Priyanka.
\newblock Do female politicians matter for female labor market outcomes?
  evidence from state legislative elections in india.
\newblock \emph{Labour Economics}, 64:\penalty0 101822, 2020.

\bibitem[Saez et~al.(2019)Saez, Schoefer, and Seim]{saez2019payroll}
Emmanuel Saez, Benjamin Schoefer, and David Seim.
\newblock Payroll taxes, firm behavior, and rent sharing: Evidence from a young
  workers' tax cut in sweden.
\newblock \emph{American Economic Review}, 109\penalty0 (5):\penalty0
  1717--1763, 2019.

\bibitem[Santoleri et~al.(2022)Santoleri, Mina, Di~Minin, and
  Martelli]{santoleri2022causal}
Pietro Santoleri, Andrea Mina, Alberto Di~Minin, and Irene Martelli.
\newblock The causal effects of r\&d grants: Evidence from a regression
  discontinuity.
\newblock \emph{Review of Economics and Statistics}, pages 1--42, 2022.

\bibitem[Small et~al.(2017)Small, Tan, Ramsahai, Lorch, and
  Brookhart]{small2017instrumental}
Dylan~S Small, Zhiqiang Tan, Roland~R Ramsahai, Scott~A Lorch, and M~Alan
  Brookhart.
\newblock Instrumental variable estimation with a stochastic monotonicity
  assumption.
\newblock 2017.

\bibitem[Sojourner et~al.(2015)Sojourner, Frandsen, Town, Grabowski, and
  Chen]{sojourner2015impacts}
Aaron~J Sojourner, Brigham~R Frandsen, Robert~J Town, David~C Grabowski, and
  Min~M Chen.
\newblock Impacts of unionization on quality and productivity: Regression
  discontinuity evidence from nursing homes.
\newblock \emph{ILR Review}, 68\penalty0 (4):\penalty0 771--806, 2015.

\bibitem[S{\o}rensen(2023)]{sorensen2023educated}
Rune~J S{\o}rensen.
\newblock Educated politicians and government efficiency: Evidence from
  norwegian local government.
\newblock \emph{Journal of Economic Behavior \& Organization}, 210:\penalty0
  163--179, 2023.

\bibitem[Tan(2023)]{tan2023consequences}
Brandon~Joel Tan.
\newblock The consequences of letter grades for labor market outcomes and
  student behavior.
\newblock \emph{Journal of Labor Economics}, 41\penalty0 (3):\penalty0
  565--588, 2023.

\bibitem[Torrione et~al.(2024)Torrione, Arduini, and
  Forastiere]{torrione2024regression}
Elena~Dal Torrione, Tiziano Arduini, and Laura Forastiere.
\newblock Regression discontinuity designs under interference.
\newblock \emph{arXiv preprint arXiv:2410.02727}, 2024.

\bibitem[Vaghul and Zipperer(2016)]{vaghul2016historical}
Kavya Vaghul and Ben Zipperer.
\newblock Historical state and sub-state minimum wage data.
\newblock 2016.

\bibitem[Valentim and Dinas(2024)]{valentim2024does}
Vicente Valentim and Elias Dinas.
\newblock Does party-system fragmentation affect the quality of democracy?
\newblock \emph{British Journal of Political Science}, 54\penalty0
  (1):\penalty0 152--178, 2024.

\bibitem[Wang and Young(2022)]{wang2022unionization}
Sean Wang and Samuel Young.
\newblock Unionization, employer opposition, and establishment closure.
\newblock \emph{Essays on Employment and Human Capital, PhD diss. MIT}, 2022.

\bibitem[Yitzhaki(1996)]{yitzhaki1996using}
Shlomo Yitzhaki.
\newblock On using linear regressions in welfare economics.
\newblock \emph{Journal of Business \& Economic Statistics}, 14\penalty0
  (4):\penalty0 478--486, 1996.

\bibitem[Zigler and Papadogeorgou(2021)]{zigler2021bipartite}
Corwin~M Zigler and Georgia Papadogeorgou.
\newblock Bipartite causal inference with interference.
\newblock \emph{Statistical science: a review journal of the Institute of
  Mathematical Statistics}, 36\penalty0 (1):\penalty0 109, 2021.

\end{thebibliography}

\appendix
\begin{center}
{\huge\textbf{Appendix}}{\huge\par}
\par\end{center}

\setcounter{table}{0} \renewcommand{\thetable}{A\arabic{table}}
\setcounter{figure}{0} \renewcommand{\thefigure}{A\arabic{figure}}
\renewcommand\theHtable{Appendix.\thetable}\renewcommand\theHfigure{Appendix.\thefigure}

\section{\protect\label{sec:Appx-Proof}Proof of Proposition \ref{prop:identification}}

We start with the lower-level solution. Assuming that the population
coefficients on $r_{j}$ and $r_{j}^{+}$ in the first-stage and reduced-form
of (\ref{eq:model-j-lowerlevel}) have some limits as $h\to0$, including
them in the specification does not affect the set of estimands, as
both of these variables have no variation in the limit. Thus, the
estimand limit is
\[
\lim_{h\downarrow0}\beta_{h}^{\ell}=\frac{\lim_{r\downarrow0}\expec{s_{j}Y_{\i(j)}\mid r_{j}=r}-\lim_{r\uparrow0}\expec{s_{j}Y_{\i(j)}\mid r_{j}=r}}{\lim_{r\downarrow0}\expec{s_{j}X_{\i(j)}\mid r_{j}=r}-\lim_{r\uparrow0}\expec{s_{j}X_{\i(j)}\mid r_{j}=r}}.
\]
We can expand $\expec{Y_{\i(j)}\mid r_{j}=r}=\expec{Y_{\i(j)}(X_{\i(j)}(1,\boldsymbol{z}_{\i(j)-j}))\mid r_{j}=r}$
for $r\ge0$, with analogous expressions for other terms. Thus, by
Assumption 3a, 
\[
\beta_{h}^{\ell}=\frac{\expec{s_{j}Y_{\i(j)}(X_{\i(j)}(1,\boldsymbol{z}_{\i(j)-j}))\mid r_{j}=0}-\expec{s_{j}Y_{\i(j)}(X_{\i(j)}(0,\boldsymbol{z}_{\i(j)-j}))\mid r_{j}=0}}{\expec{s_{j}X_{\i(j)}(1,\boldsymbol{z}_{\i(j)-j})\mid r_{j}=0}-\expec{s_{j}X_{\i(j)}(0,\boldsymbol{z}_{\i(j)-j})\mid r_{j}=0}},
\]
where conditioning on $r_{j}=0$ is well-defined by Assumption 4.
By Assumption 5, the denominator is positive. Moreover, by standard
arguments (e.g., \citet{yitzhaki1996using}), it is a convexly-weighted
average of $\partial Y_{i}(x)/\partial x$ (or the discrete analog
thereof when $X_{i}$ is discrete) across both units and values of
treatment.

By Proposition 1, the upper-level solution only differs by residualizing
$Y_{i}$ and $X_{i}$ on $Q_{i}$. By Assumption 3b, we have $Q_{i}\approx\left(s_{j},0,0\right)$
conditionally on $-h\le r_{j}\le h$ as $h\to0$. Invoking Assumption
3b again and assuming again that the population coefficients on $Q_{i}$
in the first-stage and reduced-form of (\ref{eq:model-i}) converge,
this implies that residualizing on $Q_{i}$ makes no difference in
(\ref{eq:model-j-equiv}), provided an intercept is included in that
subunit-level specification. This concludes the proof.

\section{Data Appendix}

Our analysis is at the level of state-industry cells for each decennial
census year from 1960–2010 (i.e., each year that divides by ten).
We begin by discussing the data construction steps for our three data
sources separately: decennial censuses, CPS, the unionization elections
data. We then turn to how we merge them into a consistent cell-level
panel; this involves creating a harmonized industry classification
and computing cell-level variables, such as inequality measures.

\paragraph{Decennial census data.}

Census public-use micro-data were downloaded from the IPUMS USA website.
For 1960, 1980, 1990, and 2000, we used the 5\% “state” samples.
For 1970, we combined the 2\% “state” and 2\% “metro”
samples. To represent 2010, we combine the American Community Survey
(ACS) samples for 2009–2013. All of these samples are nationally representative
for the US population.

We assign each individual to their state of residence. While state
of employment could be more relevant for our analysis, it is not available
in some of the samples. The state of residence is observed in most
samples; the only exception is the 1970 metro samples which do not
report the state for the 17\% of individuals living in metropolitan
areas and other county groups that cross state boundaries. We duplicate
observations for those individuals and adjust their projection weights
to equal the original weight multiplied by each state's proportion
of the metropolitan area or county group's total population.\footnote{We used \citet{ipums_1970_countygroups} to identify the county composition
of each metropolitan area and county group, and obtained the 1970
county populations from IPUMS NHGIS.} We discuss the assignment of workers to industries below.

For each worker, we observe their pre-tax annual labor income (variable
\texttt{INCWAGE}). Top-coded income values are multiplied by 1.5 \citep{autor2008trends}.
We then compute the hourly wage as the income dividing by the weeks
of work and the average number of hours per week.

We measure workers' education and potential work experience using
the most detailed education question (\texttt{EDUCD}). The groupings
defined by this question vary across samples, sometimes closer aligned
with the years of schooling and other times with the highest degree
obtained. In all cases, it is sufficiently detailed to determine whether
the individual has exactly a high school degree, exactly a college
degree, or other education levels above and below. It is also sufficient
to measure years of schooling, at least approximately, which we use
define potential work experience as the age minus 6 and minus the
years of schooling.

We classify a worker's occupation as managerial if the occupation
variable (\texttt{OCC1990}) belongs to the groups “Executive, Administrative,
and Managerial” and “Management Related”.

We impose several sample restrictions. First, we focus on 18–65 year-old
adults with 0–48 years of potential experience. Second, we only keep
workers who have been employed for all 52 weeks during the year preceding
the survey and worked at least 20 hours a week on average.\footnote{Since 1980, the variable is based on the \texttt{UHRSWORK} variable
that measures “usual hours worked per week {[}in the last year{]}.”
Unfortunately, this variable is not available in the 1970 and the
1960 samples, so for those years we use the \texttt{HRSWORK2} variable
that measures “hours worked last week” instead.} Individuals who worked fewer weeks likely changed jobs within the
year, making it more difficult to correctly assign them to an industry.
Third, we require that workers earn at least half their state's minimum
hourly wage (obtained from \citet{vaghul2016historical}). Fourth,
we consider only workers in the private sector who are not self-employed;
those are the workers who can unionize through NLRB unionization elections.

\paragraph{CPS data.}

To capture variables not available in the census—unionization rates
and fringe benefits—we leverage CPS samples, downloaded from IPUMS
CPS.\footnote{With the exception of the May supplement for 1979, which was downloaded
from NBER.} To measure unionization rates, we use CPS MORG (Merged Outgoing Rotation
Group) samples from 1983–2015, along with the May supplement to the
CPS in 1977–1980. The surveys ask workers whether they are union members
(\texttt{UNION} variable). For fringe benefits, we use CPS ASEC (Annual
Social and Economic Supplements) from 1980–2014, along with the May
supplement in 1979. The workers are asked if they are covered by an
employer-sponsored pension plan (\texttt{PENSION} variable) and whether
they enrolled in their employer's group health plan during the previous
year (\texttt{INCLUGH} variable). We use the same four individual-level
sample restrictions as in the decennial census. Imputed answers were
dropped.

To increase sample size when measuring changes in the rates of unionization,
pension, and insurance coverage between adjacent census years, we
pool data from multiple years of the CPS. Where possible, we compare
the five years after the end of the decade to five years before the
beginning of the decade. For instance, the differenced outcome for
the decade of the 2000s is calculated as the change between 1986–90
and 2000–04. We deviate from this rule in the 1980s because of data
availability; Appendix Table \ref{table:years-of-data} provides the
details.

\paragraph{Unionization elections.}

We use the NLRB (National Labor Relations Board) unionization election
data maintained by Hank Farber and John-Paul Ferguson. We keep observations
from 1961—the earliest available year—until the beginning of 2009,
after which industry variables are no longer available. The original
dataset includes 250,063 elections.

Each election is attributed to a geographic location (directly including
the state), an industry, and a decade. We generally define decades
as starting in the year that ends with 0 and finishing in the year
which ends with 9, e.g. 1980–1989. The earliest and latest decades
are slightly shorter because, as just mentioned, the data start in
1961 and end early in 2009. We discuss industry classifications below.

We observe the number of eligible voters in the bargaining unit, the
number of votes cast for and against the union, as well as whether
the union has been formed. In most cases, the union is formed when
the vote share (out of the total votes cast) strictly exceeds 50\%,
but in a small fraction of cases this is not the case, perhaps because
of the reversal during the appeal process. We further observe whether
the election is a union decertification election. In the main analysis,
we do not distinguish between certification and decertification elections,
counting workers in re-certified unions as newly unionized; we make
other choices in a robustness check, as well as in Section \ref{subsec:Application-Magnitudes}.

While we use the full set of union elections to construct the treatment
variable,\footnote{Except a small fraction that we cannot attribute to an industry. This
explains the slightly smaller total number of elections of 232,267
in Table \ref{table:summary_lower}. We include the entire set of
250,063 elections in the counterfactual analysis of Section \ref{subsec:Application-Magnitudes}.} our instrument is constructed only from the elections that satisfy
several criteria. Of course, it only includes close elections, defined
by the vote share in favor of the union falling within a bandwidth
around the cutoff of 50\%. We make two additional restrictions. First,
following common practice in the unionization RD literature, elections
with fewer than 20 votes were excluded \citep{dinardo2004economic,lee2012long,frandsen_2021}.
Second, we drop elections that ended in a tie. Although by the National
Labor Relations Act a tie is equivalent to a union loss, including
those elections in the sample may lead to bias. Elections with an
even number of votes will mechanically have a much higher chance of
resulting in a close union loss than in a close union win. For example,
consider an election with 20 votes given bandwidth of 10\%. There
are there outcomes resulting in a close loss (8-12, 9-11, and 10-1)
but only two outcomes resulting in a close win (12-8 and 11-9). Excluding
the tie, we prevent a negative mechanical correlation between the
election result and workplace size.\footnote{In support of this argument, \citet{wang2022unionization} provides
empirical evidence that tied elections are very different from elections
where the union lost or won by a slightly larger margin.}

Third, following \citet{frandsen2017party,frandsen_2021}, the instrument
construction also excludes elections that ended with a one-vote margin
in either direction. \citet{frandsen2017party,frandsen_2021} documented
the imbalance of covariates and unequal frequency of union wins and
losses in such elections. Figure \ref{fig:votes_diff} confirms this
in our dataset, showing approximately 18,000 “missing” workers
in elections that ended with exactly a one-vote margin in favor of
the unions. A presumed explanation is that published unionization
result may in some cases report vote counts post appeals and other
legal process, and that the tendency to appeal and the acceptance
rate of those appeals is heterogeneous between close wins and close
losses.

Table \ref{table:summary_lower} presents summary statistics for the
full set of unionization elections, as well as the restricted set
for the instrument construction. The average vote share for the unions
and the share of union wins are close to 50\% in both samples. The
average number of votes in the treatment (instrument) sample is 54.9
(98.3), and the share of manufacturing establishments is 43\% (52\%).
In both samples, over half of the elections occurred in the 1960s
and 1970s, which constitute only 40\% of the study period, reflecting
the decline in unionization since 1980.

\paragraph{Industry classification.}

To construct both labor market outcomes and new unionization rates
by state-industry cells, we had to create an industry classification
harmonized both across data sources and over time. The classifications
in all our data sources are based, in one way or another, on SIC codes
before around 2000 and on NAICS codes thereafter. Thus, we also created
two harmonized classifications for the two subperiods, resulting in
70 industries based on 2-digit SIC87 codes and 73 industries based
on 3-digit NAICS97 codes covering the entire private sector of the
economy. We use our SIC-based classification for the 1960s–1990s and
the NAICS-based classification for the 2000s.

For union elections filed before the end of 1998 (which covers most
elections that took place before the end of 1999), the NLRB data report
the 3-digit SIC code (with some changes in the classification over
time discussed below). The decennial censuses up until 1990 and the
CPS up until 2002 use a different classification—variable \texttt{IND1990}—that
was harmonized over time by IPUMS.\footnote{Note that the harmonization process conducted by IPUMS follows a majority
rule, assigning each original industry code to the harmonized industry
that the majority of observations would have been assigned to. This
process may lead to some inconsistencies between the census and the
unionization data in earlier decades.} This classification is loosely based on on 3-digit SIC87 codes. To
minimize measurement error, we convert \texttt{IND1990} industries
to 2-digit SIC87 codes by leveraging crosswalks created by David Dorn
\citep{autor2019work}. Specifically, he published many-to-one crosswalks
from \texttt{IND1990} to a slightly aggregated classification developed
by him called \texttt{IND1990ddx}, as well as from 4-digit SIC87 codes
to \texttt{IND1990ddx}. While the resulting mapping from \texttt{IND1990}
to the 3-digit SIC87 codes that we observe in the NLRB data would
be many-to-many, the mapping to 2-digit SIC87 codes is mostly many-to-one.
In the rare cases where this is not the case, we merge those 2-digit
SIC87 codes together, resulting in the 70 industry codes in our classification.

We make adjustments for the changes in the SIC classifications over
time in the NLRB data. While SIC87 codes are reported since 1987,
the SIC72 classification is used between 1972 and 1986. In those years,
we only keep the elections in industries for which over 95\% of national
employment mapped to one of our (significantly more aggregated) codes.\footnote{This calculation relies on the economic censuses of 1987 \citep{united19931987},
which report total employment data according to both SIC72 and SIC87.} Elections prior to 1972 use even earlier SIC schemes. We drop all
elections that did not align with a single code in our classification
using the crosswalk used by \citet{eckert2021imputing}.

The NAICS-based concordance is more straightforward. For union elections
filed since 1999, the NLRB data report the 3-digit NAICS code, also
with some changes in the classification over time. The decennial censuses
of 2000 and 2010 and the CPS starting in 2003 report the variable
\texttt{INDNAICS} which is closely based on NAICS97 codes and harmonized
over time. Specifically, its first three digits coincide with 3-digit
NAICS97, except for “not specified” industries which we drop
and a few other industries which we merge together.\footnote{We dropped codes 3MS “Not specified industries” and 4MS “Not
specified retail trade.” An example of industries that we merged
is construction, which has only one \texttt{INDNAICS} code (code 22)
but multiple 3-digit NAICS codes.} We also merge a few codes where the NAICS codes changed over time
in the decennial censuses and CPS in a way that is not many-to-one
from the newer codes to NAICS97, ultimately resulting in our 73-industry
classification. In total, across the entire period of our study, we
were able to match 92.9\% of unionization elections and 99\% of census
records to our industry codes.

The transition from SIC to NAICS industries creates additional complications
when measuring outcomes in 10-year changes. For decennial census variables,
this affects the change from 1990 to 2000. The 2000 census, however,
includes the \texttt{IND1990} variable converted from \texttt{INDNAICS}
using a crosswalk designed by IPUMS. We therefore compute the 1990
to 2000 change using our SIC-based classification. While the conversion
could introduce noise in differenced outcomes, we find it reassuring
that employment changes do not exhibit substantially higher volatility
in the 1990s than in the decade before or after.\footnote{A small issue also arises from the fact that about one-third of the
unionization elections that happened in 1999 are recorded with NAICS
codes. We exclude these elections from the RDA variables construction.}

Industry transitions in the CPS, which happened between 2002 and 2003,
require more care. This affects differenced outcomes not only in the
1990s (measured as 2000–2004 minus 1986–1990) but also in the 2000s
(measured as 2010–2014 minus 1996–2000; see Table \ref{table:years-of-data}).
The conversion from \texttt{INDNAICS} codes to \texttt{IND1990} for
2003–04 was done by IPUMS, and we use it to compute the changes over
the 1990s in our SIC-based classification. In the other direction,
we apply the IPUMS crosswalk ourselves, converting \texttt{IND1990}
to \texttt{INDNAICS} for 1996–2000 to compute the changes over the
2000s using our NAICS-based classification.

\paragraph{State-by-industry panel and variable construction.}

Once all of our datasets have harmonized state and industry information
and are assigned to a decade (in the NLRB data) or a census year (in
the decennial censuses and the CPS), we compute differenced outcomes
in the panel of state-level cells by decade. Since our industry classification
changes in 2000, we compute all outcomes for the year 2000 using both
classifications.

The computation of the treatment, instrument, and the RDA controls
is described in detail in the main text. We drop a small number of
cell-decades in which there were more workers involved in unionization
attempts than the total number of workers at the beginning of the
decade according to the census, as those cells probably contain miscoding
of the industry variable or inconsistency between the census and NLRB
coding.

Most outcomes are straightforward. The union density, pension coverage,
and insurance coverage rates are fractions of the corresponding answers
in the CPS, using projection weights. In the decennial censuses, the
Gini index and the top-10\% income share are calculated based on the
total labor income. The log 90-10 wage ratio and the variance of log
wages, along with average wages for different workforce segments,
are computed based from log hourly wages. All measures are weighted
by projection weights. Employment is measured as the sum of weights
within each cell.

To estimate the college wage premium by cell and year, we broadly
follow \citet{autor2008trends} and \citet{farber2021unions}. We
select the sample or workers with either exactly a high-school degree
or at least a college degree (or more). We then estimate the following
regression by OLS with projection weights:

\begin{align}
\log Wage_{o} & =\gamma_{s(o)i(o)t(o)}College_{o}+\beta_{s(o)t(o)}'X_{o}+\text{error}_{o}\label{eq:college=000020premium}
\end{align}
where $o$ stands for an individual observation that belongs to state
$s(o)$, industry $i(o)$, and year $t(o)$, $College_{o}$ is the
dummy of having at least a college degree, and $X_{o}$ include the
following set of demographics variables: a female dummy, a non-white
dummy, a quartic term in experience, and interaction terms of the
female dummy with both the non-white dummy and the quartic in experience
term. Like \citet{farber2021unions}, we allow the demographic coefficients
to vary by state and decade. However, we restrict them to be the same
across industries within each state-decade because the sample size
would be insufficient for a fully-flexible specification. The estimates
of $\gamma_{sit}$ are the estimated college wage premiums.

For the inequality variables (except the college wage premium), overall
wages and employment variables, and all CPS variables, we replace
a cell-decade observation as missing if it represents less than 1,000
workers after applying projection weights or is based on less than
10 raw observations, either at the beginning or at the end of the
decade. For census variables based on a subgroup, we require to have
10 raw observations for that subgroup: e.g., we only compute the average
wage of college-educated workers in a cell if there are at least 10
college-educated respondents in that cell. For the college wage premium,
we require at least 10 observations for both high school graduates
and workers with a college degree or more.

\section*{Appendix Tables}

{\footnotesize{}
\begin{table}[H]
\begin{centering}
\caption{Published Papers Suitable for the RDA Design \protect\label{table:rda=000020papers}}
\medskip{}
\par\end{centering}
\begin{centering}
{\footnotesize{}%
\begin{tabular}{llcc}
\toprule 
{\footnotesize Citation} & {\footnotesize Journal} & {\footnotesize Equations} & {\footnotesize Tables}\tabularnewline
\midrule
\multicolumn{4}{c}{{\footnotesize\textbf{Panel A: Difference-in-Differences Specifications}}}\tabularnewline
{\footnotesize\citet{besley2003political}} & {\footnotesize Journal of Economic Literature} &  & \tabularnewline
{\footnotesize\citet{cooper2010corporate}} & {\footnotesize The Journal of Finance} &  & \tabularnewline
{\footnotesize\citet{saez2019payroll}} & {\footnotesize American Economic Review} &  & \tabularnewline
{\footnotesize\citet{akcigit2023connecting}} & {\footnotesize Econometrica} &  & \tabularnewline
 &  &  & \tabularnewline
\multicolumn{4}{c}{{\footnotesize\textbf{Panel B: Aggregated Outcomes, Upper-Level Specifications}}}\tabularnewline
{\footnotesize\citet{clots2011women}} & {\footnotesize Journal of Public Economics} & {\footnotesize 3,4} & {\footnotesize 7}\tabularnewline
{\footnotesize\citet{clots2012female}} & {\footnotesize American Economic Journal: Applied Economics} & {\footnotesize 1,2} & {\footnotesize 2,3}\tabularnewline
{\footnotesize\citet{bhalotra2014health}} & {\footnotesize American Economic Journal: Economy Policy} & {\footnotesize 1,2} & {\footnotesize 2}\tabularnewline
{\footnotesize\citet{bhalotra2014religion}} & {\footnotesize Journal of Economic Behavior \& Organization} & {\footnotesize 1,2} & {\footnotesize 2,3}\tabularnewline
{\footnotesize\citet{folke2014shades}} & {\footnotesize Journal of the European Economic Association} & {\footnotesize 4,5} & {\footnotesize 3}\tabularnewline
{\footnotesize\citet{akey2015valuing}} & {\footnotesize The Review of Financial Studies} & {\footnotesize –} & {\footnotesize 7}\tabularnewline
{\footnotesize\citet{nellis2016parties}} & {\footnotesize Quarterly Journal of Political Science} & {\footnotesize 2,3} & {\footnotesize 3}\tabularnewline
{\footnotesize\citet{nellis2018secular}} & {\footnotesize American Political Science Review} & {\footnotesize 2,3} & {\footnotesize 2}\tabularnewline
{\footnotesize\citet{campante2018long}} & {\footnotesize Quarterly Journal of Economics} & {\footnotesize 7} & {\footnotesize 4}\tabularnewline
{\footnotesize\citet{hyytinen2018public}} & {\footnotesize American Political Science Review} & {\footnotesize 1,4} & {\footnotesize 3}\tabularnewline
{\footnotesize\citet{azoulay2019public}} & {\footnotesize The Review of Economic Studies} & {\footnotesize 3.3, 3.4} & {\footnotesize 8}\tabularnewline
{\footnotesize\citet{priyanka2020female}} & {\footnotesize Labour Economics} & {\footnotesize 1,2} & {\footnotesize 3}\tabularnewline
{\footnotesize\citet{bhalotra2021religion}} & {\footnotesize Journal of Development Economics} & {\footnotesize 1,2} & {\footnotesize 2}\tabularnewline
{\footnotesize\citet{merilainen2022political}} & {\footnotesize The Economic Journal} & {\footnotesize 2,4} & {\footnotesize 3}\tabularnewline
{\footnotesize\citet{bahar2023innovation}} & {\footnotesize Management Science} & {\footnotesize 6,7} & {\footnotesize 8}\tabularnewline
{\footnotesize\citet{sorensen2023educated}} & {\footnotesize Journal of Economic Behavior \& Organization} & {\footnotesize 1,2} & {\footnotesize 2}\tabularnewline
{\footnotesize\citet{baskaran2024young}} & {\footnotesize Journal of Economic Behavior \& Organization} & {\footnotesize 1,2} & {\footnotesize 1}\tabularnewline
{\footnotesize\citet{geys2024public}} & {\footnotesize The Journal of Politics} & {\footnotesize 1,4} & {\footnotesize Fig. 2}\tabularnewline
{\footnotesize\citet{valentim2024does}} & {\footnotesize British Journal of Political Science} & {\footnotesize –} & {\footnotesize 3}\tabularnewline
 &  &  & \tabularnewline
\multicolumn{4}{c}{{\footnotesize\textbf{Panel C: Aggregated Outcomes, Lower-Level Specifications}}}\tabularnewline
{\footnotesize\citet{ferreira2014does}} & {\footnotesize Journal of Public Economics} & {\footnotesize 4} & {\footnotesize 11}\tabularnewline
{\footnotesize\citet{beach2017gridlock}} & {\footnotesize American Economic Journal: Economy Policy} & {\footnotesize 1} & {\footnotesize 3}\tabularnewline
{\footnotesize\citet{dell2018nation}} & {\footnotesize Quarterly Journal of Economics} & {\footnotesize 1} & {\footnotesize 2,3}\tabularnewline
{\footnotesize\citet{de2020politics}} & {\footnotesize Journal of Politics} & {\footnotesize –} & {\footnotesize Fig. 7}\tabularnewline
{\footnotesize\citet{carozzi2022political}} & {\footnotesize American Economic Journal: Applied Economics} & {\footnotesize 2,3} & {\footnotesize 2}\tabularnewline
{\footnotesize\citet{baskaran2023women}} & {\footnotesize The Review of Economics and Statistics} & {\footnotesize 2} & {\footnotesize 3}\tabularnewline
{\footnotesize\citet{tan2023consequences}} & {\footnotesize Journal of Labor Economics} & {\footnotesize 2} & {\footnotesize 2–4}\tabularnewline
{\footnotesize\citet{de2024partisanship}} & {\footnotesize American Journal of Political Science} & {\footnotesize –} & {\footnotesize Fig. 4}\tabularnewline
 &  &  & \tabularnewline
\multicolumn{4}{c}{{\footnotesize\textbf{Panel D: Spillovers}}}\tabularnewline
{\footnotesize\citet{clark2009performance}} & {\footnotesize Journal of Political Economy} & {\footnotesize 2} & {\footnotesize 4}\tabularnewline
{\footnotesize\citet{dahl2014peer}} & {\footnotesize American Economic Review} & {\footnotesize 3–5} & {\footnotesize 1}\tabularnewline
{\footnotesize\citet{isen2014local}} & {\footnotesize Journal of Public Economics} & {\footnotesize 5,6} & {\footnotesize 3–5}\tabularnewline
{\footnotesize\citet{baskaran2018does}} & {\footnotesize American Economic Journal: Economic Policy} & {\footnotesize 2} & {\footnotesize 9}\tabularnewline
{\footnotesize\citet{bhalotra2018pathbreakers}} & {\footnotesize The Economic Journal} & {\footnotesize 1} & {\footnotesize 3}\tabularnewline
{\footnotesize\citet{dube2019fairness}} & {\footnotesize American Economic Review} & {\footnotesize 9} & {\footnotesize 3}\tabularnewline
{\footnotesize\citet{fu2019estimation}} & {\footnotesize Econometrica} & {\footnotesize 4} & {\footnotesize Fig. 4}\tabularnewline
{\footnotesize\citet{johnson2020regulation}} & {\footnotesize American Economic Review} & {\footnotesize 2} & {\footnotesize 2}\tabularnewline
{\footnotesize\citet{santoleri2022causal}} & {\footnotesize The Review of Economics and Statistics} & {\footnotesize 1} & {\footnotesize 7}\tabularnewline
{\footnotesize\citet{jones2022factor}} & {\footnotesize American Economic Review} & {\footnotesize 3} & {\footnotesize 7,8}\tabularnewline
{\footnotesize\citet{mora2023peer}} & {\footnotesize The Review of Economics and Statistics} & {\footnotesize 3,4} & {\footnotesize 1}\tabularnewline
{\footnotesize\citet{baskaran2024women}} & {\footnotesize Journal of Economic Growth} & {\footnotesize 1} & {\footnotesize 4}\tabularnewline
{\footnotesize\citet{dechezlepretre2023tax}} & {\footnotesize American Economic Journal: Economic Policy} & {\footnotesize 5–7} & {\footnotesize 6}\tabularnewline
 &  &  & \tabularnewline
\multicolumn{4}{c}{{\footnotesize\textbf{Panel E: RD Aggregation over Time}}}\tabularnewline
{\footnotesize\citet{cellini2010value}} & {\footnotesize Quarterly Journal of Economics} & {\footnotesize 12} & {\footnotesize 4}\tabularnewline
{\footnotesize\citet{darolia2013integrity}} & {\footnotesize Journal of Public Economics} & {\footnotesize 4,5} & {\footnotesize 4–8}\tabularnewline
{\footnotesize\citet{martorell2016investing}} & {\footnotesize Journal of Public Economics} & {\footnotesize 3} & {\footnotesize 3}\tabularnewline
{\footnotesize\citet{lang2018assessing}} & {\footnotesize Journal of Public Economics} & {\footnotesize 11} & {\footnotesize 3}\tabularnewline
{\footnotesize\citet{baron2022school}} & {\footnotesize American Economic Journal: Economy Policy} & {\footnotesize 2} & {\footnotesize 5}\tabularnewline
{\footnotesize\citet{kuliomina2021personal}} & {\footnotesize European Journal of Political Economy} & {\footnotesize –} & {\footnotesize 4}\tabularnewline
{\footnotesize\citet{biasi2024works}} & {\footnotesize NBER Working Paper} & {\footnotesize 4} & {\footnotesize 3}\tabularnewline
\bottomrule
\end{tabular}}\medskip{}
\par\end{centering}
{\footnotesize\emph{Notes:}}{\footnotesize{} This table lists published
papers featuring research designs suitable for RDA. We have organized
these papers into the five categories as in Sections \ref{subsec:literature-DiD}–\ref{subsec:literature-lower}
and \ref{subsec:Spillover-Settings}–\ref{subsec:Dynamic-RD}, although
some papers may belong to more than one category. The “Equations”
column specifies the regression specification most suitable for the
RDA analysis, while the “Tables” column references the tables
and figures containing the main results from these specifications.
There are no references to equations and tables in Panel A since those
papers do not use RDA-type specifications in settings where that could
be possible.}{\footnotesize\par}
\end{table}
}{\footnotesize\par}

\begin{table}[H]
\begin{centering}
\caption{Unionization Data Summary Statistics \protect\label{table:summary_lower}}
\medskip{}
\par\end{centering}
\begin{centering}
{\small{}%
\begin{tabular}{lcccccc}
\toprule 
\multirow{2}{*}{{\small Variable}} & \multicolumn{3}{c}{{\small Treatment sample}} & \multicolumn{3}{c}{{\small Instrument sample}}\tabularnewline
\cmidrule(l){2-4}\cmidrule(l){5-7}
 & {\small All} & {\small Union wins} & {\small Union losses} & {\small All} & {\small Union wins} & {\small Union losses}\tabularnewline
\midrule 
{\small Number of elections} & {\small 232267} & {\small 116467} & {\small 115800} & {\small 118817} & {\small 53024} & {\small 61244}\tabularnewline
 &  &  &  &  &  & \tabularnewline
{\small Union win} & {\small 0.500} & {\small 0.995} & {\small 0.001} & {\small 0.457} & {\small 0.996} & {\small 0.001}\tabularnewline
{\small Union vote share} & {\small 0.538} & {\small 0.780} & {\small 0.295} & {\small 0.511} & {\small 0.729} & {\small 0.318}\tabularnewline
{\small Number of votes} & {\small 54.926} & {\small 46.003} & {\small 63.900} & {\small 98.359} & {\small 89.113} & {\small 109.930}\tabularnewline
{\small Number of eligible voters} & {\small 64.804} & {\small 57.874} & {\small 71.774} & {\small 116.123} & {\small 113.187} & {\small 123.262}\tabularnewline
 &  &  &  &  &  & \tabularnewline
{\small Years 1961–1969} & {\small 0.226} & {\small 0.261} & {\small 0.191} & {\small 0.226} & {\small 0.263} & {\small 0.193}\tabularnewline
{\small Years 1970–1979} & {\small 0.326} & {\small 0.316} & {\small 0.335} & {\small 0.314} & {\small 0.297} & {\small 0.327}\tabularnewline
{\small Years 1980–1989} & {\small 0.199} & {\small 0.173} & {\small 0.226} & {\small 0.195} & {\small 0.172} & {\small 0.217}\tabularnewline
{\small Years 1990–1999} & {\small 0.150} & {\small 0.141} & {\small 0.159} & {\small 0.162} & {\small 0.153} & {\small 0.169}\tabularnewline
{\small Years 2000–2009} & {\small 0.099} & {\small 0.109} & {\small 0.089} & {\small 0.103} & {\small 0.115} & {\small 0.094}\tabularnewline
{\small Manufacturing industry} & {\small 0.430} & {\small 0.410} & {\small 0.449} & {\small 0.520} & {\small 0.500} & {\small 0.539}\tabularnewline
\bottomrule
\end{tabular}}\medskip{}
\par\end{centering}
{\footnotesize\emph{Notes:}}{\footnotesize{} The table reports mean
characteristics for union elections included in our sample. The set
of elections used to construct the treatment variable (“treatment
sample”) contains all elections, while the set used for the instrument
construction (“instrument sample”) contains only elections with
the vote share for the union of $50\pm10\%$, at least 20 votes, and
vote margin of at least 2 votes. Union wins are defined as strictly
more than 50\% of votes in favor of the union.}{\footnotesize\par}

{\footnotesize\emph{Sources: }}{\footnotesize NLRB data.}{\footnotesize\par}
\end{table}

{\footnotesize{}
\begin{table}[H]
\begin{centering}
\caption{Within-Between Variance Decomposition of Log Wages, Census Data \protect\label{table:VarDecomp}}
\medskip{}
\par\end{centering}
\begin{centering}
{\small{}%
\begin{tabular}{lcccc}
\toprule 
\multirow{2}{*}{{\small Year}} & \multicolumn{3}{c}{{\small Var(log hourly wage)}} & {\small Industry}\tabularnewline
\cmidrule{2-4}
 & {\small Total} & {\small Within cells} & {\small Between cells} & {\small classification}\tabularnewline
\midrule 
{\small 1960} & {\small 0.262} & {\small 0.192} & {\small 0.071} & {\small SIC}\tabularnewline
{\small 1970} & {\small 0.283} & {\small 0.222} & {\small 0.061} & {\small SIC}\tabularnewline
{\small 1980} & {\small 0.284} & {\small 0.228} & {\small 0.056} & {\small SIC}\tabularnewline
{\small 1990} & {\small 0.361} & {\small 0.293} & {\small 0.068} & {\small SIC}\tabularnewline
{\small 2000} & {\small 0.387} & {\small 0.326} & {\small 0.061} & {\small SIC}\tabularnewline
 &  &  &  & \tabularnewline
{\small 2000} & {\small 0.387} & {\small 0.321} & {\small 0.066} & {\small NAICS}\tabularnewline
{\small 2010} & {\small 0.437} & {\small 0.344} & {\small 0.093} & {\small NAICS}\tabularnewline
\bottomrule
\end{tabular}}\medskip{}
\par\end{centering}
{\footnotesize\emph{Notes:}}{\footnotesize{} The table presents a variance
decomposition of log wages, broadly following \citet{helpman2016trade}.
For each census year, we decompose the overall variance in log wages
among workers in our Census sample into within and between (state-industry)
cell components. In 2000, we perform the decomposition twice, using
two industry coding frameworks: SIC-based (used in our main analysis
until 2000) and NAICS-based (used from 2000 onwards). Census projection
weights are used.}{\footnotesize\par}

{\footnotesize\emph{Sources:}}{\footnotesize{} Decennial Census; authors'
calculations.}{\footnotesize\par}
\end{table}
}{\footnotesize\par}

\begin{table}[H]
\begin{centering}
\caption{Instrument Balance, 15\% Bandwidth\protect\label{table:balance15}}
\medskip{}
\par\end{centering}
\begin{centering}
{\small{}%
\begin{tabular}{lccccc}
\toprule 
 & {\small$\text{NewUnions}$} &  & \multicolumn{3}{c}{{\small$Z$}}\tabularnewline
\cmidrule(l){2-2}\cmidrule(l){4-6}
 & {\small (1)} &  & {\small (2)} & {\small (3)} & {\small (4)}\tabularnewline
\midrule 
{\small Log avg hourly wage$_{t-1}$} & {\small 1.056} &  & {\small 0.188} & {\small -0.155} & {\small -0.013}\tabularnewline
 & {\small (0.517)} &  & {\small (0.127)} & {\small (0.070)} & {\small (0.032)}\tabularnewline
{\small Log college premium$_{t-1}$} & {\small 4.452} &  & {\small 2.331} & {\small -0.002} & {\small 0.046}\tabularnewline
 & {\small (0.405)} &  & {\small (0.169)} & {\small (0.091)} & {\small (0.045)}\tabularnewline
{\small Log 90/10 ratio$_{t-1}$} & {\small 3.895} &  & {\small 1.297} & {\small 0.052} & {\small -0.005}\tabularnewline
 & {\small (0.630)} &  & {\small (0.250)} & {\small (0.133)} & {\small (0.062)}\tabularnewline
{\small Gini index$_{t-1}$} & {\small -58.022} &  & {\small -20.529} & {\small -1.533} & {\small 0.015}\tabularnewline
 & {\small (6.424)} &  & {\small (1.915)} & {\small (1.034)} & {\small (0.456)}\tabularnewline
{\small Top 10\% share$_{t-1}$} & {\small 40.646} &  & {\small 13.759} & {\small 1.036} & {\small 0.058}\tabularnewline
 & {\small (6.474)} &  & {\small (1.796)} & {\small (0.983)} & {\small (0.433)}\tabularnewline
{\small Var(log hourly wage)$_{t-1}$} & {\small -7.027} &  & {\small -2.406} & {\small 0.204} & {\small -0.162}\tabularnewline
 & {\small (1.486)} &  & {\small (0.387)} & {\small (0.219)} & {\small (0.100)}\tabularnewline
{\small Log employment$_{t-1}$} & {\small -0.158} &  & {\small -0.095} & {\small -0.002} & {\small 0.000}\tabularnewline
 & {\small (0.051)} &  & {\small (0.019)} & {\small (0.010)} & {\small (0.005)}\tabularnewline
{\small Share with college degree$_{t-1}$} & {\small -4.510} &  & {\small -1.497} & {\small 0.153} & {\small 0.090}\tabularnewline
 & {\small (0.609)} &  & {\small (0.158)} & {\small (0.089)} & {\small (0.050)}\tabularnewline
{\small Share white$_{t-1}$} & {\small -0.527} &  & {\small -0.302} & {\small -0.456} & {\small -0.049}\tabularnewline
 & {\small (0.561)} &  & {\small (0.241)} & {\small (0.132)} & {\small (0.051)}\tabularnewline
{\small Share male$_{t-1}$} & {\small -0.869} &  & {\small -0.556} & {\small -0.078} & {\small 0.037}\tabularnewline
 & {\small (0.516)} &  & {\small (0.170)} & {\small (0.082)} & {\small (0.036)}\tabularnewline
 &  &  &  &  & \tabularnewline
{\small\# significant controls} & {\small 9} &  & {\small 9} & {\small 3} & {\small 0}\tabularnewline
{\small Partial $R^{2}$} & {\small 0.1353} &  & {\small 0.1049} & {\small 0.0017} & {\small 0.0000}\tabularnewline
{\small Partial $F$-statistic} & {\small 292.7} &  & {\small 226.4} & {\small 6.3} & {\small 1.0}\tabularnewline
{\small Partial $F$-test, p-value} & {\small 0.0000} &  & {\small 0.0000} & {\small 0.0000} & {\small 0.4557}\tabularnewline
 &  &  &  &  & \tabularnewline
{\small RDA controls} & {\small None} &  & {\small None} & {\small$\sum_{j\in\mathcal{C}_{sit}}s_{j}$} & {\small All}\tabularnewline
{\small Observations} & {\small 12,799} &  & {\small 12,799} & {\small 12,799} & {\small 12,799}\tabularnewline
\bottomrule
\end{tabular}}\medskip{}
\par\end{centering}
{\footnotesize\emph{Notes:}}{\footnotesize{} Same as Table \ref{table:balance},
except the instrument $Z$ is constructed from close elections defined
by the bandwidth of 15\%, i.e. the union vote share of $50\pm15$
percent.}{\footnotesize\par}

{\footnotesize\emph{Sources: }}{\footnotesize Decennial Census, NLRB
data, and authors' calculations.}{\footnotesize\par}
\end{table}

{\footnotesize{}
\begin{table}[H]
\begin{centering}
\caption{Placebo Tests: The Impact of New Unionization on Predetermined Outcomes
\protect\label{table:balance_rhs}}
\medskip{}
\par\end{centering}
\begin{centering}
{\small{}%
\begin{tabular}{lccccc}
\toprule 
\multicolumn{6}{c}{{\small\textbf{Panel A: Lagged Outcomes in Levels }}{\small\medskip{}
}}\tabularnewline
 & {\small College} & \multirow{2}{*}{{\small Log 90/10}} & {\small Gini} & \multirow{2}{*}{{\small Top 10\% share}} & \multirow{2}{*}{{\small Var(Log Wage)}}\tabularnewline
 & {\small premium} &  & {\small coefficient} &  & \tabularnewline
{\small$Z_{sit}$} & {\small 0.046} & {\small 0.052} & {\small -0.019} & {\small -0.018} & {\small 0.013}\tabularnewline
 & {\small (0.207)} & {\small (0.225)} & {\small (0.059)} & {\small (0.047)} & {\small (0.077)}\tabularnewline
 & {\small{[}-0.736,0.760{]} \medskip{}
} & {\small{[}-0.482,1.137{]}} & {\small{[}-0.211,0.235{]}} & {\small{[}-0.170,0.179{]}} & {\small{[}-0.225,0.313{]}}\tabularnewline
{\small Observations} & {\small 10,716} & {\small 12,896} & {\small 12,896} & {\small 12,896} & {\small 12,896}\tabularnewline
 &  &  &  &  & \tabularnewline
\multicolumn{6}{c}{{\small\textbf{Panel B: Lagged Outcomes in Differences}}{\small\medskip{}
}}\tabularnewline
 & {\small$\Delta$College} & \multirow{2}{*}{{\small$\Delta$Log 90/10}} & {\small$\Delta$Gini} & \multirow{2}{*}{{\small$\Delta$Top 10\% share}} & \multirow{2}{*}{{\small$\Delta$Var(Log Wage)}}\tabularnewline
 & {\small premium} &  & {\small coefficient} &  & \tabularnewline
{\small$Z_{sit}$} & {\small -0.344} & {\small 0.491} & {\small 0.038} & {\small -0.008} & {\small 0.098}\tabularnewline
 & {\small (0.335)} & {\small (0.237)} & {\small (0.060)} & {\small (0.056)} & {\small (0.082)}\tabularnewline
 & {\small{[}-1.547,0.852{]}\medskip{}
} & {\small{[}-0.279,1.391{]}} & {\small{[}-0.045,0.391{]}} & {\small{[}-0.122,0.309{]}} & {\small{[}-0.08,0.460{]}}\tabularnewline
{\small Observations} & {\small 6,467} & {\small 7,706} & {\small 7,706} & {\small 7,706} & {\small 7,706}\tabularnewline
 &  &  &  &  & \tabularnewline
\multicolumn{6}{c}{{\small\textbf{Panel C: Covariates }}{\small\medskip{}
}}\tabularnewline
 & {\small log(Avg wage)} & {\small log(Empl.)} & {\small Share college} & {\small Share white} & {\small Share male}\tabularnewline
{\small$Z_{sit}$} & {\small -0.016} & {\small -0.587} & {\small -0.033} & {\small -0.062} & {\small 0.096}\tabularnewline
 & {\small (0.178)} & {\small (1.180)} & {\small (0.065)} & {\small (0.112)} & {\small (0.095)}\tabularnewline
 & {\small{[}-0.472,1.098{]}\medskip{}
} & {\small{[}-7.612,4.531{]}} & {\small{[}-0.286,0.250{]}} & {\small{[}-0.330,0.586{]}} & {\small{[}-0.299,0.611{]}}\tabularnewline
{\small Observations} & {\small 12,896} & {\small 12,896} & {\small 12,896} & {\small 12,896} & {\small 12,896}\tabularnewline
\bottomrule
\end{tabular}}{\small\par}
\par\end{centering}
\begin{centering}
\medskip{}
\par\end{centering}
{\footnotesize\emph{Notes: }}{\footnotesize Placebo tests based on
OLS estimates corresponding to the reduced-form of our main upper-level
IV specification as in Table \ref{table:unions_inequilty}, Panel
A. See notes to Table \ref{table:unions_inequilty} for details about
data and variables construction. The outcomes are measured at the
beginning of the decade in which the instrument is measured in Panels
A and C and as a 10-year lagged difference in Panel B.}{\footnotesize\par}

{\footnotesize\emph{Sources:}}{\footnotesize{} Decennial Census and
NLRB unionization data; authors' calculations.}{\footnotesize\par}
\end{table}
}{\footnotesize\par}

{\footnotesize{}
\begin{sidewaystable}[H]
\begin{centering}
\caption{The Impact of New Unionization on 10-Year Changes in Inequality: Robustness
Checks\protect\label{table:unions_inequilty-Robustness}}
\medskip{}
\par\end{centering}
\begin{centering}
{\small{}%
\begin{tabular}{lccccc}
\toprule 
 & {\small$\Delta$Log college premium} & {\small$\Delta$Log 90/10} & {\small$\Delta$Gini coeff.} & {\small$\Delta$Top 10\% Share} & {\small$\Delta$Var(Log Wage)}\tabularnewline
 & {\small (1)} & {\small (2)} & {\small (3)} & {\small (4)} & {\small (5)}\tabularnewline
\midrule 
\multicolumn{6}{c}{{\small\textbf{Panel A: Baseline}}{\small{} \medskip{}
}}\tabularnewline
{\small Share newly unionized} & {\small -0.314} & {\small -0.459} & {\small -0.176} & {\small -0.143} & {\small -0.255}\tabularnewline
 & {\small (0.285)} & {\small (0.225)} & {\small (0.061)} & {\small (0.055)} & {\small (0.084)}\tabularnewline
 & {\small{[}-0.921,0.416{]}} & {\small{[}-0.933,0.122{]}} & {\small{[}-0.325,-0.035{]}} & {\small{[}-0.279,-0.020{]}} & {\small{[}-0.463,-0.054{]}}\tabularnewline
 &  &  &  &  & \tabularnewline
\multicolumn{6}{c}{{\small\textbf{Panel B: 15\% Bandwidth}}{\small{} \medskip{}
}}\tabularnewline
{\small Share newly unionized} & {\small -0.361} & {\small -0.281} & {\small -0.123} & {\small -0.101} & {\small -0.198}\tabularnewline
 & {\small (0.211)} & {\small (0.151)} & {\small (0.041)} & {\small (0.038)} & {\small (0.057)}\tabularnewline
 & {\small{[}-0.856,0.095{]}} & {\small{[}-0.613,0.085{]}} & {\small{[}-0.222,-0.030{]}} & {\small{[}-0.193,-0.018{]}} & {\small{[}-0.33,-0.062{]}}\tabularnewline
 &  &  &  &  & \tabularnewline
\multicolumn{6}{c}{{\small\textbf{Panel C: Excluding Union Decertification Elections}}{\small{}
\medskip{}
}}\tabularnewline
{\small Share newly unionized} & {\small -0.332} & {\small -0.758} & {\small -0.193} & {\small -0.147} & {\small -0.334}\tabularnewline
 & {\small (0.322)} & {\small (0.268)} & {\small (0.070)} & {\small (0.063)} & {\small (0.099)}\tabularnewline
 & {\small{[}-1.062,0.498{]}} & {\small{[}-0.987,0.169{]}} & {\small{[}-0.341,-0.023{]}} & {\small{[}-0.286,-0.003{]}} & {\small{[}-0.480,-0.038{]}}\tabularnewline
 &  &  &  &  & \tabularnewline
\multicolumn{6}{c}{{\small\textbf{Panel D: State, Year, Industry Fixed Effects}}{\small{}
\medskip{}
}}\tabularnewline
{\small Share newly unionized} & {\small -0.393} & {\small -0.466} & {\small -0.165} & {\small -0.136} & {\small -0.220}\tabularnewline
 & {\small (0.275)} & {\small (0.237)} & {\small (0.062)} & {\small (0.055)} & {\small (0.089)}\tabularnewline
 & {\small{[}-0.930,0.341{]}} & {\small{[}-0.993,0.103{]}} & {\small{[}-0.311,-0.017{]}} & {\small{[}-0.269,-0.008{]}} & {\small{[}-0.433,-0.009{]}}\tabularnewline
 &  &  &  &  & \tabularnewline
\multicolumn{6}{c}{{\small\textbf{Panel E: No Fixed Effects}}{\small\medskip{}
}}\tabularnewline
{\small Share newly unionized} & {\small -0.416} & {\small -0.486} & {\small -0.180} & {\small -0.155} & {\small -0.261}\tabularnewline
 & {\small (0.335)} & {\small (0.265)} & {\small (0.070)} & {\small (0.062)} & {\small (0.102)}\tabularnewline
 & {\small{[}-1.126,0.473{]}} & {\small{[}-1.089,0.152{]}} & {\small{[}-0.336,-0.018{]}} & {\small{[}-0.297,-0.013{]}} & {\small{[}-0.503,-0.016{]}}\tabularnewline
 &  &  &  &  & \tabularnewline
\multicolumn{6}{c}{{\small\textbf{Panel F: Unweighted}}{\small{} \medskip{}
}}\tabularnewline
{\small Share newly unionized} & {\small -0.230} & {\small -0.518} & {\small -0.239} & {\small -0.187} & {\small -0.339}\tabularnewline
 & {\small (0.334)} & {\small (0.249)} & {\small (0.081)} & {\small (0.077)} & {\small (0.114)}\tabularnewline
 & {\small{[}-1.738,0.503{]}} & {\small{[}-1.089,0.400{]}} & {\small{[}-0.451,0.007{]}} & {\small{[}-0.421,0.014{]}} & {\small{[}-0.665,-0.035{]}}\tabularnewline
\bottomrule
\end{tabular}}\medskip{}
\par\end{centering}
{\footnotesize\emph{Notes: }}{\footnotesize Panel A reports upper-level
IV estimates of the effects of new unionization rates on wage inequality,
replicating Panel A of Table \ref{table:unions_inequilty}. The following
columns deviate from Panel A in different ways, one at a time. Panel
B redefines the instrument and RDA controls using a bandwidth of 15\%
instead of 10\%. Panel C reports results excluding decertification
elections from the unionization elections sample when constructing
the treatment, instrument, and RDA controls. Panel D includes separate
state, industry, and year fixed effects instead of state-year and
industry-year fixed effects. Panel E reports estimates without any
fixed effects. Panel F presents unweighted estimation results, instead
of weighting state-industry-decade observations by beginning-of-decade
employment.}{\footnotesize\par}

{\footnotesize\emph{Sources:}}{\footnotesize{} Decennial Census and
NLRB unionization data; authors' calculations.}{\footnotesize\par}
\end{sidewaystable}
}{\footnotesize\par}

{\footnotesize{}
\begin{table}[H]
\begin{centering}
\caption{The Impact of New Unionization on 10-Year Employment Changes \protect\label{table:emp}}
\medskip{}
\par\end{centering}
\begin{centering}
{\small{}%
\begin{tabular}{lccccc}
\toprule 
 & {\small All} & {\small HS} & {\small Non-College} & {\small College} & {\small College+}\tabularnewline
 & {\small (1)} & {\small (2)} & {\small (3)} & {\small (4)} & {\small (5)}\tabularnewline
\midrule 
\multicolumn{6}{c}{{\small\textbf{Panel A: Upper-Level Estimator}}}\tabularnewline
{\small Share newly unionized} & {\small -0.336} & {\small 0.195} & {\small -0.148} & {\small -1.782} & {\small -1.719}\tabularnewline
 & {\small (0.474)} & {\small (0.489)} & {\small (0.459)} & {\small (0.895)} & {\small (0.829)}\tabularnewline
 & {\small{[}-1.638,0.701{]}} & {\small{[}-1.184,1.204{]}} & {\small{[}-1.381,0.872{]}} & {\small{[}-4.161,0.214{]}} & {\small{[}-3.961,0.085{]}}\tabularnewline
 &  &  &  &  & \tabularnewline
{\small Mean outcome} & {\small 10.987} & {\small 9.960} & {\small 10.678} & {\small 8.943} & {\small 9.292}\tabularnewline
{\small Mean treatment} & {\small 2.76\%} & {\small 2.76\%} & {\small 2.76\%} & {\small 2.76\%} & {\small 2.76\%}\tabularnewline
{\small RDA controls} & {\small$\checkmark$} & {\small$\checkmark$} & {\small$\checkmark$} & {\small$\checkmark$} & {\small$\checkmark$}\tabularnewline
{\small Industry-decade and state-decade FE} & {\small$\checkmark$} & {\small$\checkmark$} & {\small$\checkmark$} & {\small$\checkmark$} & {\small$\checkmark$}\tabularnewline
{\small Observations} & {\small 12,910} & {\small 12,910} & {\small 12,910} & {\small 12,723} & {\small 12,792}\tabularnewline
 &  &  &  &  & \tabularnewline
\multicolumn{6}{c}{{\small\textbf{Panel B: Lower-Level Estimator}}}\tabularnewline
{\small Share newly unionized} & {\small 0.025} & {\small 0.417} & {\small 0.154} & {\small -1.086} & {\small -1.003}\tabularnewline
 & {\small (0.395)} & {\small (0.429)} & {\small (0.386)} & {\small (0.813)} & {\small (0.753)}\tabularnewline
 & {\small{[}-1.029,0.807{]}} & {\small{[}-0.736,1.236{]}} & {\small{[}-0.853,0.938{]}} & {\small{[}-3.402,0.347{]}} & {\small{[}-3.220,0.277{]}}\tabularnewline
 &  &  &  &  & \tabularnewline
{\small Mean outcome} & {\small 10.618} & {\small 9.687} & {\small 10.432} & {\small 8.162} & {\small 8.519}\tabularnewline
{\small Mean treatment} & {\small 9.54} & {\small 9.54} & {\small 9.54} & {\small 9.53} & {\small 9.54}\tabularnewline
{\small RDA controls} & {\small$\checkmark$} & {\small$\checkmark$} & {\small$\checkmark$} & {\small$\checkmark$} & {\small$\checkmark$}\tabularnewline
{\small Industry-decade and state-decade FE} & {\small$\checkmark$} & {\small$\checkmark$} & {\small$\checkmark$} & {\small$\checkmark$} & {\small$\checkmark$}\tabularnewline
{\small Observations} & {\small 31,256} & {\small 31,256} & {\small 31,256} & {\small 31,167} & {\small 31,207}\tabularnewline
\bottomrule
\end{tabular}}\medskip{}
\par\end{centering}
{\footnotesize\emph{Notes: }}{\footnotesize This table reports IV estimates
from the same specifications as in Table \ref{table:unions_inequilty}
but for different outcomes. The outcome in column 1 is the 10-year
change in the total log employment in the state-industry cell. The
other columns consider changes in log employment by education group:
those with exactly a high-school degree (column 2), any level of education
below a college degree (column 3), exactly a college degree (column
4), and college degree or more (column 5).}{\footnotesize\par}

{\footnotesize\emph{Sources:}}{\footnotesize{} Decennial Census and
NLRB unionization data; authors' calculations.}{\footnotesize\par}
\end{table}
}{\footnotesize\par}

{\footnotesize{}
\begin{sidewaystable}[H]
\begin{centering}
\caption{The Impact of New Unionization on 10-Year Wage Changes: Robustness
Checks \protect\label{table:robustness_decomposition}}
\medskip{}
\par\end{centering}
\begin{centering}
{\small{}%
\begin{tabular}{lcccccccc}
\toprule 
 & {\small All Workers} & {\small College} & {\small HS} & {\small P90} & {\small P50} & {\small P10} & {\small Mgmt} & {\small Non-Mgmt}\tabularnewline
 & {\small (1)} & {\small (2)} & {\small (3)} & {\small (4)} & {\small (5)} & {\small (6)} & {\small (7)} & {\small (8)}\tabularnewline
\midrule 
\multicolumn{9}{c}{{\small\textbf{Panel A: Baseline}}{\small{} \medskip{}
}}\tabularnewline
{\small Share newly unionized} & {\small -0.348} & {\small -0.449} & {\small -0.105} & {\small -0.282} & {\small -0.129} & {\small 0.176} & {\small -0.917} & {\small -0.181}\tabularnewline
 & {\small (0.150)} & {\small (0.349)} & {\small (0.142)} & {\small (0.180)} & {\small (0.145)} & {\small (0.166)} & {\small (0.399)} & {\small (0.134)}\tabularnewline
 & {\small{[}-0.762,-0.032{]}} & {\small{[}-1.221,0.468{]}} & {\small{[}-0.442,0.227{]}} & {\small{[}-0.735,0.138{]}} & {\small{[}-0.533,0.188{]}} & {\small{[}-0.283,0.497{]}} & {\small{[}-1.854,0.058{]}} & {\small{[}-0.561,0.091{]}}\tabularnewline
 &  &  &  &  &  &  &  & \tabularnewline
\multicolumn{9}{c}{{\small\textbf{Panel B: 15\% Bandwidth}}{\small{} \medskip{}
}}\tabularnewline
{\small Share newly unionized} & {\small -0.208} & {\small -0.340} & {\small 0.079} & {\small -0.144} & {\small -0.052} & {\small 0.136} & {\small -0.467} & {\small -0.077}\tabularnewline
 & {\small (0.102)} & {\small (0.252)} & {\small (0.098)} & {\small (0.122)} & {\small (0.098)} & {\small (0.115)} & {\small (0.276)} & {\small (0.093)}\tabularnewline
 & {\small{[}-0.447,0.031{]}} & {\small{[}-0.970,0.215{]}} & {\small{[}-0.139,0.307{]}} & {\small{[}-0.424,0.161{]}} & {\small{[}-0.288,0.193{]}} & {\small{[}-0.13,0.396{]}} & {\small{[}-1.123,0.11{]}} & {\small{[}-0.293,0.133{]}}\tabularnewline
 &  &  &  &  &  &  &  & \tabularnewline
\multicolumn{9}{c}{{\small\textbf{Panel C: Excluding Union Decertification Elections}}{\small{}
\medskip{}
}}\tabularnewline
{\small Share newly unionized} & {\small -0.436} & {\small -0.444} & {\small -0.091} & {\small -0.417} & {\small -0.260} & {\small 0.341} & {\small -0.626} & {\small -0.188}\tabularnewline
 & {\small (0.174)} & {\small (0.449)} & {\small (0.180)} & {\small (0.203)} & {\small (0.163)} & {\small (0.208)} & {\small (0.480)} & {\small (0.154)}\tabularnewline
 & {\small{[}-0.857,-0.046{]}} & {\small{[}-1.461,0.502{]}} & {\small{[}-0.523,0.212{]}} & {\small{[}-0.847,0.118{]}} & {\small{[}-0.627,0.169{]}} & {\small{[}-0.381,0.469{]}} & {\small{[}-2.470,-0.131{]}} & {\small{[}-0.581,0.139{]}}\tabularnewline
 &  &  &  &  &  &  &  & \tabularnewline
\multicolumn{9}{c}{{\small\textbf{Panel D: State, Year, Industry Fixed Effects}}{\small{}
\medskip{}
}}\tabularnewline
{\small Share newly unionized} & {\small -0.223} & {\small -0.316} & {\small 0.029} & {\small -0.189} & {\small 0.006} & {\small 0.277} & {\small -0.629} & {\small -0.091}\tabularnewline
 & {\small (0.205)} & {\small (0.335)} & {\small (0.209)} & {\small (0.223)} & {\small (0.207)} & {\small (0.218)} & {\small (0.376)} & {\small (0.197)}\tabularnewline
 & {\small{[}-0.76,0.232{]}} & {\small{[}-1.011,0.565{]}} & {\small{[}-0.504,0.497{]}} & {\small{[}-0.76,0.318{]}} & {\small{[}-0.526,0.475{]}} & {\small{[}-0.279,0.727{]}} & {\small{[}-1.492,0.307{]}} & {\small{[}-0.632,0.316{]}}\tabularnewline
 &  &  &  &  &  &  &  & \tabularnewline
\multicolumn{9}{c}{{\small\textbf{Panel E: No Fixed Effects}}{\small\medskip{}
}}\tabularnewline
{\small Share newly unionized} & {\small -0.323} & {\small -0.434} & {\small 0.034} & {\small -0.242} & {\small -0.079} & {\small 0.244} & {\small -0.543} & {\small -0.145}\tabularnewline
 & {\small (0.312)} & {\small (0.382)} & {\small (0.224)} & {\small (0.336)} & {\small (0.318)} & {\small (0.293)} & {\small (0.440)} & {\small (0.222)}\tabularnewline
 & {\small{[}-1.085,0.392{]}} & {\small{[}-1.266,0.561{]}} & {\small{[}-0.506,0.531{]}} & {\small{[}-1.058,0.538{]}} & {\small{[}-0.831,0.643{]}} & {\small{[}-0.449,0.865{]}} & {\small{[}-1.541,0.536{]}} & {\small{[}-0.723,0.331{]}}\tabularnewline
 &  &  &  &  &  &  &  & \tabularnewline
\multicolumn{9}{c}{{\small\textbf{Panel F: Unweighted}}{\small{} \medskip{}
}}\tabularnewline
{\small Share newly unionized} & {\small -0.384} & {\small -0.352} & {\small -0.193} & {\small -0.283} & {\small 0.087} & {\small 0.236} & {\small -0.848} & {\small -0.137}\tabularnewline
 & {\small (0.163)} & {\small (0.365)} & {\small (0.181)} & {\small (0.185)} & {\small (0.130)} & {\small (0.151)} & {\small (0.549)} & {\small (0.132)}\tabularnewline
 & {\small{[}-1.057,0.012{]}} & {\small{[}-2.102,0.583{]}} & {\small{[}-0.747,0.278{]}} & {\small{[}-0.817,0.407{]}} & {\small{[}-0.629,0.357{]}} & {\small{[}-0.412,0.691{]}} & {\small{[}-2.323,0.377{]}} & {\small{[}-0.828,0.086{]}}\tabularnewline
\bottomrule
\end{tabular}}\medskip{}
\par\end{centering}
{\footnotesize\emph{Notes: }}{\footnotesize This table performs the
robustness checks of Table \ref{table:unions_inequilty-Robustness}
for the outcomes of Table \ref{table:unions_inequilty=000020decomposition10}.
Please refer to those tables for the specifications and the data construction.}{\footnotesize\par}

{\footnotesize\emph{Sources:}}{\footnotesize{} Decennial Census and
NLRB unionization data; authors' calculations.}{\footnotesize\par}
\end{sidewaystable}
}{\footnotesize\par}

\begin{table}[H]
\caption{Definitions of Decades\protect\label{table:years-of-data}}
\medskip{}

\begin{centering}
{\small{}%
\begin{tabular}{ccccc}
\toprule 
\multirow{3}{*}{{\small Decade}} & \multirow{2}{*}{{\small NLRB}} & {\small Decennial Census} & {\small CPS MORG} & {\small CPS ASEC}\tabularnewline
 &  & {\small and ACS} & {\small (for union density)} & {\small (for fringe benefits)}\tabularnewline
 & {\small (1)} & {\small (2)} & {\small (3)} & {\small (4)}\tabularnewline
\midrule
{\small 1960s} & {\small 1961–69} & {\small 1970 minus 1960} & {\small n/a} & {\small n/a}\tabularnewline
{\small 1970s} & {\small 1970–79} & {\small 1980 minus 1970} & {\small n/a} & {\small n/a}\tabularnewline
{\small 1980s} & {\small 1980–89} & {\small 1990 minus 1980} & {\small 1990–1994 minus 1977–1980$^{b}$} & {\small 1990–1994 minus 1979–1981$^{b}$}\tabularnewline
{\small 1990s} & {\small 1990–99} & {\small 2000 minus 1990} & {\small 2000–2004 minus 1986–1990} & {\small 2000–2004 minus 1986–1990}\tabularnewline
{\small 2000s} & {\small 2000–09$^{a}$} & {\small 2009–2013 minus 2000} & {\small 2010–2014 minus 1996–2000} & {\small 2010–2014 minus 1996–2000}\tabularnewline
\bottomrule
\end{tabular}}\medskip{}
\par\end{centering}
{\footnotesize\emph{Notes:}}{\footnotesize{} For each decade in the
data, column 1 reports the years of NLRB data used to compute the
new unionization rate. Columns 2–4 report the years in the decennial
census (plus ACS) data, CPS MORG, and CPS ASEC data used to compute
differenced outcomes. For example, the change over 1990s in the CPS
MORG is computed as the average of 2000–2004 minus the average of
1986–1990.}{\footnotesize\par}

$^{a}$ {\footnotesize The NLRB data for the year of 2009 cover only
the beginning of year.}{\footnotesize\par}

$^{b}$ {\footnotesize The May supplement to the CPS replaces CPS MORG
for years 1977–1980 and replaces CPS ASEC for the year of 1979.}{\footnotesize\par}
\end{table}

\newpage{}

\section*{Appendix Figures}

\begin{figure}[H]
\caption{Within-Between Variance Decomposition of Log Wages, CPS Data\protect\label{fig:Theil-Decomposition}}

\begin{centering}
\medskip{}
\par\end{centering}
\begin{centering}
\includegraphics[width=0.75\columnwidth]{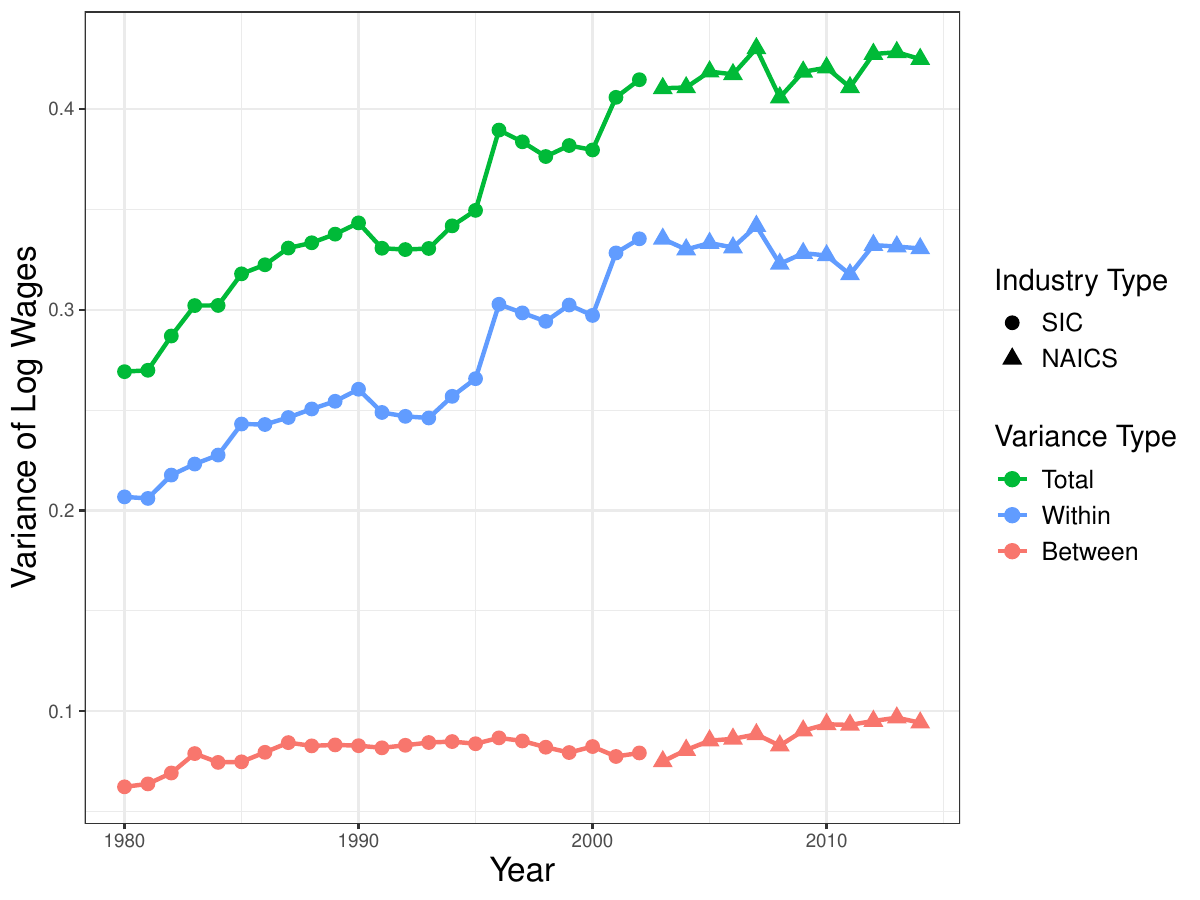}\medskip{}
\par\end{centering}
{\footnotesize\emph{Notes:}}{\footnotesize{} The plot presents a variance
decomposition of log wages, broadly following \citet{helpman2016trade},
using CPS ASEC samples. In each year from 1980–2014, we decompose
the overall variance in log wages among workers into within and between
(state-industry) cell components, using projection weights. The industry
coding framework shifts from SIC-based to NAICS-based in 2003.}{\footnotesize\par}

{\footnotesize\emph{Sources:}}{\footnotesize{} CPS ASEC samples; authors'
calculations.}{\footnotesize\par}
\end{figure}

\begin{figure}[H]
\caption{The Counterfactual with Unionization Remaining at the 1960s Level:
Other Inequality Outcomes\protect\label{fig:counterfactual_apdx}}

\begin{centering}
\medskip{}
\par\end{centering}
\begin{centering}
\includegraphics[width=1\columnwidth]{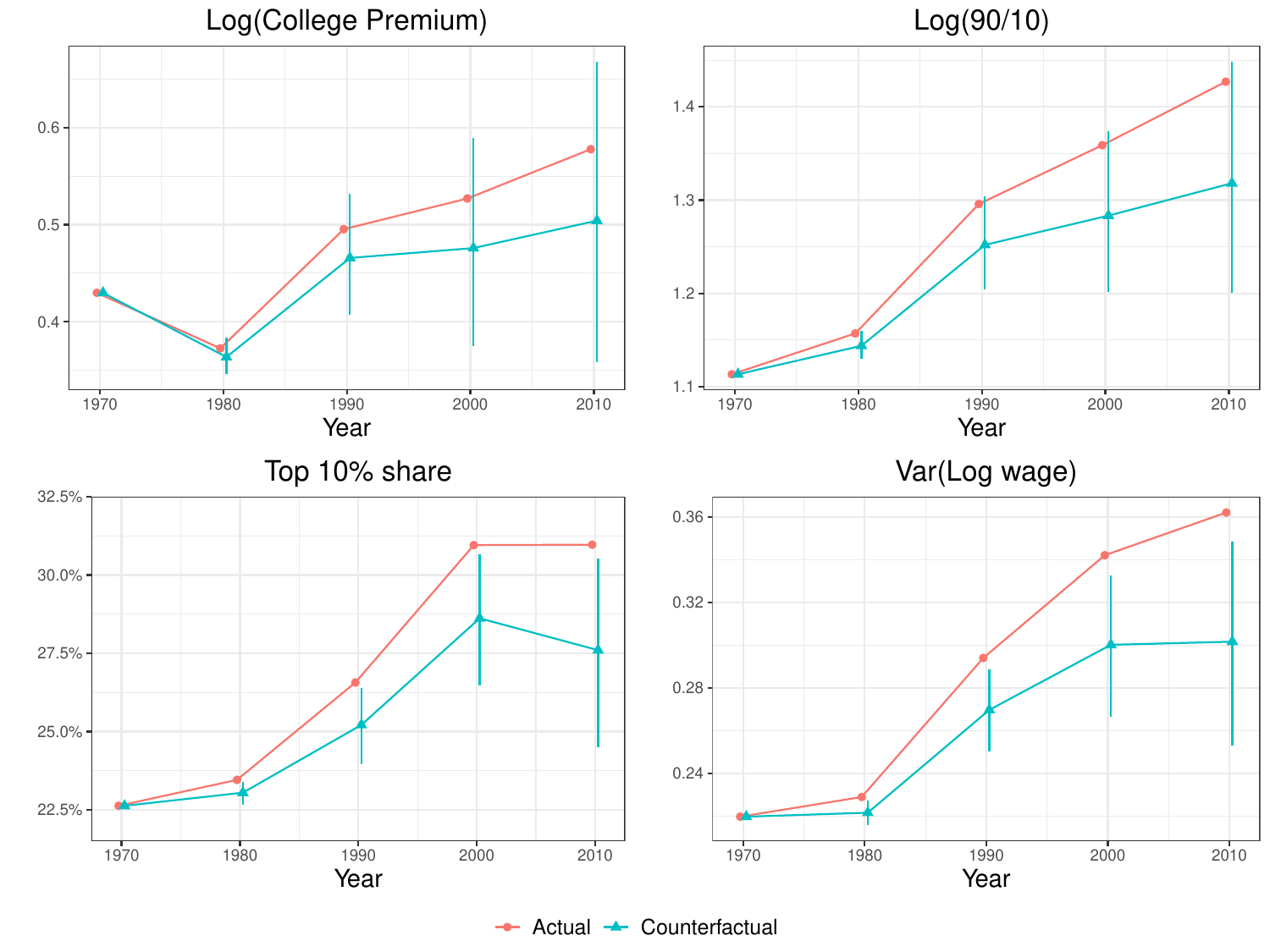}\medskip{}
\par\end{centering}
{\footnotesize\emph{Notes:}}{\footnotesize{} Each panel presents a measure
of actual within (state-industry) cell inequality (averaged across
cells with beginning-of-decade employment weights) by census year
from 1970–2010, as well as its counterfactual value had the rate of
new unionization stayed at the 1960s level. The counterfactual is
calculated by adding to the actual measure the product of the treatment
effect estimated in Table \ref{table:unions_inequilty} and the cumulative
shortfall in the rate of new unionization between the actual and counterfactual
scenarios (presented in the first panel of Figure \ref{fig:counterfactual}).
Confidence intervals for the counterfactuals are based on the 95\%
robust bias-correct confidence intervals for the corresponding treatment
effects.}{\footnotesize\par}

{\footnotesize\emph{Sources: }}{\footnotesize Decennial Census, NLRB
unionization data; authors' calculations.}{\footnotesize\par}
\end{figure}

\begin{figure}[H]
\caption{Distribution of Total Eligible Voters by Union Vote Count Margin\protect\label{fig:votes_diff}}

\begin{centering}
\medskip{}
\par\end{centering}
\begin{centering}
\includegraphics[width=0.75\columnwidth]{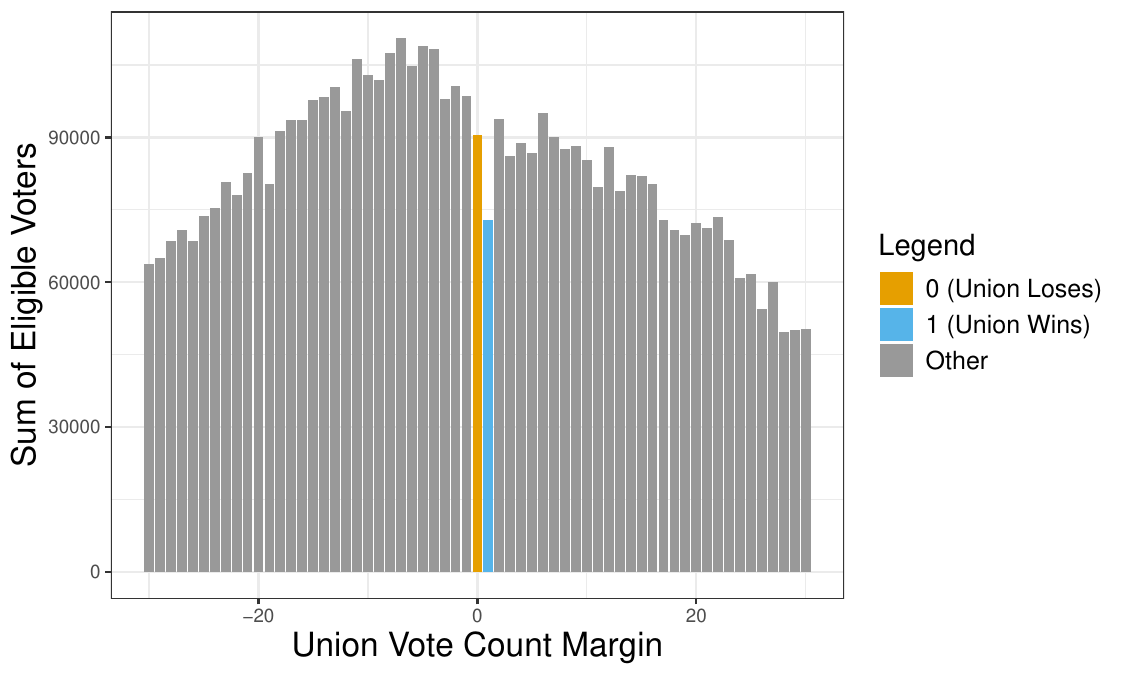}\medskip{}
\par\end{centering}
{\footnotesize\emph{Notes: }}{\footnotesize For each margin between
-30 and 30 votes, the histogram shows the total number of eligible
workers in workplaces where unionization elections ended with this
vote margin (i.e., the gap between the votes in favor and against
the union). Tied elections and elections in which the union wins by
one vote are highlighted. The time rage is 1961-2009.}{\footnotesize\par}

{\footnotesize\emph{Sources:}}{\footnotesize{} NLRB unionization data;
authors' calculations.}{\footnotesize\par}
\end{figure}

\end{document}